\documentclass[11pt,a4paper]{article}
\usepackage[bookmarks=true,colorlinks=true,citecolor=blue,linkcolor=red,urlcolor=magenta]{hyperref}
\usepackage[affil-sl,auth-lg]{authblk}
\usepackage{subfigure}
\usepackage{graphicx}
\graphicspath{{figures/}}
\usepackage{mathrsfs}
\usepackage{multirow}
\usepackage{url}
\usepackage{amsmath}

\numberwithin{equation}{section}

\title{\Huge \bf On the role of longitudinal momenta in high energy hadron-hadron scattering}

\author[1]{Igor~Sharf}
\author[2,1]{Andrii~Tykhonov}
\author[1]{Grygorii~Sokhrannyi}
\author[2,1]{Maksym~Deliyergiyev}
\author[1]{Natalia~Podolyan}
\author[1,3]{Vitaliy~Rusov \thanks{\it E-mail: siiis@te.net.ua}}

\affil[1]{Department of Theoretical and Experimental Nuclear Physics, Odessa National Polytechnic University, Shevchenko av. 1, 65044 Odessa, Ukraine}

\affil[2]{Department of Experimental Particle Physics, Jozef Stefan Institute, Jamova 39, SI-1000 Ljubljana, Slovenia}

\affil[3]{Department of Mathematics, Bielefeld University, Universitatsstrasse 25, 33615 Bielefeld, Germany}

\begin{document}







\maketitle

\begin{abstract}
We demonstrate a new method for the calculation of inelastic scattering cross-section, which in contrary to the Regge-based methods takes into account the energy momentum conservation law. By virtue of this method it was shown that the main contribution to integral expressing inelastic scattering cross-sections comes not from the multi-Regge domain. In particular this leads to the fact that accounting of longitudinal momenta contribution to virtualities is sufficient and results in the new mechanism of cross-section growth. The necessity of taking into account the large number of interference contributions is shown and the approximate method for this purpose is developed. By considering the interference contributions from a single fitting constant achieved a qualitative agreement of the total and inelastic cross sections with experimental data.
\end{abstract}


\section{Introduction}
\begin{figure}
  \centering
  \subfigure[]{
  \includegraphics[scale=0.6]{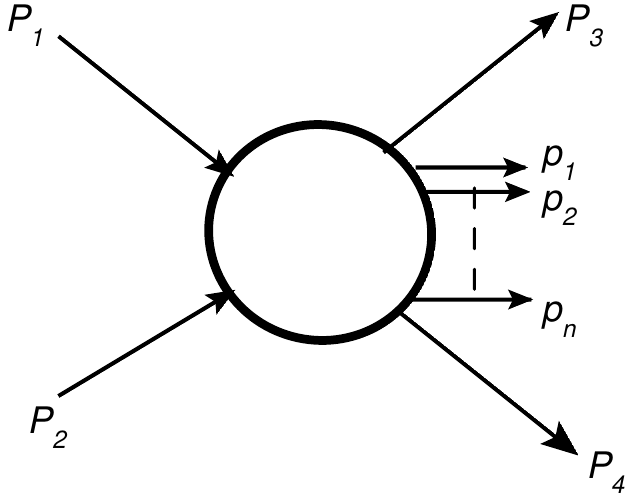} 
   \label{fig:Fig_1a} 
 }    
  \subfigure[]{
  \includegraphics[scale=0.6]{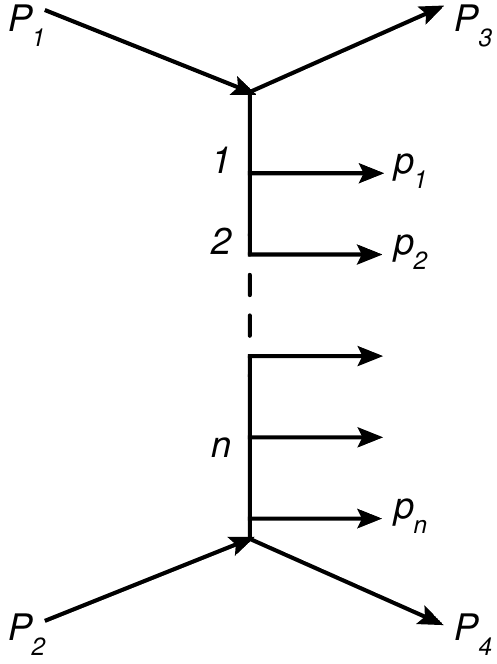} 
   \label{fig:Fig_2} 
 } 
  \caption{ \ref{fig:Fig_1a} - a general view of an inelastic scattering diagram, \ref{fig:Fig_2} - an elementary inelastic scattering diagram in the multi-peripheral model (``the comb")}
  \label{fig:Fig_1}
\end{figure}

Despite the fact the multi-peripheral model \cite{springerlink:10.1007/BF02781901} has been used for description of hadron scattering for a long time, formal difficulties, which appear in calculating of inelastic scattering cross-section, in our opinion, are not overcame until now. These difficulties are caused by the fact that inelastic scattering cross-section with production of a given number of secondary particles in the finite state (Fig.\ref{fig:Fig_1}) is described by the multidimensional integral of scattering amplitude squared modulus over the phase volume of finite state:
\begin{eqnarray}
 {\sigma _n} = \frac{1}{{4n!\sqrt {{{({P_1}{P_2})}^2} - {{({M_1}{M_2})}^2}} }}\int {\frac{{d{{\vec P}_3}}}{{2{P_{30}}{{(2\pi )}^3}}}} \frac{{d{{\vec P}_4}}}{{2{P_{40}}{{(2\pi )}^3}}}\prod\limits_{k = 1}^n {\frac{{d{{\vec p}_k}}}{{2{p_{0k}}{{(2\pi )}^3}}}   } \nonumber\\
 \times {\left| {T(n,{p_1},{p_2},...,{p_n},{P_1},{P_2},{P_3},{P_4})} \right|^2} {\delta ^{\left( 4 \right)}}\left (  {{P_3} + {P_4} + \sum\limits_{k = 1}^n {{p_k}}  - {P_1} - {P_2}} \right) \label{eq:eq_part1_1}  
\end{eqnarray}
where $M_{1}$ and $M_{2}$ are the masses of colliding particles with four-momentums $P_{1}$ and $P_{2}$; 
\\$T(n, p_1, p_2, \ldots, p_n, P_{1}, P_{2}, P_3, P_4)$ is scattering amplitude corresponding to inelastic process shown in Fig.\ref{fig:Fig_1};
$\delta^{(4)}$ is a four-dimensional delta function describing the conservation laws of energy and three momentum components in this process. Here it is also assumed that particles with four-momentums $P_3$ and $P_4$ are the same sorts as $P_1$ and $P_2$, respectively, and $n$ secondary particles with four-momentums $p_1, p_2,\ldots, p_n$ are identical.

Since scattering amplitude is, in general, not a product of functions of some variables, and also due to the complexity of integration domain, the multidimensional integral in Eq.\ref{eq:eq_part1_1} is not a product of smaller-dimensional ones. In considered inelastic process this domain of phase space of finite state particles is determined by the energy-momentum conservation law. As a result, the integration limits for one variable depend on the values of others. In order to overcome these difficulties one usually deals with the multi-Regge kinematics \cite{Collins:111502,Nikitin:113716,bfkl_1976,Lipatov:2008,Byckling:100542,Ter-Martirosyan,levin_2,KozlovNSU_2007}. The disadvantages of this approach we discuss in details in Section \ref{sec:discussion}. 

The ultimate goal of this paper is to develop a new approach which is based on well-known Laplace's method \cite{DeBruijn:225131} for the case, when scattering amplitude is set of multi-peripheral diagrams Fig.\ref{fig:Fig_2} within the framework of the perturbation theory. The essence of this method consists in finding the constrained maximum point of scattering amplitude squared modulus in Eq.\ref{eq:eq_part1_1} under four conditions imposed by $\delta^{(4)}$-function of Eq.\ref{eq:eq_part1_1}. Then, expressing the scattering amplitude squared modulus as $|T|^2=exp(ln(|T|^2))$, it is possible to expand the exponent of the exponential function in Taylor series about a point of the the constrained maximum, coming to nothing more than quadratic items. After that we obtain Gaussian integral, whose calculation is reduced to computation of matrix determinant of second derivatives with respect to $ln(|T|^2)$.  

The implementation of the aforementioned approach enables us to calculate the inelastic scattering cross section without the use of peculiar  approximations of multi-Regge kinematics. As the result, we can draw the conclusion that the longitudinal momenta, which are usually neglected, play the significant role in the behavior of hadron inelastic scattering cross-section with energy $\sqrt s$ growth.

At the same time, according to the Wick's theorem, the scattering amplitude is the sum of diagrams of all possible orders of external lines attaching to the diagram in Fig.\ref{fig:Fig_2} (interference terms). In order to take into account these interference contributions one needs to modify the aforementioned  procedure in the way, which will be outlined further in the paper.

As will be shown, the value of inelastic scattering amplitude increases at the maximum point due to decrease of virtualities, which correspond to internal lines of ``combs". Question, is this increase responsible for the total scattering cross-section growth, which is observed in the experiment, is also the subject of presented paper.

The paper is structured as follows. In Section \ref{sec:symmetry_properties} we set ourselves the problem of finding the constrained maximum point of multi-peripheral scattering amplitude under the condition of energy-momentum conservation. Furthermore we consider some simplifications, based on the scattering amplitude symmetry properties. Afterwards, the analytical solution of the constrained maximum problem is given in Section \ref{sec:analytical-maximum}. The motivation of examination of the interference contributions at the calculation of inelastic scattering cross-section is discussed in Section \ref{sec:interference-contrib}. The calculation of inelastic scattering cross section, taking into account the contribution of all the interference terms, with the application of Laplace method at the relatively small number of final state particles is presented in Section \ref{sec:Laplace-method}. The approximate method for calculation of inelastic scattering cross section at any number of final state particles is developed in Section \ref{sec:analytical_interference}. Summary and conclusions are given in Section \ref{sec:discussion}.

\section{The constrained maximum problem for the scattering amplitude in multi-peripheral model.}
\label{sec:symmetry_properties}
First, before we pass to the constrained maximization problem, let's examine the following simplifications. According to Feynman diagram technique, the expression for scattering amplitude, which corresponds to a diagram in Fig.\ref{fig:Fig_2}, has a form:
\begin{eqnarray}
 T\left( {n,{P_3},{P_4},{p_1},{p_2},...,{p_n},{P_1},{P_2}} \right) = {\left( { - ig{{\left( {2\pi } \right)}^4}} \right)^2} {\left( { - i\lambda {{\left( {2\pi } \right)}^4}} \right)^n}{\left( {\frac{{ - i}}{{{{\left( {2\pi } \right)}^4}}}} \right)^{n + 1}}
\nonumber\\ 
\times A(n,{P_3},{P_4},{p_1},{p_2},...,{p_n},{P_1},{P_2})
  \label{eq:eq_part1_2}
\end{eqnarray} 
with
\begin{eqnarray}
\begin{array}{l}
\mbox{\fontsize{11}{12}\selectfont $   A\left( {n,{P_3},{P_4},{p_1},{p_2}, \cdots ,{p_n},{P_1},{P_2}} \right) = \frac{1}{{{m^2} - {{\left( {{P_1} - {P_3}} \right)}^2} - i\varepsilon }}\frac{1}{{{m^2} - {{\left( {{P_1} - {P_3} - {p_1}} \right)}^2} - i\varepsilon }}\frac{1}{{{m^2} - {{\left( {{P_1} - {P_3} - {p_1} - {p_2}} \right)}^2} - i\varepsilon }} $}\\  \\
\mbox{\fontsize{11}{12}\selectfont $   \cdots  \cdots  \cdots \frac{1}{{{m^2} - {{\left( {{P_1} - {P_3} - {p_1} - {p_2} -  \cdots  - {p_{n - 1}}} \right)}^2} - i\varepsilon }}\frac{1}{{{m^2} - {{\left( {{P_1} - {P_3} - {p_1} - {p_2} -  \cdots  - {p_{n - 1}} - {p_n}} \right)}^2} - i\varepsilon }}$} \\ 
 \end{array}
\label{eq:eq_part1_3}
\end{eqnarray} 
where $g$ is a coupling constant in the outermost vertices of the diagram; $\lambda$ is a coupling constant in all other vertices; $m$ is the mass of virtual particle field and also secondary particles. As in the original version of multi-peripheral model \cite{springerlink:10.1007/BF02781901}, pions are taken both as virtual and secondary particles. It was assumed that the particle masses with four-momentums $P_1$, $P_2$, $P_3$, $P_4$ are equal, i.e., $M_1=M_2=M_3=M_4=M$, where $M$ is the proton mass. Note that the concrete choice of numerical value of mass $M$ has no importance for the results presented in the paper.

As it was noted in \cite{Byckling:100542}, for the most ratios of particle masses in the initial and final state, the 
virtual particles four-momentums on the diagram of Fig.\ref{fig:Fig_2} are space-like, i.e., their scalar squares are negative in Minkowski space. The negativity of scalar squares of virtual four-momentums at the given mass configuration $M_1=M_2=M_3=M_4=M$ is easy to prove (see \cite{part1}, page 30).

Since the virtual particle squared four-momentums $(P_1-P_3)^2$, $(P_1-P_3-p_1)^2$, $(P_1-P_3-p_1-p_2-\ldots-p_{n-1}-p_n)^2$ are negative at the physical values of four-momentums of final state particles, denominators in Eq\ref{eq:eq_part1_1} are equal to zero nowhere in the physical region. Therefore it is possible to reduce $i\epsilon$ to zero before all calculations.

Due to negativity of virtual particle squared four-momentums the magnitude Eq.\ref{eq:eq_part1_3} is real and positive. Therefore the search of the constrained maximum point of scattering amplitude squared modulus reduces to the search of the constrained maximum point of function Eq.\ref{eq:eq_part1_3}. Hereinafter we'll refer the expression Eq.\ref{eq:eq_part1_3} as well as Eq.\ref{eq:eq_part1_2}, which differs from it by constant factor, as scattering amplitude, for short.

Let us examine Eq.\ref{eq:eq_part1_3} in c.m.s. of colliding particles $P_1$ and $P_2$. In such a frame of reference the initial and finite states have some symmetry, which is possible to use for solving the constrained maximum problem. In particular, the consideration of symmetries makes it possible to reduce the search of the constrained maximum of scattering amplitude to the search of the maximum of its restriction on a certain subset of physical process domain shown in Fig.\ref{fig:Fig_2}. This restriction is the function of substantially smaller number of independent variables than the initial amplitude.

For the further discussion of these symmetries and related simplifications, it would be convenient at first to take into account the conservation laws, expressing the scattering amplitude as a function of independent variables only. After decomposition of the three-dimensional particle momentums in c.m.s. frame to components, which are parallel $p_{k\parallel}$ and orthogonal $\vec{k}_{k\perp}$ to collision axis, and lets name them longitudinal and transversal momentums, respectively. 

Energy of each particles in the finite state can be expressed by their momentum using the mass shell conditions, having $n+2$ particles in finite state Fig.\ref{fig:Fig_2}, that give us $3(n+2)$ momentum components of these particles. Since we are looking for a constrained extremum, it is necessary to take into account four relations, which express an energy-momentum conservation law. It will result in the fact that amplitude Eq.\ref{eq:eq_part1_3} can be represented as a function of $3n+2$ independent variables. The first $3n$ variables we choose are longitudinal and transverse components of momentums $\vec{p}_1,\vec{p}_2,\ldots,\vec{p}_n$ of particles produced along the``comb`` in Fig.\ref{fig:Fig_2}. The other two variables are the transverse components of momentum $\vec{P}_{3\perp}$.

If $z$-axis coincides with momentum direction $\vec{P}_1$ in c.m.s. and $x$ and $y$ axes are the coordinate axes in the plane of transverse momentums, the conservation laws look like
\begin{eqnarray}%
\begin{array}{l}
 {P_{30}} + {P_{40}} = \sqrt s  - \left( {{p_{10}} + {p_{20}} + ... + {p_{n0}}} \right) \\
 {P_{3\parallel }} + {P_{4\parallel }} =  - \left( {{p_{1\parallel }} + {p_{2\parallel }} + ... + {p_{\parallel }}} \right) \\
 {P_{4 \bot x}} =  - \left( {{p_{1 \bot x}} + {p_{2 \bot x}} + ... + {p_{n \bot x}} + {P_{3 \bot x}}} \right) \\
 {P_{4 \bot y}} =  - \left( {{p_{1 \bot y}} + {p_{2 \bot y}} + ... + {p_{n \bot y}} + {P_{4 \bot y}}} \right) \\ 
\end{array}
\label{eq:eq_part1_4}
\end{eqnarray}
where
\begin{eqnarray}%
\begin{array}{l}
 s = {\left( {{P_1} + {P_2}} \right)^2} \\
 {p_{k0}} = \sqrt {{m^2} + {{\left( {{p_{k\parallel }}} \right)}^2} + {{\left( {{p_{k \bot x}}} \right)}^2} + {{\left( {{p_{k \bot y}}} \right)}^2}} \\
 {P_{30}} = \sqrt {{M^2} + {{\left( {{P_{3\parallel }}} \right)}^2} + {{\left( {{P_{3 \bot x}}} \right)}^2} + {{\left( {{P_{3 \bot y}}} \right)}^2}} \\
 {P_{40}} = \sqrt {{M^2} + {{\left( {{P_{4\parallel }}} \right)}^2} + {{\left( {{P_{4 \bot x}}} \right)}^2} + {{\left( {{P_{4 \bot y}}} \right)}^2}} \\ 
\end{array} 
\label{eq:eq_part1_5}
\end{eqnarray}
Let's enter the following denotations:
\begin{equation}
\begin{array}{l}
E_p \equiv \sqrt s - \left( p_{10} + p_{20} + \ldots p_{n0}  \right)  \\
P_{\parallel p} \equiv p_{1\parallel} + p_{2\parallel} + \ldots + p_{n\parallel} \\
P_{px} \equiv p_{1\perp x} + p_{2\perp x} + \ldots + p_{n\perp x}  \\
P_{py} \equiv p_{1\perp y} + p_{2\perp y} + \ldots + p_{n\perp y}  \\
\end{array}
\label{eq:eq_part1_6}
\end{equation} 

Then, solving the system Eq.\ref{eq:eq_part1_4} for the unknown $P_{3\parallel}, P_{4\parallel}, P_{4{\perp}x}, P_{4{\perp}y}$ gives:
\begin{eqnarray}
\begin{array}{l}
\mbox{\fontsize{12}{12}\selectfont $  {P_{3\parallel }} = \frac{{E_p^2 - {P_{\parallel p}}^2 - \vec P_{p \bot }^2 - 2\left( {{{\vec P}_{p \bot }} \cdot {{\vec P}_{3 \bot }}} \right)}}{{2\left( {E_p^2 - {P_{\parallel p}}^2} \right)}}\left( { - {P_{\parallel p}} \pm {E_p}\sqrt {1 - \frac{{4\left( {{M^2} + {{\left( {{{\vec P}_{3 \bot }}} \right)}^2}} \right)\left( {E_p^2 - {P_{\parallel p}}^2} \right)}}{{{{\left( {E_p^2 - {P_{\parallel p}}^2 - \vec P_{p \bot }^2 - 2\left( {{{\vec P}_{p \bot }} \cdot {{\vec P}_{3 \bot }}} \right)} \right)}^2}}}} } \right)  $ }\\ \\
\mbox{\fontsize{12}{12}\selectfont $ {P_{4\parallel }} = \frac{{E_p^2 - {P_{\parallel p}}^2 - \vec P_{p \bot }^2 - 2\left( {{{\vec P}_{p \bot }} \cdot {{\vec P}_{3 \bot }}} \right)}}{{2\left( {E_p^2 - {P_{\parallel p}}^2} \right)}}\left( { - {P_{\parallel p}} \mp {E_p}\sqrt {1 - \frac{{4\left( {{M^2} + {{\left( {{{\vec P}_{3 \bot }}} \right)}^2}} \right)\left( {E_p^2 - {P_{\parallel p}}^2} \right)}}{{{{\left( {E_p^2 - {P_{\parallel p}}^2 - \vec P_{p \bot }^2 - 2\left( {{{\vec P}_{p \bot }} \cdot {{\vec P}_{3 \bot }}} \right)} \right)}^2}}}} } \right)  $ }\\ 
 \end{array}
\label{eq:eq_part1_7}
\end{eqnarray}
where $\vec{P}_{p{\perp}}=(P_{px},P_{py})$

Substituting expression for $P_{3\parallel}$ (Eq.\ref{eq:eq_part1_7}) into Eq.\ref{eq:eq_part1_5} and resulting $P_{30}$ into Eq.\ref{eq:eq_part1_3} give us the scattering amplitude as a function of independent variables only, for which the conversation laws of all components of energy-momentum four-vector are taken into account. Below, referring to Eq.\ref{eq:eq_part1_3}, we assume that these substitutions have already been done. Taking into consideration this fact, we designate the amplitude Eq.\ref{eq:eq_part1_3} as
\begin{eqnarray}%
 A\left( {n,{{\vec P}_{3 \bot }},{{\vec p}_{1 \bot }},{{\vec p}_{2 \bot }},...,{{\vec p}_{n \bot }},{p_{1\parallel }},{p_{2\parallel }},...,{p_{n\parallel }}} \right) \label{eq:eq_part1_8}
\end{eqnarray}%
numerating only independent variables in the argument list.

As it was shown in \cite{part1}, for the further search of the constrained maximum we can limit ourselves to reduction of the scattering amplitude on a subset of the values of its independent arguments, which corresponds to zero values of transverse momentums of all particles in the final state. This reduction is a function of the longitudinal components of momentum $p_{1\parallel},p_{2\parallel},\ldots,p_{n\parallel}$, which we designate as $A_{\parallel}(n,p_{1\parallel},p_{2\parallel},\ldots,p_{n\parallel})$. Then from Eq.\ref{eq:eq_part1_3} we get:
\begin{eqnarray}%
\begin{array}{l}
 A{\,_\parallel }\left( {n,{p_{1\parallel }},{p_{2\parallel }}, \cdots ,{p_{n\parallel }}} \right) = {\left( {{m^2} - {{\left( {{P_{10}} - {P_{30}}} \right)}^2} + {{\left( {{P_{1\parallel }} - {P_{3\parallel }}} \right)}^2}} \right)^{ - 1}} \times  \\ 
  \times \prod\limits_{l = 1}^n {{{\left( {{m^2} - {{\left( {{P_{10}} - {P_{30}} - \sum\limits_{k = 1}^l {{p_{k0}}} } \right)}^2} + {{\left( {{P_{1\parallel }} - {P_{3\parallel }} - \sum\limits_{k = 1}^l {{p_{k\parallel }}} } \right)}^2}} \right)}^{ - 1}}} \\ 
 \end{array}
\label{eq:eq_part1_10}
\end{eqnarray}
where $p_{k0}=\sqrt{m^2+(p_{k\parallel})^2}$, $P_{30}=\sqrt{M^2+(P_{3\parallel})^2}$.

At the same time, assuming that all transverse momentums equal to zero, we have from Eq.\ref{eq:eq_part1_7}:
\begin{eqnarray}%
\begin{array}{l}
 {P_{3\parallel }} = \frac{1}{2}{\left({P_{\parallel p}} \pm {E_p}\sqrt {1 - \frac{{4{M^2}}}{{{{\left( {{E_p}} \right)}^2} - {{\left( {{P_{\parallel p}}} \right)}^2}}}} \right)} \\
{P_{4\parallel }} = \frac{1}{2}{\left({P_{\parallel p}} \mp {E_p}\sqrt {1 - \frac{{4{M^2}}}{{{{\left( {{E_p}} \right)}^2} - {{\left( {{P_{\parallel p}}} \right)}^2}}}} \right)} \\
\end{array}
\label{eq:eq_part1_11}
\end{eqnarray}

Moreover, we have in c.m.s., $P_{10}={\sqrt{s} / 2}$, $P_{1\parallel}=\sqrt{{s/4}-M^2}$, where $s$ is determined by Eq.\ref{eq:eq_part1_5}.

For the further analysis it is convenient to switch from longitudinal momentums of secondary particles to rapidities $y_k$ defined by following relation:
\begin{eqnarray}%
{p_{k\parallel }} = m \cdot sh\left( {{y_k}} \right),\quad k = 1,\;2,...,n 
\label{eq:eq_part1_12}
\end{eqnarray}

\begin{figure}
  \centering
  \includegraphics[scale=0.7]{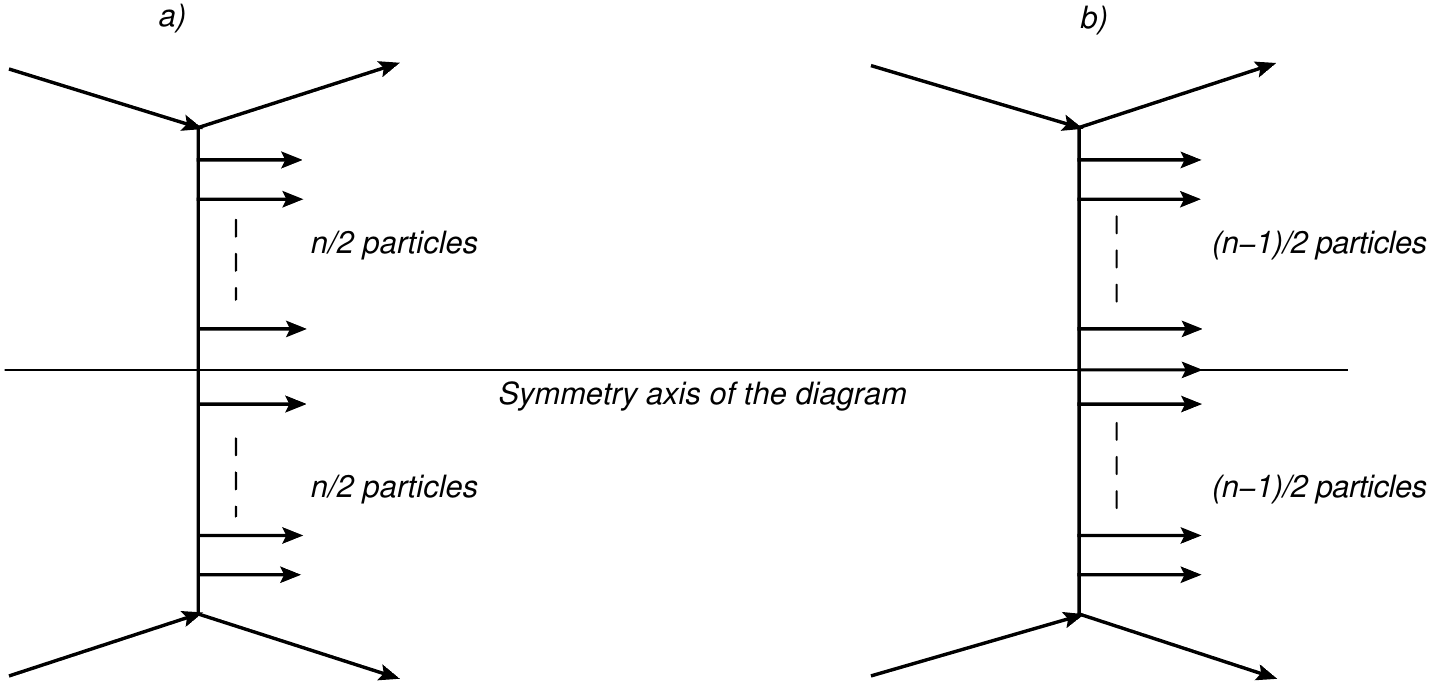}  
  \caption{An elementary inelastic scattering diagram in the multi-peripheral model with even (a) and with odd (b) number of particles on the ``the comb" and it symmetry axis.}
  \label{fig:Fig_3}
\end{figure}

Then function $A_{\parallel}$ can be written as $A_{\parallel}=A_{\parallel}(n,y_1,y_2,\ldots,y_n)$. The initial state in c.m.s is symmetric with respect to changes in positive direction of collision axis. In addition, those type of the diagrams presented in Fig.\ref{fig:Fig_2} have an axis of symmetry shown in Fig.\ref{fig:Fig_3} for the case of even number (left) and for the case of odd number (right) of secondary particles.

It follows that for extremum search \cite{part1} we can examine the further reduction of scattering amplitude which is defined by the following equations
\begin{eqnarray}%
 {A_0}\left( {n,{y_1},{y_2},...,{y_{\frac{n}{2}}}} \right)  = {A_\parallel }\left( {n,{y_1},{y_2},...,{y_{\frac{n}{2}}}, - {y_{\frac{n}{2} - 1}},..., - {y_2}, - {y_1}} \right) 
\label{eq:eq_part1_21}
\end{eqnarray}
and at odd $n$
\begin{eqnarray}%
 {A_0}\left( {n,{y_1},{y_2},...,{y_{\frac{{n - 1}}{2}}}} \right) = {A_\parallel }\left( {n,{y_1},{y_2},...,{y_{\frac{{n - 1}}{2} + 1}} = 0, - {y_{\frac{{n - 1}}{2} + 1}}, - {y_{\frac{{n - 1}}{2}}},..., - {y_1}} \right)  
\label{eq:eq_part1_22}
\end{eqnarray}

Considering the formula Eq.\ref{eq:eq_part1_10} on subset, where reduction $A_0$ is considered, we will have ${P_p} =0$ by virtue of  Eq.\ref{eq:eq_part1_6}. And therefore instead of Eq.\ref{eq:eq_part1_11} we have the following expressions:
\begin{eqnarray}%
 {P_{3\parallel }} = \frac{{{E_p}}}{2}\sqrt {1 - \frac{{4{M^2}}}{{{{\left( {{E_p}} \right)}^2}}}}; \quad \quad {P_{4\parallel }} = \frac{{ - {E_p}}}{2}\sqrt {1 - \frac{{4{M^2}}}{{{{\left( {{E_p}} \right)}^2}}}}
\label{eq:eq_part1_23}
\end{eqnarray}
or
\begin{eqnarray}%
 {P_{3\parallel }} = \frac{{{-E_p}}}{2}\sqrt {1 - \frac{{4{M^2}}}{{{{\left( {{E_p}} \right)}^2}}}}; \quad \quad {P_{4\parallel }} = \frac{{{E_p}}}{2}\sqrt {1 - \frac{{4{M^2}}}{{{{\left( {{E_p}} \right)}^2}}}}
\label{eq:eq_part1_24}
\end{eqnarray}

Let us take into account that if we decompose all scalar square terms in denominators of Eq.\ref{eq:eq_part1_3} they will include the following difference $P_{1\parallel}-P_{3\parallel}=\sqrt{{s/4}-M^2}-P_{3\parallel}$ and negative Eq.\ref{eq:eq_part1_24}, chosen as the $P_{3\parallel}$ in the end will give us greater value in the denominator than in the choice of Eq.\ref{eq:eq_part1_23}. Therefore it is naturally to suppose that the main contribution to cross-section Eq.\ref{eq:eq_part1_1} gives the range of constrained maximum point  determined by the scattering amplitude, where $P_{3\parallel}$ and $P_{4\parallel}$ are given by Eq.\ref{eq:eq_part1_23}, but not by Eq.\ref{eq:eq_part1_24}. Hence, considering the expression Eq.\ref{eq:eq_part1_21} and Eq.\ref{eq:eq_part1_22} for the reduction of the scattering amplitude at zero transverse momentum region, and performing further transformations, we assume that the magnitude $P_{3\parallel}$ and related with it quantity $P_{30}=\sqrt{M^2+(P_{3\parallel})^2}$ are expressed in terms of the longitudinal momenta of secondary particles by the relation Eq.\ref{eq:eq_part1_23}. 

Applying symmetry relation and conversation of energy, it follows that on subset, on which the considered amplitude reduction $A_0$ is defined, the energy corresponding to the line connecting $n/2$ and $n/2+1$ vertices of the diagram in Fig.\ref{fig:Fig_2} is equal to zero in case of even number of particles at any values of independent variables (on which $A_0$ depends). Similarly, for an odd number of particles the energy transferred along the line, which joins $(n-1)/2$ and $(n-1)/2+1$ verticies, is equal to $m/2$. The corresponding proof is given in \cite{part1}.

Taking into account these results, the reduction of $A_0$ for the diagram in Fig.\ref{fig:Fig_2} with even number of particles can be written in the form, which is convenient for the further numerical and analytical calculations:
\begin{eqnarray}%
\begin{array}{l}
 {A_0}\left( {n,{y_1},{y_2},...,{y_{\frac{n}{2}}}} \right) = {\left( {{m^2} - {{\left( {\sum\limits_{k = 1}^{\frac{n}{2}} {{E_k}} } \right)}^2} + {{\left( {{S_M}} \right)}^2}} \right)^{ - 2}} {\left( {{m^2} + {{\left( {{S_M} - \sum\limits_{k = 1}^{\frac{n}{2}} {{p_{k\parallel}}} } \right)}^2}} \right)^{ - 1}}\\
 \times \prod\limits_{j = 2}^{\frac{n}{2}} {{{\left( {{m^2} - {{\left( {\sum\limits_{k = j}^{\frac{n}{2}} {{E_k}} } \right)}^2} + {{\left( {{S_M} - \sum\limits_{k = 1}^{j - 1} {{p_{k\parallel}}} } \right)}^2}} \right)}^{ - 2}}} \\
\end{array}
\label{eq:eq_part1_25}
\end{eqnarray}
where ${S_M} = \sqrt {{s}/{4} - {M^2}}  - {P_{3\parallel }}$, ${E_k} = m\cdot ch\left( {{y_k}} \right)$ and ${p_{k\parallel}}$ defined by Eq.\ref{eq:eq_part1_12}.

The similar expression in case of odd number of particles in comb looks like:
\begin{eqnarray}%
\begin{array}{l}
{A_0}\left( {n,{y_1},{y_2},...,{y_{\frac{n-1}{2}}}} \right)= {\left( {{m^2} - {{\left( {\frac{m}{2} + \sum\limits_{k = 1}^{\frac{{n - 1}}{2}} {{E_k}} } \right)}^2} + {{\left( {{S_M}} \right)}^2}} \right)^{ - 2}}  \\ 
\times{\left( {{m^2} - {{\left( {\frac{m}{2}} \right)}^2} + {{\left( {{S_M} - \sum\limits_{k = 1}^{\frac{{n - 1}}{2}} {{p_{k \parallel}}} } \right)}^2}} \right)^{ - 2}}\\ 
\times \prod\limits_{j = 2}^{\frac{{n - 1}}{2}} {{{\left( {{m^2} - {{\left( {\frac{m}{2} + \sum\limits_{k = j}^{\frac{{n - 1}}{2}} {{E_k}} } \right)}^2} + {{\left( {{S_M} - \sum\limits_{k = 1}^{j - 1} {{p_{k \parallel}}} } \right)}^2}} \right)}^{ - 2}}} 
\end{array}
\label{eq:eq_part1_26}
\end{eqnarray}

As it follows from Eqs.\ref{eq:eq_part1_25}, \ref{eq:eq_part1_26}, it is convenient for the further calculations to make all quantities dimensionless by the mass $m$. In dimensionless form, these relations were used for numerical and analytical solution of the extremum for the reduction of the scattering amplitude $A_0$. The numerical solution of the constrained extremum problem for these expressions is described in detail in \cite{part1}. 

Summary results of the current  section: It has been shown \cite{part1} that multi-peripheral scattering amplitude indeed has a point of constrained maximum under conditions imposed by the energy-momentum conservation law. The aforementioned symmetry relations for the constrained maxim point takes place. The rapidities of final state particles at the constrained maximum point produce an arithmetic progression. The difference of two adjacent rapidities on the "comb" increase with the growth of energy $\sqrt s$. The main conclusion is that the absolute values of scalar squares of the final state particles four-momentums at the constrained maximum point decrease with the growth of energy $\sqrt s$, which leads to the growth of scattering amplitude at the constrained maximum point with the growth of energy.

As will be shown further, this results in a new mechanism of inelastic scattering cross-section growth which has not been taken into account before.

\section{Analytical solution of the constrained extremum problem as the approximation of equal-denominators}
\label{sec:analytical-maximum}

We first consider in more detail the case of even number of particles $n$ in the diagram of Fig.\ref{fig:Fig_2}. The scattering amplitude reduction $A_0(n, y_1,y_2,\ldots,y_{n/2})$ defined by Eq.\ref{eq:eq_part1_25} undimensioned by mass $m$ in this case looks like
\begin{eqnarray}%
\begin{array}{l}
 {A_0}\left( {n,{y_1},{y_2},...,{y_{\frac{n}{2}}}} \right) = {\left( {1 - {{\left( {\sum\limits_{k = 1}^{\frac{n}{2}} {ch\left( {{y_k}} \right)} } \right)}^2} + {{\left( {{P_{13\parallel }}} \right)}^2}} \right)^{ - 2}} {\left( {1 + {{\left( {{P_{13\parallel }} - \sum\limits_{k = 1}^{\frac{n}{2}} {sh\left( {{y_k}} \right)} } \right)}^2}} \right)^{ - 1}} \\
\times \prod\limits_{j = 2}^{\frac{n}{2}} {{{\left( {1 - {{\left( {\sum\limits_{k = j}^{\frac{n}{2}} {ch\left( {{y_k}} \right)} } \right)}^2} + {{\left( {{P_{13\parallel }} - \sum\limits_{k = 1}^{j - 1} {sh\left( {{y_k}} \right)} } \right)}^2}} \right)}^{ - 2}}} \\
\end{array}
\label{eq:eq_part1_33}
\end{eqnarray}
Here $P_{1\parallel}=\sqrt{{s/4}-M^2}$  and instead of $\sqrt{s}$ and $M$ we use their dimensionless by mass $m$ values. Since we search for the constrained extremum, under condition of energy-momentum conservation, it is assumed that $P_3$ is described by Eq.\ref{eq:eq_part1_23}, in which again all values are undimensioned by mass $m$. In particular, taking into account the symmetry relation, normalization and the introduction of rapidity (see Eq.\ref{eq:eq_part1_12}), we obtain for $E_p$ instead Eq.\ref{eq:eq_part1_6}:
\begin{eqnarray}%
 {E_p} = \sqrt s  - 2\sum\limits_{k = 1}^{\frac{n}{2}} {ch\left( {{y_k}} \right)} 
\label{eq:eq_part1_34}
\end{eqnarray}

For further calculations we use the following denotation
\begin{eqnarray}%
 E = \sum\limits_{k = 1}^{\frac{n}{2}} {ch\left( {{y_k}} \right)},\\
\Delta P = {P_{1\parallel }} - {P_{3\parallel }} 
\label{eq:eq_part1_35}
\end{eqnarray}

Note that due to Eqs.\ref{eq:eq_part1_23},\ref{eq:eq_part1_34} and Eq.\ref{eq:eq_part1_35} quantity $\Delta P$ depends on rapidity as the composite function of $E$, which is denoted as $\Delta P(E)$ and the first term of Eq.\ref{eq:eq_part1_33} depends on rapidity only via $E$. 

Instead of looking for the maximum of function $A_0(n, y_1,y_2,\ldots,y_{n/2})$ we can look for the maximum of its logarithm, which we define as $L$:
\begin{eqnarray}%
\begin{array}{l}
 L =  - 2\ln \left( {1 - {{\left( E \right)}^2} + {{\left( {\Delta P\left( E \right)} \right)}^2}} \right)
 - \ln \left( {1 + {{\left( {\Delta P\left( E \right) - \sum\limits_{k = 1}^{\frac{n}{2}} {sh\left( {{y_k}} \right)} } \right)}^2}} \right) \\
 - 2\sum\limits_{j = 2}^{\frac{n}{2}} {\ln \left( {1 - {{\left( {\sum\limits_{k = j}^{\frac{n}{2}} {ch\left( {{y_k}} \right)} } \right)}^2} + {{\left( {\Delta P\left( E \right) - \sum\limits_{k = 1}^{j - 1} {sh\left( {{y_k}} \right)} } \right)}^2}} \right)} 
 
\end{array}
\label{eq:eq_part1_36}
\end{eqnarray}

In addition, we make following denotations:
\begin{eqnarray}%
\begin{array}{l}
 {Z_1} = 1 - {\left( E \right)^2} + {\left( {\Delta P\left( E \right)} \right)^2}\\
 {Z_j} = 1 - {\left( {\sum\limits_{k = j}^{\frac{n}{2}} {ch\left( {{y_k}} \right)} } \right)^2}  + {\left( {\Delta P\left( E \right) - \sum\limits_{k = 1}^{j - 1} {sh\left( {{y_k}} \right)} } \right)^2},j = 1,2,...,\frac{n}{2} \\
 {Z_{\frac{n}{2} + 1}} = 1 + {\left( {\Delta P\left( E \right) - \sum\limits_{k = 1}^{\frac{n}{2}} {sh\left( {{y_k}} \right)} } \right)^2} \\
\end{array}
\label{eq:eq_part1_37}
\end{eqnarray}

Since after taking into account Eq.\ref{eq:eq_part1_7}, all variables of function $A_0(n, y_1,y_2,\ldots,y_{n/2})$ and hence logarithm became independent, then the extreme point can be found under condition that partial derivatives with respect to all variables are equal to zero. The equations for the extreme point problem can be written down in a form:
\begin{eqnarray}%
 \frac{{\partial L}}{{\partial {y_1}}} = \frac{{\partial L}}{{\partial E}}sh\left( {{y_1}} \right) + 4ch\left( {{y_1}} \right)\sum\limits_{j = 2}^{\frac{n}{2}} {\frac{{\Delta P\left( E \right) - \sum\limits_{k = 1}^{j - 1} {sh\left( {{y_k}} \right)} }}{{{Z_j}}}} + 2ch\left( {{y_1}} \right)\frac{{\Delta P\left( E \right) - \sum\limits_{k = 1}^{\frac{n}{2}} {sh\left( {{y_k}} \right)} }}{{{Z_{\frac{n}{2} + 1}}}} = 0 \label{eq:eq_part1_38}
\end{eqnarray}
\begin{eqnarray}%
\begin{array}{l}
\frac{{\partial L}}{{\partial {y_l}}} = \frac{{\partial L}}{{\partial E}}sh\left( {{y_l}} \right) + 4sh\left( {{y_l}} \right)\sum\limits_{j = 2}^l {\frac{{\sum\limits_{k = j}^{\frac{n}{2}} {ch\left( {{y_k}} \right)} }}{{{Z_j}}}} \\
+ 4ch\left( {{y_l}} \right)\sum\limits_{j = l + 1}^{\frac{n}{2}} {\frac{{\Delta P\left( E \right) - \sum\limits_{k = 1}^{j - 1} {sh\left( {{y_k}} \right)} }}{{{Z_j}}}} + 2ch({y_l})\frac{{\Delta P(E) - \sum\limits_{k = 1}^{\frac{n}{2}} {sh({y_k})} }}{{{Z_{\frac{n}{2} + 1}}}} = 0
\end{array}
\label{eq:eq_part1_39}
\end{eqnarray}
\begin{eqnarray}%
 \frac{{\partial L}}{{\partial {y_{\frac{n}{2}}}}} = \frac{{\partial L}}{{\partial E}}sh\left( {{y_{\frac{n}{2}}}} \right) + 4sh\left( {{y_{\frac{n}{2}}}} \right)\sum\limits_{j = 2}^{\frac{n}{2}} {\frac{{\sum\limits_{k = j}^{\frac{n}{2}} {ch\left( {{y_k}} \right)} }}{{{Z_j}}}} + 2ch\left( {{y_{\frac{n}{2}}}} \right)\frac{{\Delta P\left( E \right) - \sum\limits_{k = 1}^{\frac{n}{2}} {sh\left( {{y_k}} \right)} }}{{{Z_{\frac{n}{2} + 1}}}} = 0 
\label{eq:eq_part1_40}
\end{eqnarray}

where $l = 2,\;3,\;...\;\frac{n}{2} - 1$. Eqs.\ref{eq:eq_part1_38}-\ref{eq:eq_part1_40} form the system of equations for the extreme point search. An approximate solution of this system is the purpose of this section. The simplification of this system of equations can be attained in approximation, which we call "the equal-denominators approximation". Detailed justification for this approximation is given in \cite{part1}. Thus, for the further analysis of that system of equations at the maximum point we use approximation, in which all the denominators are equal to each other. We'll define their approximate common value as $Z$, i.e.,
\begin{eqnarray}%
{Z_j} \approx Z,j = 1,2, \cdots ,\frac{n}{2} + 1
\label{eq:eq_part1_41}
\end{eqnarray}

In this case the system of equations for the maximum point takes form
\begin{eqnarray}%
 \frac{Z}{2}\frac{{\partial L}}{{\partial E}} + 2\frac{{ch\left( {{y_1}} \right)}}{{sh\left( {{y_1}} \right)}}\sum\limits_{j = 2}^{\frac{n}{2}} {\left( {\Delta P\left( E \right) - \sum\limits_{k = 1}^{j - 1} {sh\left( {{y_k}} \right)} } \right)} + \frac{{ch\left( {{y_1}} \right)}}{{sh\left( {{y_1}} \right)}}\left( {\Delta P\left( E \right) - \sum\limits_{k = 1}^{\frac{n}{2}} {sh\left( {{y_k}} \right)} } \right) = 0 \label{eq:eq_part1_42}
\end{eqnarray}
\begin{eqnarray}%
\begin{array}{l}
\frac{Z}{2}\frac{{\partial L}}{{\partial E}} + 2\sum\limits_{j = 2}^l {\left( {\sum\limits_{k = j}^{\frac{n}{2}} {ch\left( {{y_k}} \right)} } \right)} + 2\frac{{ch\left( {{y_l}} \right)}}{{sh\left( {{y_l}} \right)}}\sum\limits_{j = l + 1}^{\frac{n}{2}} {\left( {\Delta P\left( E \right) - \sum\limits_{k = 1}^{j - 1} {sh\left( {{y_k}} \right)} } \right)} \\
+ \frac{{ch\left( {{y_l}} \right)}}{{sh\left( {{y_l}} \right)}}\left( {\Delta P\left( E \right) - \sum\limits_{k = 1}^{\frac{n}{2}} {sh\left( {{y_k}} \right)} } \right) = 0
\end{array}
\label{eq:eq_part1_43}
\end{eqnarray}
\begin{eqnarray}%
\frac{Z}{2}\frac{{\partial L}}{{\partial E}} + 2\sum\limits_{j = 2}^{\frac{n}{2}} {\left( {\sum\limits_{k = j}^{\frac{n}{2}} {ch\left( {{y_k}} \right)} } \right)} + \frac{{ch\left( {{y_{\frac{n}{2}}}} \right)}}{{sh\left( {{y_{\frac{n}{2}}}} \right)}}\left( {\Delta P\left( E \right) - \sum\limits_{k = 1}^{\frac{n}{2}} {sh\left( {{y_k}} \right)} } \right) = 0 
\label{eq:eq_part1_44}
\end{eqnarray}

From approximation Eq.\ref{eq:eq_part1_41}, in particular, we obtain $Z_{n/2}\approx Z_{n/2+1}$. Taking into account Eq.\ref{eq:eq_part1_37} will result in identity:
\begin{eqnarray}%
 \Delta P\left( E \right) - \sum\limits_{k = 1}^{{n \mathord{\left/
 {\vphantom {n 2}} \right.
 \kern-\nulldelimiterspace} 2}} {sh\left( {{y_k}} \right)}  = \frac{1}{{2sh\left( {{y_{{n \mathord{\left/
 {\vphantom {n 2}} \right.
 \kern-\nulldelimiterspace} 2}}}} \right)}} 
\label{eq:eq_part1_45}
\end{eqnarray}
Inserting Eq.\ref{eq:eq_part1_45} to the system of Eqs.\ref{eq:eq_part1_42}-\ref{eq:eq_part1_44} we will get
\begin{eqnarray}%
2\frac{{ch\left( {{y_1}} \right)}}{{sh\left( {{y_1}} \right)}}\sum\limits_{j = 2}^{\frac{n}{2}} {\left( {\frac{1}{{2sh\left( {{y_{\frac{n}{2}}}} \right)}} + \sum\limits_{k = j}^{\frac{n}{2}} {sh\left( {{y_k}} \right)} } \right)} + \frac{Z}{2}\frac{{\partial L}}{{\partial E}} + \frac{{ch\left( {{y_1}} \right)}}{{2sh\left( {{y_{\frac{n}{2}}}} \right)sh\left( {{y_1}} \right)}} = 0  \label{eq:eq_part1_46}
\end{eqnarray}
\begin{eqnarray}
\begin{array}{l}
\frac{Z}{2}\frac{{\partial L}}{{\partial E}} + 2\sum\limits_{j = 2}^l {\left( {\sum\limits_{k = j}^{\frac{n}{2}} {ch\left( {{y_k}} \right)} } \right)}  + 2\frac{{ch\left( {{y_l}} \right)}}{{sh\left( {{y_l}} \right)}}\sum\limits_{j = l + 1}^{\frac{n}{2}} {\left( {\frac{1}{{2sh\left( {{y_{\frac{n}{2}}}} \right)}} + \sum\limits_{k = j}^{\frac{n}{2}} {sh\left( {{y_k}} \right)} } \right)}
+ \frac{{ch\left( {{y_l}} \right)}}{{2sh\left( {{y_{\frac{n}{2}}}} \right)sh\left( {{y_l}} \right)}} = 0
\end{array}
\label{eq:eq_part1_47}
\end{eqnarray} 
\begin{eqnarray}
\frac{Z}{2}\frac{{\partial L}}{{\partial E}} + 2\sum\limits_{j = 2}^{\frac{n}{2}} {\left( {\sum\limits_{k = j}^{\frac{n}{2}} {ch\left( {{y_k}} \right)} } \right)}  + \frac{{ch\left( {{y_{\frac{n}{2}}}} \right)}}{{2{{\left( {sh\left( {{y_{\frac{n}{2}}}} \right)} \right)}^2}}} = 0 
\label{eq:eq_part1_48}
\end{eqnarray}

Eqs.\ref{eq:eq_part1_46}-\ref{eq:eq_part1_48} form the system of equations for the search of constrained maximum point of inelastic scattering amplitude in the approximation of equal denominators Eq.\ref{eq:eq_part1_41}. To compute this system we consider Eq.\ref{eq:eq_part1_47} with $l={n\over 2}-1$:
\begin{eqnarray}
 \frac{{ch\left( {{y_{\frac{n}{2} - 1}}} \right)}}{{sh\left( {{y_{\frac{n}{2} - 1}}} \right)}}\left( {\frac{3}{{2sh\left( {{y_{\frac{n}{2}}}} \right)}} + 2sh\left( {{y_{\frac{n}{2}}}} \right)} \right)  + \frac{Z}{2}\frac{{\partial L}}{{\partial E}} + 2\sum\limits_{j = 2}^{\frac{n}{2} - 1} {\left( {\sum\limits_{k = j}^{\frac{n}{2}} {ch\left( {{y_k}} \right)} } \right)}  = 0  
\label{eq:eq_part1_49}
\end{eqnarray}

Subtracting Eq.\ref{eq:eq_part1_48} from Eq.\ref{eq:eq_part1_49} we have
\begin{eqnarray}
 \frac{{ch\left( {{y_{\frac{n}{2} - 1}}} \right)}}{{sh\left( {{y_{\frac{n}{2} - 1}}} \right)}}\left( {\frac{3}{{2sh\left( {{y_{\frac{n}{2}}}} \right)}} + 2sh\left( {{y_{\frac{n}{2}}}} \right)} \right) - 2ch\left( {{y_{\frac{n}{2}}}} \right) - \frac{{ch\left( {{y_{\frac{n}{2}}}} \right)}}{{2{{\left( {sh\left( {{y_{\frac{n}{2}}}} \right)} \right)}^2}}} = 0  
\label{eq:eq_part1_50}
\end{eqnarray}
It follows from this equation that
\begin{eqnarray}
 th\left( {{y_{\frac{n}{2} - 1}}} \right) = \frac{{\frac{3}{{2sh\left( {{y_{\frac{n}{2}}}} \right)}} + 2sh\left( {{y_{\frac{n}{2}}}} \right)}}{{2ch\left( {{y_{\frac{n}{2}}}} \right) + \frac{{ch\left( {{y_{\frac{n}{2}}}} \right)}}{{2{{\left( {sh\left( {{y_{\frac{n}{2}}}} \right)} \right)}^2}}}}} = th\left( {3{y_{\frac{n}{2}}}} \right) \label{eq:eq_part1_51}
\end{eqnarray}
Taking into account that hyperbolic tangent is a monotonous function along the whole real axis, we have from Eq.\ref{eq:eq_part1_51}:
\begin{eqnarray}
 {y_{\frac{n}{2} - 1}} = 3{y_{\frac{n}{2}}} \label{eq:eq_part1_52}
\end{eqnarray}
Note, that this result agrees with the numerical results shown in Fig.5 of Ref.\cite{part1}.  Now prove by induction that
\begin{eqnarray}
 {y_{\frac{n}{2} - k}} = \left( {2k + 1} \right){y_{\frac{n}{2}}},\;\;k = 1,\;2,...,\;\frac{n}{2} - 1 \label{eq:eq_part1_53}
\end{eqnarray}
Eq.\ref{eq:eq_part1_53} is already proved for $k=1$, since it coincides with Eq.\ref{eq:eq_part1_52}. Suppose that this equation is true for $k=1, 2,\ldots, n/2-l-1$ (i.e., at $y_{n/2-1}, y_{n/2-2},\ldots, y_{l+1}$) and prove that it is true for $k=n/2-l$ (i.e., at $y_l$). Subtracting Eq.\ref{eq:eq_part1_48} from Eq.\ref{eq:eq_part1_47} we obtain:
\begin{eqnarray}
\begin{array}{l}
 2\sum\limits_{j = l + 1}^{\frac{n}{2}} {\left( {\sum\limits_{k = j}^{\frac{n}{2}} {ch\left( {{y_k}} \right)} } \right)}  + \frac{{ch\left( {{y_{\frac{n}{2}}}} \right)}}{{2{{\left( {sh\left( {{y_{\frac{n}{2}}}} \right)} \right)}^2}}} - 2\frac{{ch\left( {{y_l}} \right)}}{{sh\left( {{y_l}} \right)}}\sum\limits_{j = l + 1}^{\frac{n}{2}} {\left( {\frac{1}{{2sh\left( {{y_{\frac{n}{2}}}} \right)}} + \sum\limits_{k = j}^{\frac{n}{2}} {sh\left( {{y_k}} \right)} } \right)}\\
 \\
 - \frac{{ch\left( {{y_l}} \right)}}{{2sh\left( {{y_{\frac{n}{2}}}} \right)sh\left( {{y_l}} \right)}} = 0
\end{array}
\label{eq:eq_part1_54}
\end{eqnarray}

Note, that sums $\sum\limits_{k=j}^{n\over 2}ch(y_k)$ and $\sum\limits_{k=j}^{n\over 2}sh(y_k)$ include only those $y_k$ with respect to $l+1\leq j \leq n/2$, which are covered by the assumption of induction. Then, from Eq.\ref{eq:eq_part1_53}  we get $\sum\limits_{k=j}^{n\over 2}sh(y_k)$. It makes possible to calculate the sums included in Eq.\ref{eq:eq_part1_54}, which after transformations has a form:
\begin{eqnarray}
th\left( {{y_l}} \right) = th\left( {\left( {2\left( {\frac{n}{2} - l} \right) + 1} \right){y_{\frac{n}{2}}}} \right) 
\label{eq:eq_part1_55}
\end{eqnarray}

so that
\begin{eqnarray}
\begin{array}{l}
{y_l} = \left( {2\left( {\frac{n}{2} - l} \right) + 1} \right){y_{\frac{n}{2}}} \\
y_{\frac{n}{2} - l} = \left( {2l + 1} \right){y_{\frac{n}{2}}} \\
\end{array}
\label{eq:eq_part1_56}
\end{eqnarray}
i.e., it coincides with proved Eq.\ref{eq:eq_part1_53}.

Thus, for the diagrams with even number of particles we have shown that in the approximation of equal denominators, the rapidities produce an arithmetic progression at the maximum point and the ratios of all rapidities to the minimum rapidity produce the sequence of odd integers.

To determine the values of rapidities, which constrainedly maximize the scattering amplitude, we also have to calculate $y_{n/2}$, in which terms all rapidities are expressed. This can be done using the same equal-denominators approximation Eq.\ref{eq:eq_part1_41} approach.

In view of Eq.\ref{eq:eq_part1_53} computing the sums in Eq.\ref{eq:eq_part1_48} we have:
\begin{eqnarray}
\frac{Z}{2}\frac{{\partial L}}{{\partial E}} + \frac{{ch\left( {\left( {n - 1} \right){y_{\frac{n}{2}}}} \right)}}{{2{{\left( {sh\left( {{y_{\frac{n}{2}}}} \right)} \right)}^2}}} = 0
 \label{eq:eq_part1_57}
\end{eqnarray}
Now we can compute derivative $\partial L/\partial E$ taking into account Eq.\ref{eq:eq_part1_36} for the magnitude $L$:
\begin{eqnarray}
\frac{{\partial L}}{{\partial E}} = \frac{{4E}}{{{Z_1}}} - 4\left( \frac{{\Delta P\left( E \right)}}{{{Z_1}}} + \sum\limits_{j = 2}^{\frac{n}{2}} {\frac{{\Delta P\left( E \right) - \sum\limits_{k = 1}^{j - 1} {sh\left( {{y_k}} \right)} }}{{{Z_j}}}} + \frac{1}{2}\frac{{\Delta P\left( E \right) - \sum\limits_{k = 1}^{\frac{n}{2}} {sh\left( {{y_k}} \right)} }}{{{Z_{\frac{n}{2} + 1}}}}  \right)\frac{{\partial \Delta P\left( E \right)}}{{\partial E}} 
\label{eq:eq_part1_58}
\end{eqnarray}
Using the equal-denominators approximation Eq.\ref{eq:eq_part1_41}, we obtain:
\begin{eqnarray}
\frac{{\partial L}}{{\partial E}} = \frac{{4E}}{{{Z_1}}} - 4\left( \frac{{\Delta P\left( E \right)}}{{{Z_1}}} + \sum\limits_{j = 2}^{\frac{n}{2}} {\frac{{\Delta P\left( E \right) - \sum\limits_{k = 1}^{j - 1} {sh\left( {{y_k}} \right)} }}{{{Z_j}}}} + \frac{1}{2}\frac{{\Delta P\left( E \right) - \sum\limits_{k = 1}^{\frac{n}{2}} {sh\left( {{y_k}} \right)} }}{{{Z_{\frac{n}{2} + 1}}}} \right)\frac{{\partial \Delta P\left( E \right)}}{{\partial E}} 
\label{eq:eq_part1_59}
\end{eqnarray}
After transformations with respect to Eq.\ref{eq:eq_part1_45} we get:
\begin{eqnarray}
 \frac{Z}{2}\frac{{\partial L}}{{\partial E}} = \frac{{sh\left( {n{y_{\frac{n}{2}}}} \right)}}{{sh\left( {{y_{\frac{n}{2}}}} \right)}} + \frac{{sh\left( {\left( {n + 1} \right){y_{\frac{n}{2}}}} \right)}}{{2s{h^2}\left( {{y_{\frac{n}{2}}}} \right)}}\frac{{\partial {P_{3\parallel }}\left( E \right)}}{{\partial E}} = 0 \label{eq:eq_part1_60}
\end{eqnarray}
Substituting Eq.\ref{eq:eq_part1_60} to Eq.\ref{eq:eq_part1_57}, we reduce the resulting equation to the form
\begin{eqnarray}
sh\left( {\left( {n + 1} \right){y_{\frac{n}{2}}}} \right)\frac{{\partial {P_{3\parallel }}\left( E \right)}}{{\partial E}} + ch\left( {\left( {n + 1} \right){y_{\frac{n}{2}}}} \right) = 0 
\label{eq:eq_part1_61}
\end{eqnarray}

Derivative ${\partial P_{3\parallel}(E) / \partial E}$ can be computed from Eq.\ref{eq:eq_part1_23} with allowance for Eq.\ref{eq:eq_part1_34} and Eq.\ref{eq:eq_part1_35}. Then the expression for $P_{3\parallel}$, which is made dimensionaless by the mass, can be written down as:
\begin{eqnarray}
 {P_{3\parallel }} = \sqrt {{{\left( {\frac{{\sqrt s }}{2} - E} \right)}^2} - {M^2}} 
\label{eq:eq_part1_62}
\end{eqnarray}
where it is assumed that $\sqrt{s}$ and particle masses $M$ at the ends of the "comb" are undimensioned by the mass $m$ (pion mass was set to $m$ and proton mass - to $M$).

Taking the derivative of Eq.\ref{eq:eq_part1_62} and substituting it into Eq.\ref{eq:eq_part1_61}, we obtain an equation, which after simple transformations looks like
\begin{eqnarray}
\frac{{\sqrt s }}{2} - E = M \cdot ch\left( {\left( {n + 1} \right){y_{\frac{n}{2}}}} \right) 
\label{eq:eq_part1_63}
\end{eqnarray}
\begin{figure}
  \centering
  \subfigure[]{
  \includegraphics[scale=0.34]{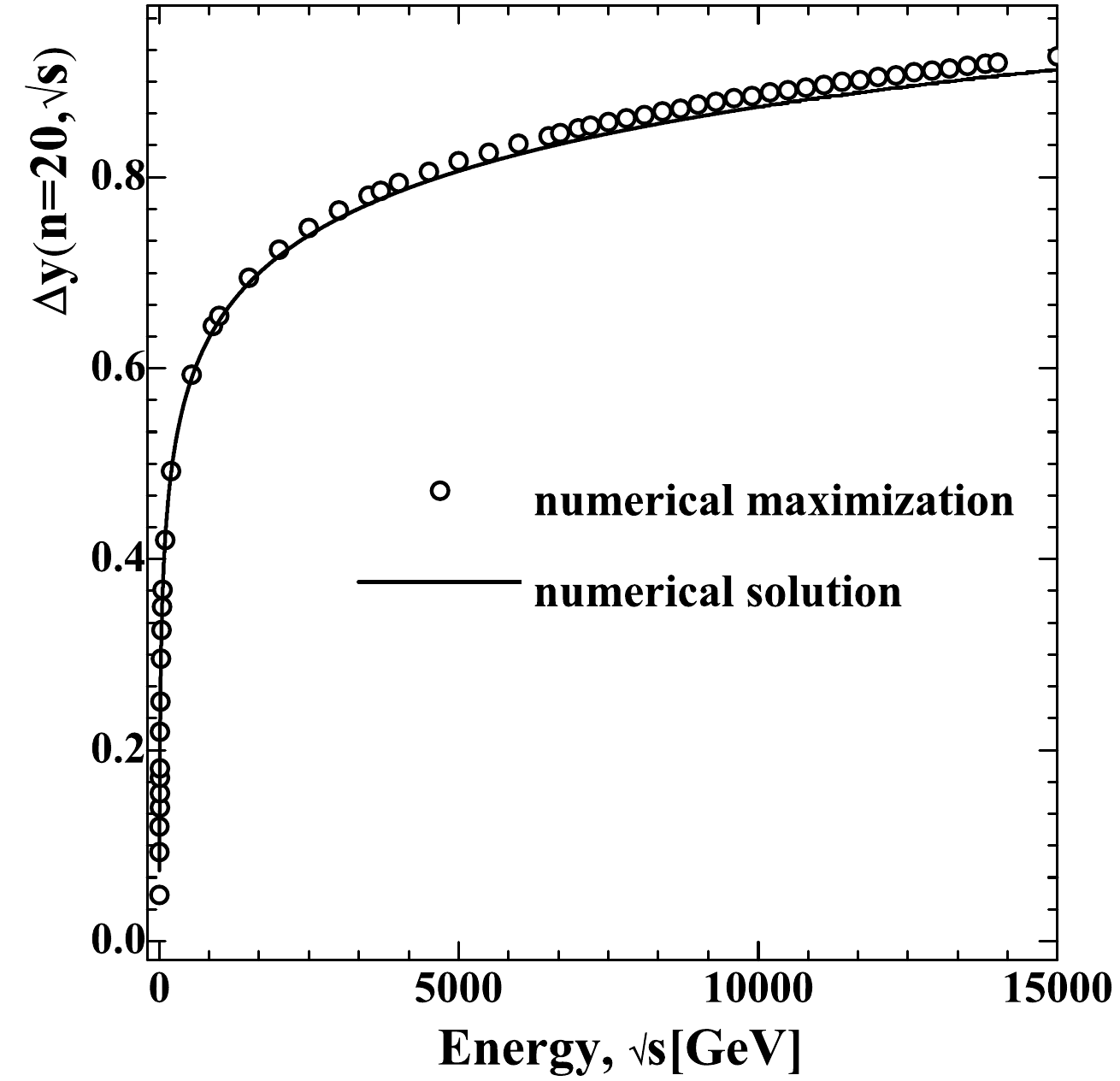} 
  \label{fig:fig_part1_10a} 
  }
  \subfigure[]{
  \includegraphics[scale=0.34]{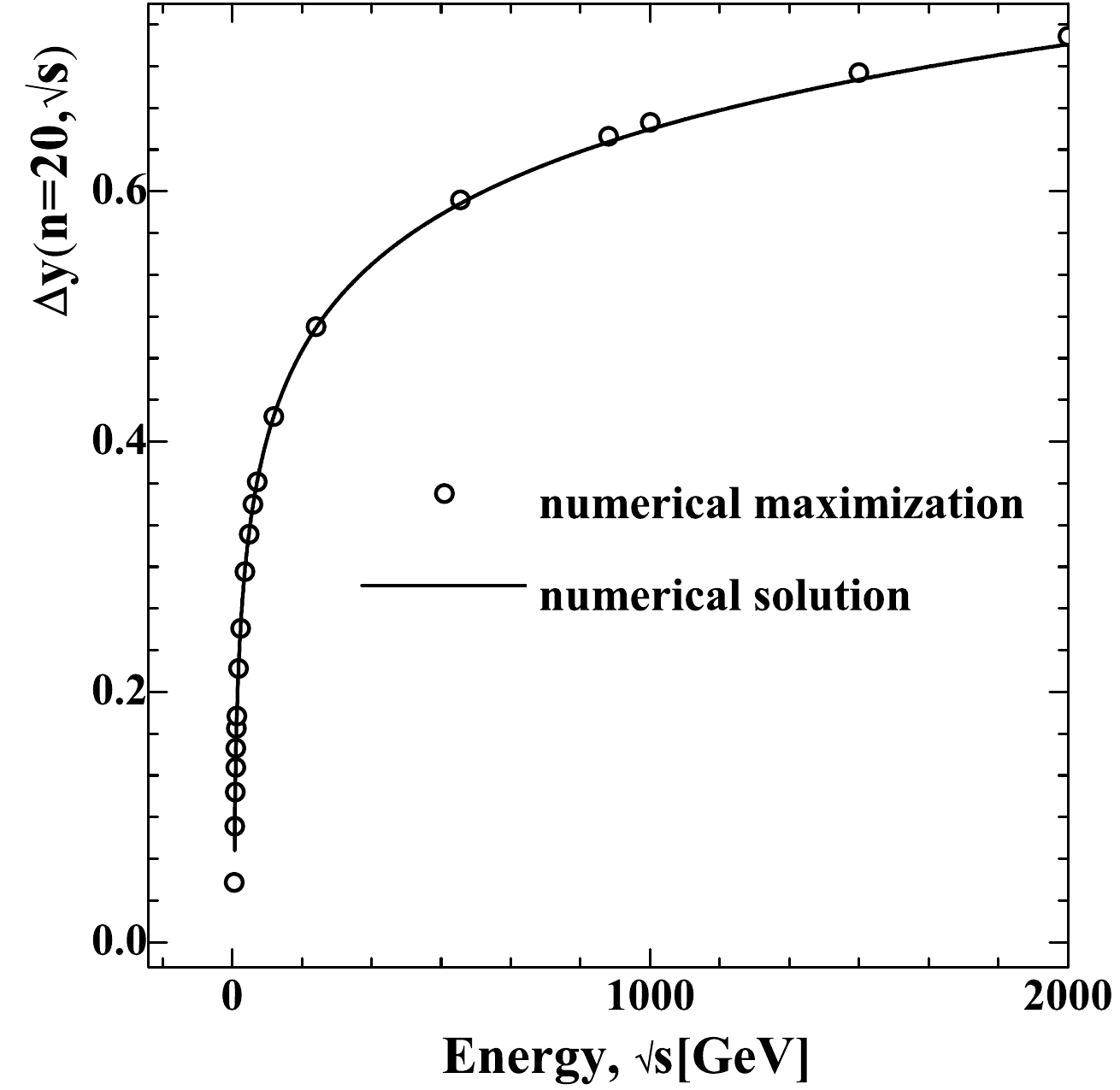}  
    \label{fig:fig_part1_10b} 
  }
  \subfigure[]{
  \includegraphics[scale=0.34]{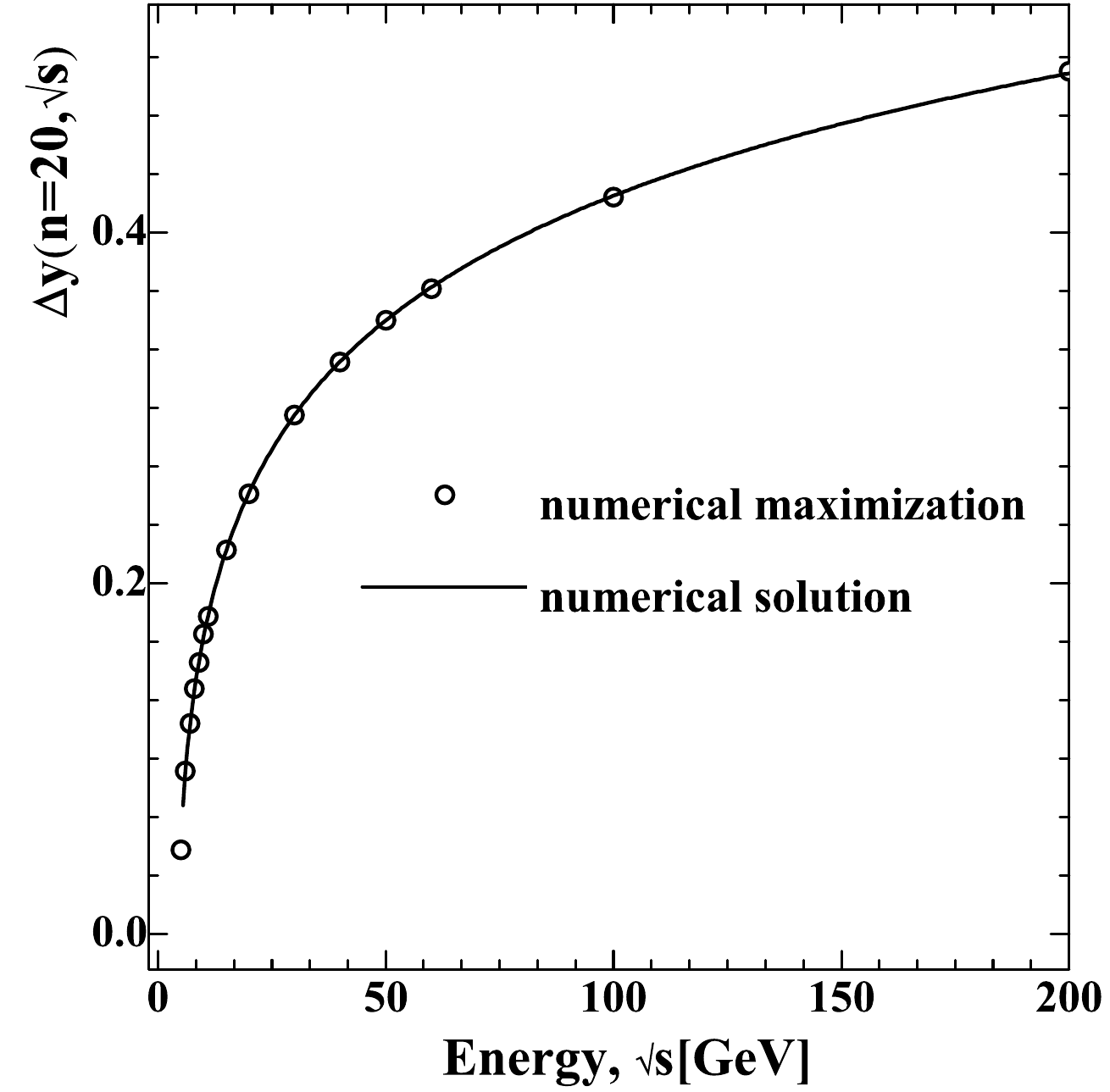} 
  \label{fig:fig_part1_10c} 
  }  
  \caption{Comparison of the results of the numerical solution of Eq.\ref{eq:eq_part1_64} (solid line) with the results of numerical maximization (circles) of the magnitude $\Delta y (n,\sqrt{s})$ at $n/2=10$ for the different energy ranges \cite{part1}, GeV: $5\div16000$ (\ref{fig:fig_part1_10a}); $5\div2000$ (\ref{fig:fig_part1_10b}); $5\div200$ (\ref{fig:fig_part1_10c}). Here, it is taken into account that $\Delta y (n,\sqrt{s})=2y_{n\over 2}$.  The good agreement of results shows the applicability of the approximation of equal denominators Eq.\ref{eq:eq_part1_41} resulting in the Eq.\ref{eq:eq_part1_63}. }
  \label{fig:fig_part1_10}
\end{figure}
Note that the rapidity corresponding to momentum $P_{3\parallel}$ is equal to $(n+1)y_{n/2}$, as it follows from Eq.\ref{eq:eq_part1_62} and Eq.\ref{eq:eq_part1_63}. This expression would be obtained from Eq.\ref{eq:eq_part1_53}, if one accepts $k=n/2$ in it, i.e., arithmetic progression Eq.\ref{eq:eq_part1_53} lengthens by one term. One might consider that rapidities of particles on the "comb" edges at the maximum point "continues" an arithmetic progression formed by the internal particles of the "comb". This fact once again indicates the close relation between the equal denominators approximation Eq.\ref{eq:eq_part1_41} and the arithmetic progression production by rapidities at the maximum point. In other words, the arithmetic progression production by rapidities at the maximum point is the consequence of equal-denominators approximation.

\begin{figure}
  \centering
  \subfigure[]{
  \includegraphics[scale=0.34]{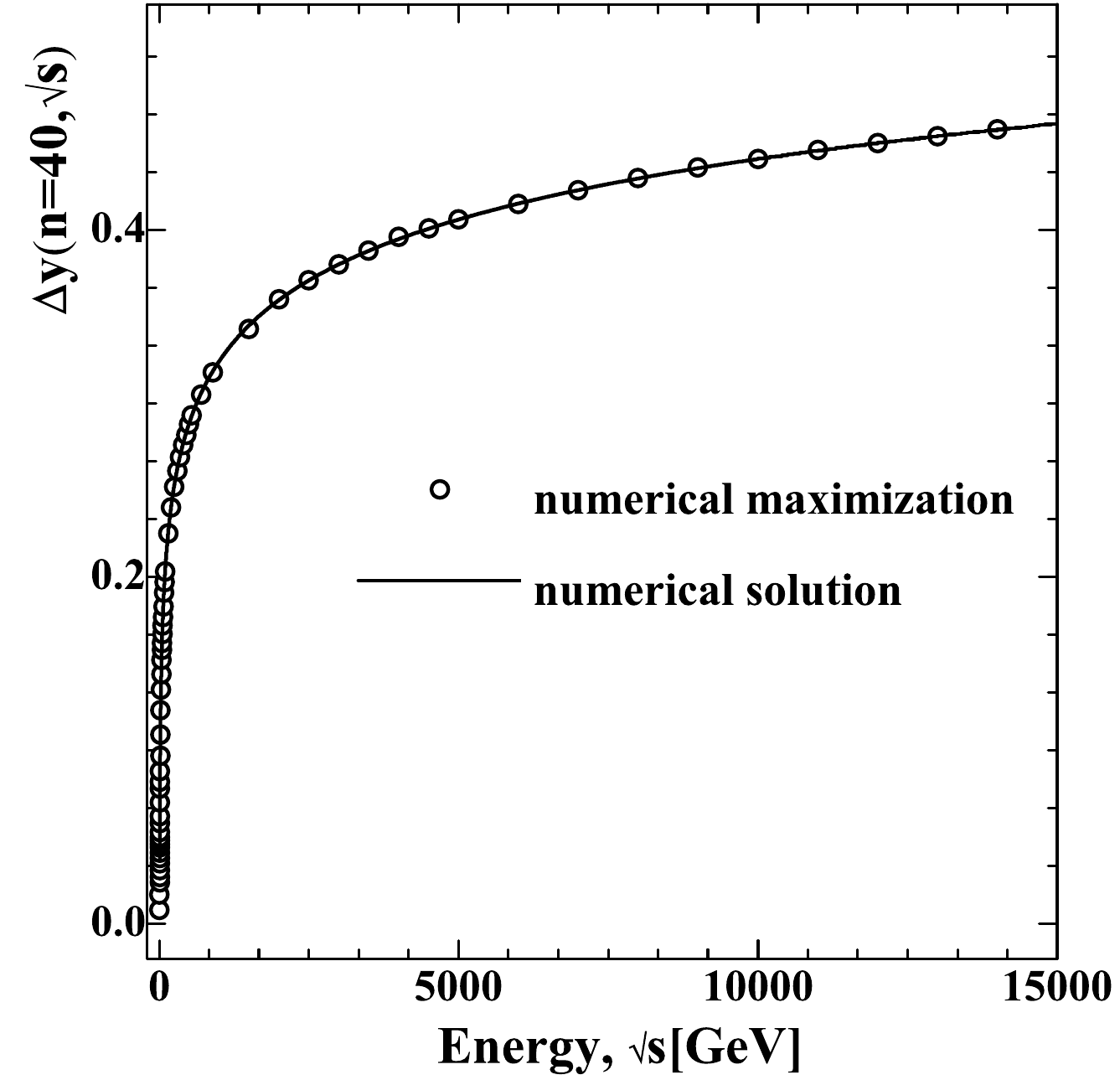} 
  \label{fig:fig_part1_11a} 
  }
  \subfigure[]{
  \includegraphics[scale=0.34]{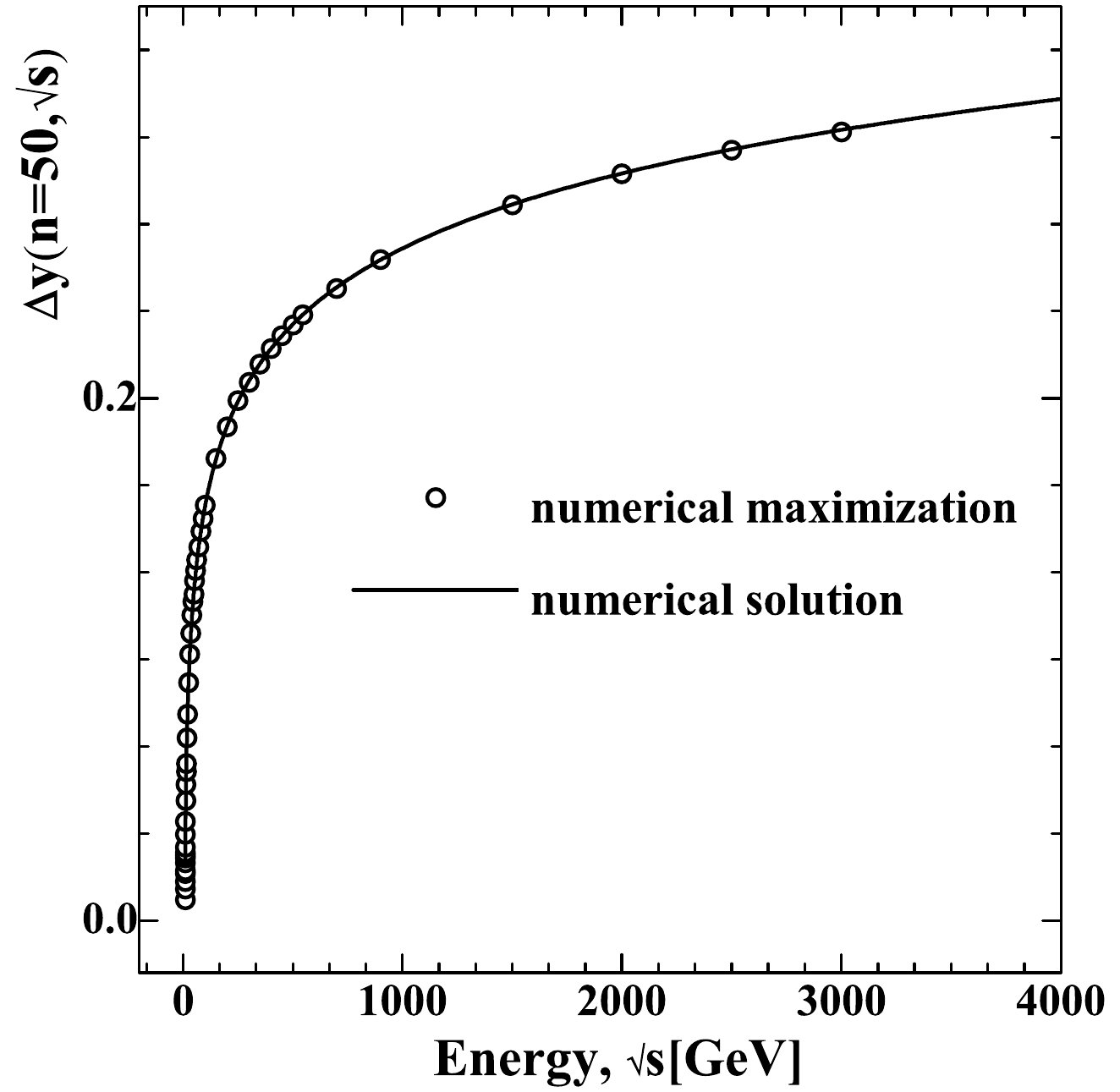}  
    \label{fig:fig_part1_11b} 
  }        
  \caption{Comparison of the results of the numerical solution of Eq.\ref{eq:eq_part1_64}(solid line) with results of numerical maximization (circles) of the magnitude $\Delta y(n, \sqrt s )$ at ${n \over 2}=20$ (\ref{fig:fig_part1_11a}) and ${n \over 2}=25$ (\ref{fig:fig_part1_11b}). Designations are the same as in Fig.\ref{fig:fig_part1_10}. The good agreement of results shows the applicability of the approximation of equal denominators Eq.\ref{eq:eq_part1_41} resulting in the Eq.\ref{eq:eq_part1_63}.}
  \label{fig:fig_part1_11}
\end{figure}

It is also possible to verify this approximation in the following way. Taking into account $E=\sum\limits_{k=1}^{n\over 2}ch(y_k)={sh(ny_{n\over 2}) / 2sh(y_{n\over 2})}$ (as it follows from Eqs.\ref{eq:eq_part1_53},\ref{eq:eq_part1_35}) we have instead of Eq.\ref{eq:eq_part1_63}:
\begin{eqnarray}
\frac{{\sqrt s }}{2} - \frac{{sh\left( {n{y_{\frac{n}{2}}}} \right)}}{{2sh\left( {{y_{\frac{n}{2}}}} \right)}} = M \cdot ch\left( {\left( {n + 1} \right){y_{\frac{n}{2}}}} \right) 
\label{eq:eq_part1_64}
\end{eqnarray}
This equation does not admit exact analytical solution, and further we will consider an approximate solution of it. However, we can verify permissibility of the approximations made above, which resulted in Eq.\ref{eq:eq_part1_64}, by the numerical solving of this equation at the different energies $\sqrt{s}$ and by it comparison with the result of numerical determination of the maximum point. The results of such comparison are shown in Fig.\ref{fig:fig_part1_10} and Fig.\ref{fig:fig_part1_11}.

As seen from Figs.\ref{fig:fig_part1_10},\ref{fig:fig_part1_11}, the "exact" numerical solution of Eq.\ref{eq:eq_part1_64} practically does not differ from the results of numerical computation. This is the evidence of the fact that equal-denominators approximation Eq.\ref{eq:eq_part1_53} is admissible approximation.
Now let us consider the approximate analytical solution of Eq.\ref{eq:eq_part1_64}. Note that function $sh(ny_{n\over 2}) / 2sh(y_{n\over 2})$ in Eq.\ref{eq:eq_part1_64} changes slowly at small values of $y_{n/2}$ and can be approximately replaced by $n/2$ at $y_{n/2}\rightarrow 0$. In this approximation we obtain the following solution:
\begin{eqnarray}
{y_{\frac{n}{2}}} = \frac{1}{{n + 1}} {\rm arccosh} \left( {\frac{{\sqrt s  - n}}{{2M}}} \right) 
\label{eq:eq_part1_65}
\end{eqnarray}

\begin{figure}
  \centering
  \subfigure[]{
  \includegraphics[scale=0.335]{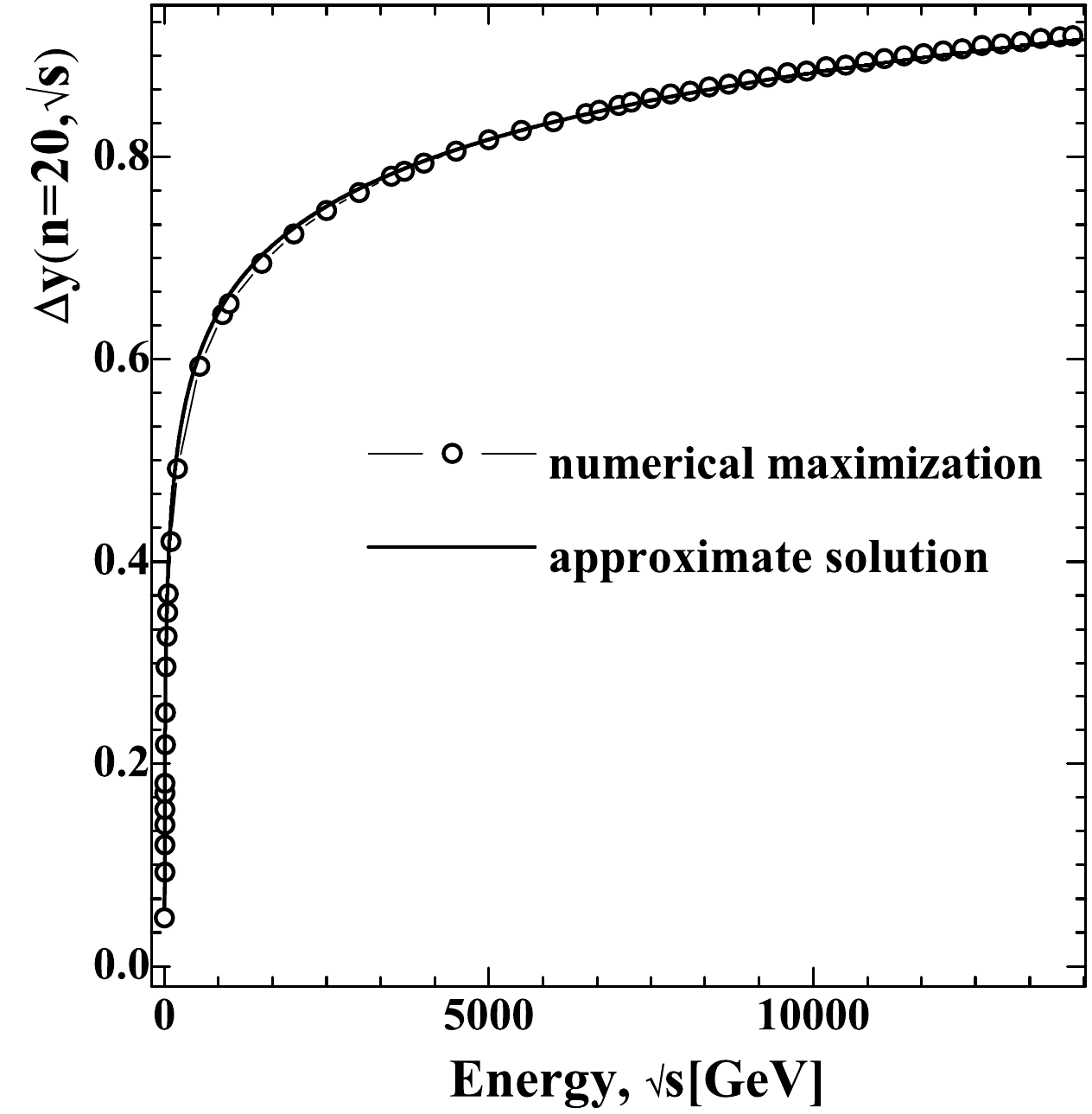} 
  \label{fig:fig_part1_12a} 
  }
  \subfigure[]{
  \includegraphics[scale=0.335]{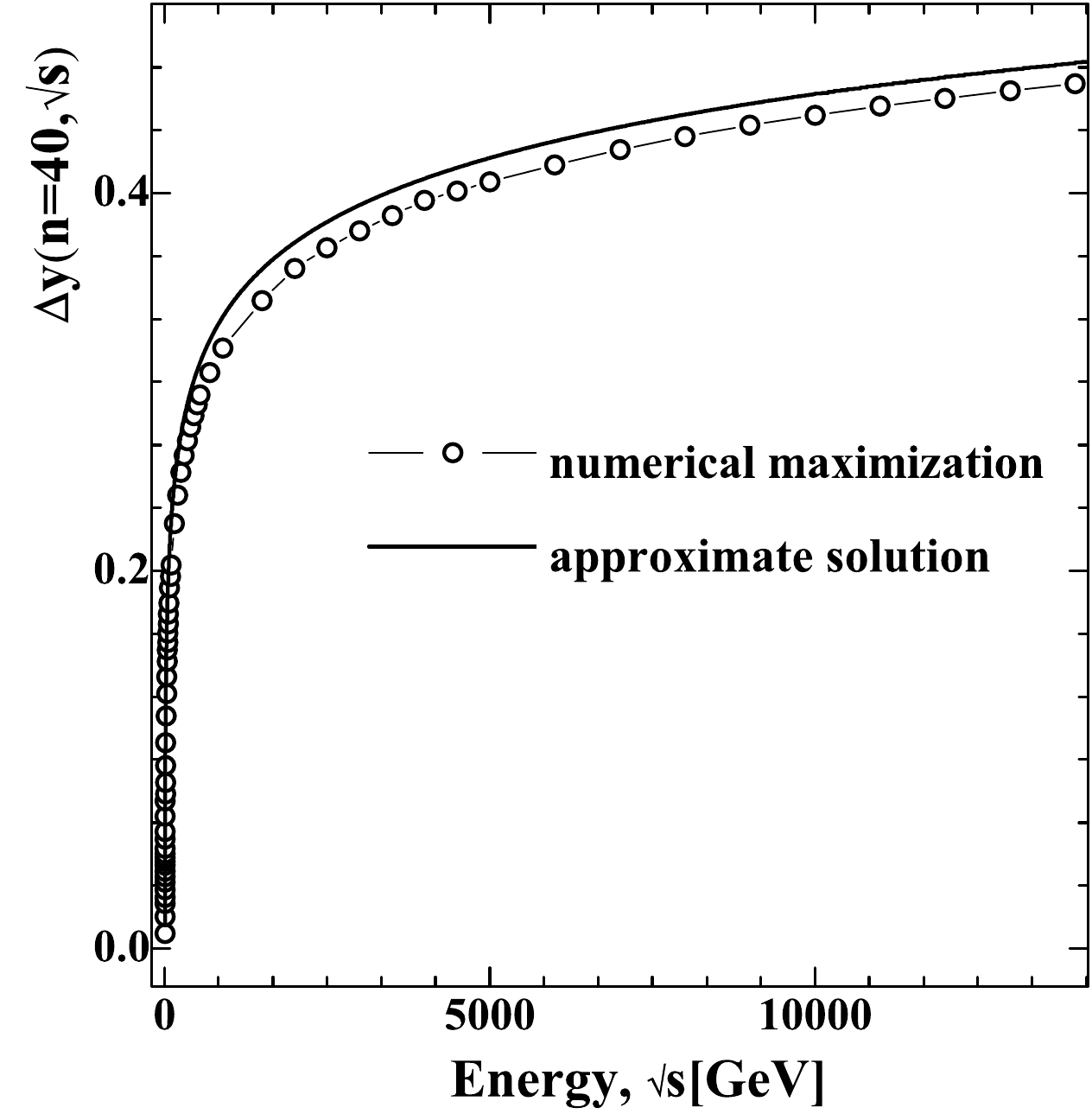} 
  \label{fig:fig_part1_12c} 
  }
  \subfigure[]{
  \includegraphics[scale=0.35]{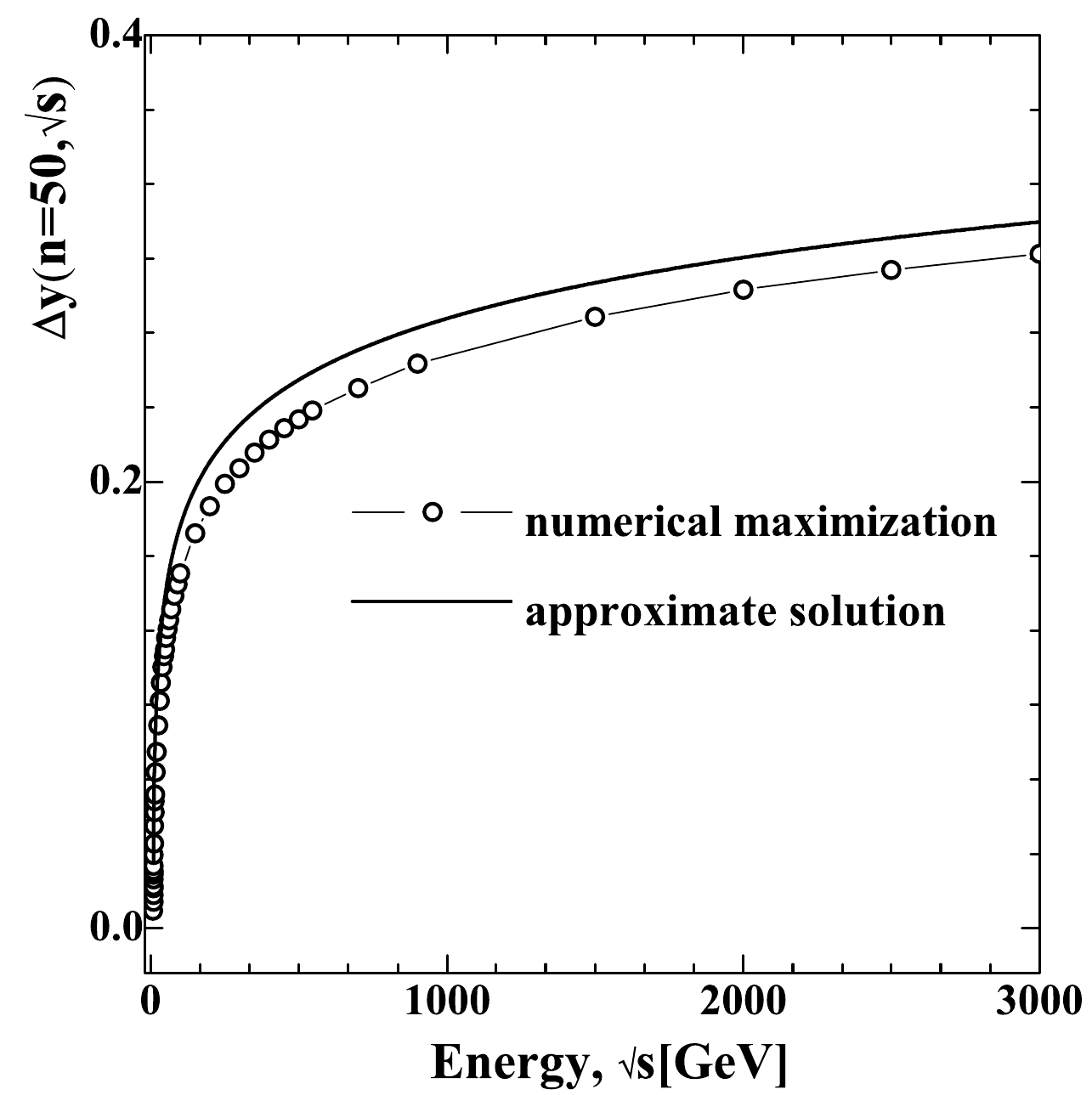} 
  \label{fig:fig_part1_12e} 
  }
  \subfigure[]{
  \includegraphics[scale=0.35]{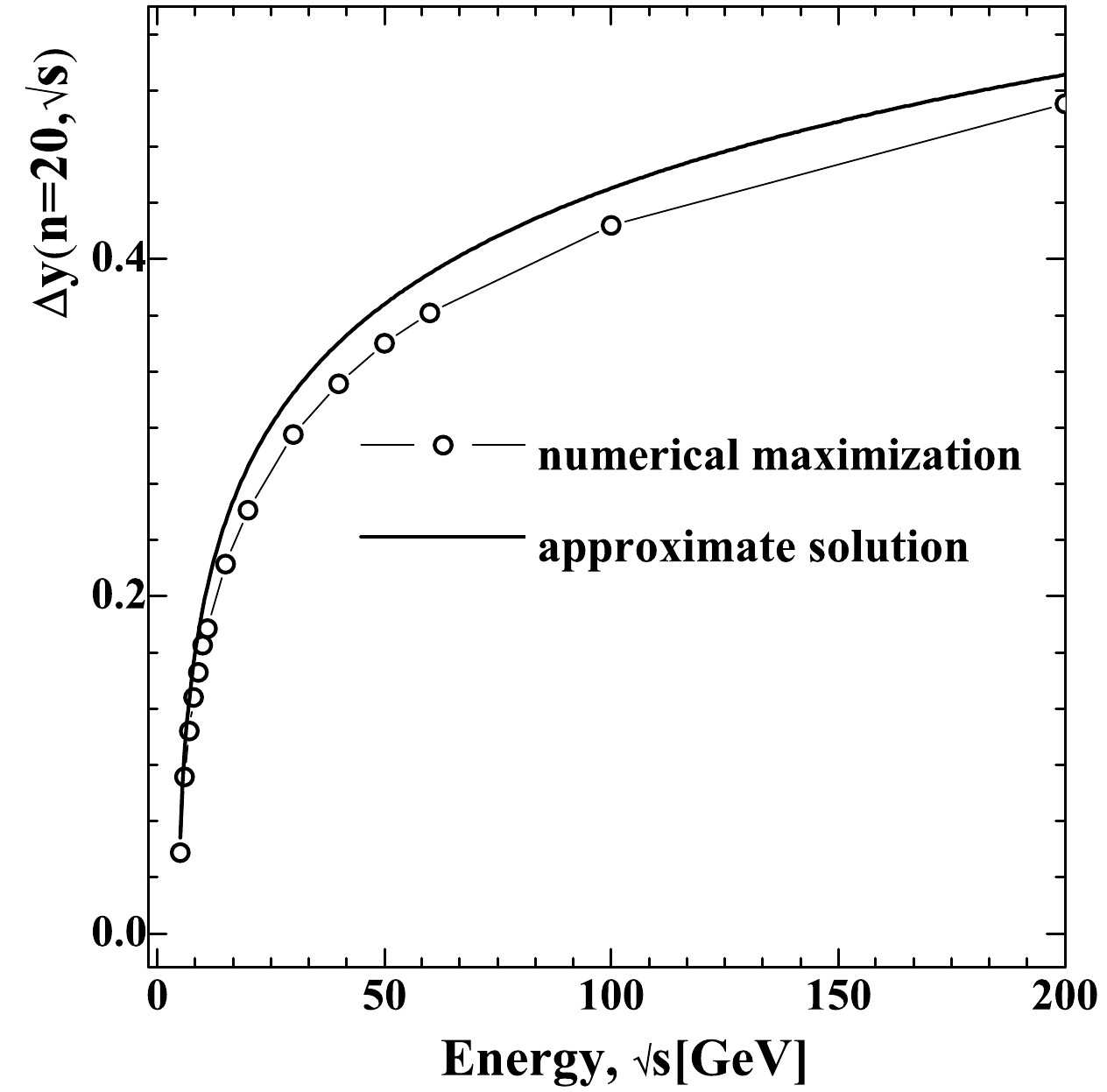}  
    \label{fig:fig_part1_12b} 
  }
  \subfigure[]{
  \includegraphics[scale=0.35]{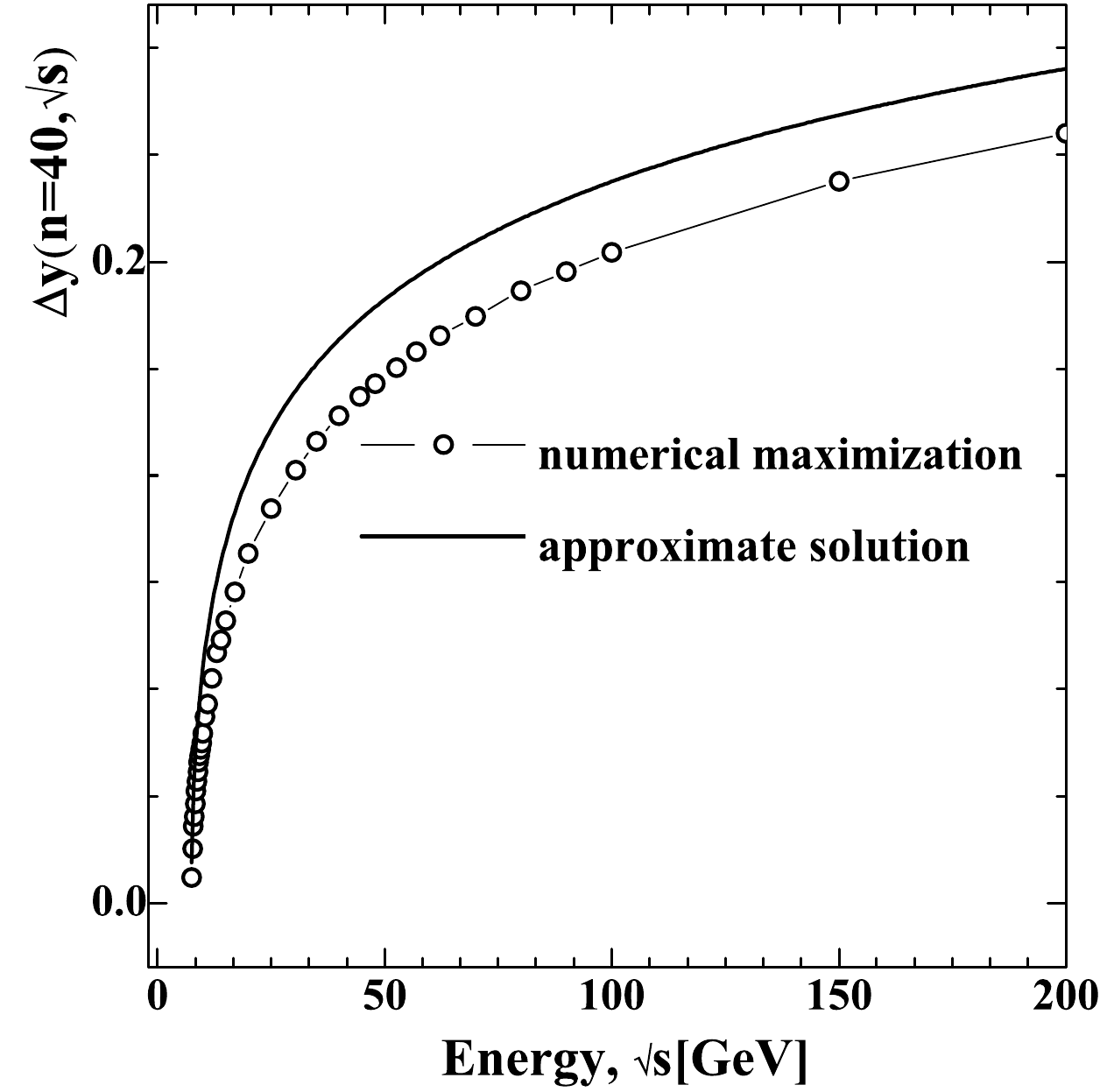} 
  \label{fig:fig_part1_12d} 
  }
  \subfigure[]{
  \includegraphics[scale=0.35]{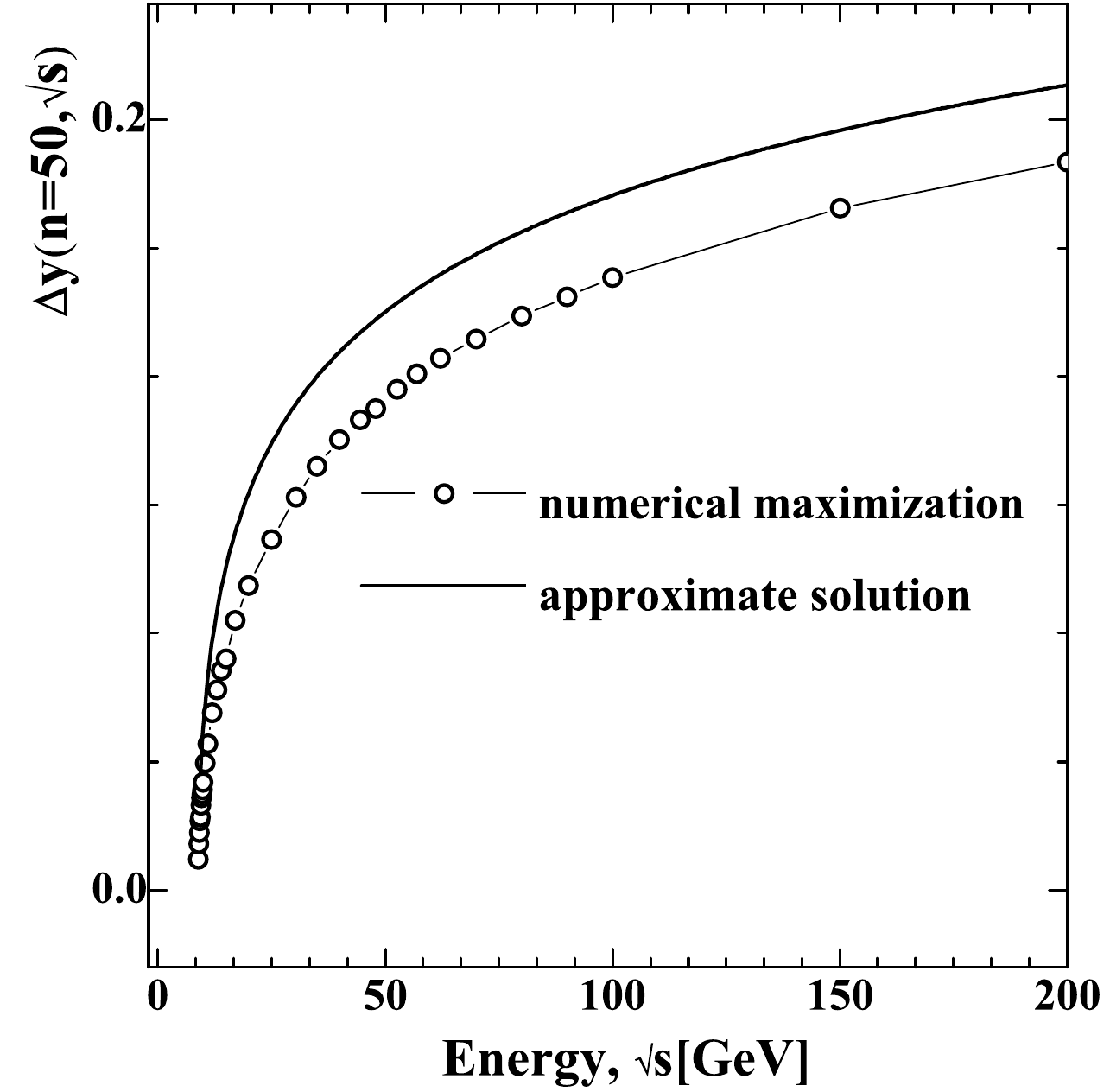} 
  \label{fig:fig_part1_12f} 
  }
    
  \caption{Comparison of the results of the approximate solution of Eq.\ref{eq:eq_part1_64} (solid line) with the results of numerical maximization (circles) of the magnitude at ${n\over 2}=10$ (\ref{fig:fig_part1_12a}), (\ref{fig:fig_part1_12b}); ${n\over 2}=20$ (\ref{fig:fig_part1_12c}), (\ref{fig:fig_part1_12d}); ${n\over 2}=25$ (\ref{fig:fig_part1_12e}), (\ref{fig:fig_part1_12f}). The range of low energies close to the threshold branch point (in dimensionless form) is shown on (\ref{fig:fig_part1_12b}), (\ref{fig:fig_part1_12d}) and (\ref{fig:fig_part1_12f}).}
  \label{fig:fig_part1_12}
\end{figure}

Comparison of approximate solution of Eq.\ref{eq:eq_part1_65} with results of numerical computation are presented in Fig.\ref{fig:fig_part1_12}. The Fig.\ref{fig:fig_part1_12} shows that Eq.\ref{eq:eq_part1_65} gives a somewhat overestimated value in comparison with numerical computation. It is natural, since using approximation ${sh(ny_{n\over 2}) / 2sh(y_{n\over 2})} \approx {n \over 2}$ in Eq.\ref{eq:eq_part1_64}, we underestimate the function $sh(ny_{n\over 2}) / 2sh(y_{n\over 2})$ and, in that way, raise too high hyperbolic cosine on the right-hand side of Eq.\ref{eq:eq_part1_64}.

Nevertheless, as evident from Fig.\ref{fig:fig_part1_12}, the absolute uncertainty of approximation Eq.\ref{eq:eq_part1_65} does not increase with the energy growth, and because $y_{n\over 2}$ itself increases the relative uncertainty decreases. We can explain it also reasoning from Eq.\ref{eq:eq_part1_64}. Since $sh(ny_{n\over 2}) / 2sh(y_{n\over 2})$ becomes small in comparison with $Mch((n+1)y_{n\over 2})$ at sufficiently high energies $\sqrt s$ (and $y_{n\over 2}$ accordingly), the accuracy of the approximation for function $sh(ny_{n\over 2}) / 2sh(y_{n\over 2})$ has no importance. By neglecting $sh(ny_{n\over 2}) / 2sh(y_{n\over 2})$ in Eq.\ref{eq:eq_part1_64} in comparison with $Mch((n+1)y_{n\over 2})$ and by neglecting $n$ in Eq.\ref{eq:eq_part1_65} in comparison with $\sqrt s$, we obtain the same result. It means that approximation Eq.\ref{eq:eq_part1_65} ensures the "correct" asymptotic of value $y_{n\over 2}$ at high $\sqrt s$.

Let us note some features of Eq.\ref{eq:eq_part1_65}. Firstly, the approximate solution of Eq.\ref{eq:eq_part1_65} undimensioned by mass $m$ has a threshold branch point at ($\sqrt s =n+2M$). This means that difference of an arithmetic progression of the rapidity has same feature, which maximizing the amplitude of inelastic process. Contribution of the considered inelastic processes to an imaginary part of the elastic scattering amplitude after calculation with the help of Laplace's method \cite{DeBruijn:225131} will be in some way expressed in terms of difference of the arithmetic progression $\Delta y$. Therefore it is possible to expect that mentioned threshold feature via difference of the arithmetic progression will be incorporated into imaginary part of the elastic scattering amplitude, which is required by unitarity condition. Note that Eq.\ref{eq:eq_part1_65} has logarithmic asymptotic at the energies substantially exceeding the threshold value and at $\Delta y=2y_{n \over 2}\sim n^{-1}$, that coincides with the results of numerical computation (see \cite{part1}).

For the following type of diagrams Fig.\ref{fig:Fig_2} with odd number of particles the whole procedure is similar to described above for diagrams with even number of particles. At first we take derivative from the logarithm of amplitude restriction Eq.\ref{eq:eq_part1_26} with respect to all rapidities $y_1,y_2,\ldots,y_{n-1 \over 2}$. After that it is possible to use the equal-denominators approximation
\begin{eqnarray}
{Z_1} \approx ... \approx {Z_{\frac{{n - 1}}{2}}} \approx {Z_{\frac{{n - 1}}{2} + 1}} = Z 
\label{eq:eq_part1_66}
\end{eqnarray}

From the condition of equality $Z_{{n-1 \over 2}-1} \approx Z_{n-1 \over 2} $ we will obtain relation similar to Eq.\ref{eq:eq_part1_45}:
\begin{eqnarray}
{P_{1\parallel }} - {P_{3\parallel }} - \sum\limits_{j = 1}^{\frac{{n - 1}}{2}} {sh\left( {{y_j}} \right)}  = \frac{{ch\left( {\frac{1}{2}{y_{\frac{{n - 1}}{2}}}} \right)}}{{2sh\left( {\frac{1}{2}{y_{\frac{{n - 1}}{2}}}} \right)}} 
\label{eq:eq_part1_67}
\end{eqnarray}

As it was done for Eq.\ref{eq:eq_part1_51} using same recipe for the derivatives of logarithm of the scattering amplitude restriction with respect to $y_{n-1 \over 2}$ and $y_{{n-1 \over 2}-1}$ and taking into account Eq.\ref{eq:eq_part1_67} we have
\begin{eqnarray}
 {y_{\frac{{n - 1}}{2} - 1}} = 2{y_{\frac{{n - 1}}{2}}} 
\label{eq:eq_part1_68}
\end{eqnarray}

Then it can be shown by induction that
\begin{eqnarray}
 \mbox{\fontsize{10}{10}\selectfont $ {y_{\frac{{n - 1}}{2} - (k - 1)}} = k{y_{\frac{{n - 1}}{2}}},\;\;k = 1,\;2,...,\;\frac{{n - 1}}{2} - 1$ } 
\label{eq:eq_part1_69}
\end{eqnarray}

Thus all the rapidities, leading to the constrained maximum of the considered scattering amplitude restriction, can be expressed in terms of $y_{n-1 \over 2}$. Repeating the calculations, which were made in order to obtain Eqs.\ref{eq:eq_part1_57}-\ref{eq:eq_part1_64}, we have for this rapidity in equal-denominators approximation:
\begin{eqnarray}
 \frac{{\sqrt s }}{2} - \frac{{sh\left( {\frac{n}{2}{y_{\frac{{n - 1}}{2}}}} \right)}}{{2sh\left( {\frac{{{y_{\frac{{n - 1}}{2}}}}}{2}} \right)}} = M \cdot ch\left( {\left( {\frac{{n - 1}}{2} + 1} \right){y_{\frac{{n - 1}}{2}}}} \right) 
\label{eq:eq_part1_70}
\end{eqnarray}

In an approximation similar to ones, which results in Eq.\ref{eq:eq_part1_65}, we get
\begin{eqnarray}
 {y_{\frac{{n - 1}}{2}}} = \frac{2}{{n + 1}} {\rm arccosh} \left( {\frac{{\sqrt s  - n}}{{2M}}} \right) 
\label{eq:eq_part1_71}
\end{eqnarray}

From Eq.\ref{eq:eq_part1_71} it is evident that in case of odd number $n$ the rapidity common difference, which constrainedly maximize the scattering amplitude, also has a threshold branch point. Note, that difference of an arithmetic progression is equal to $y_{n-1 \over 2}$ in case of odd $n$ and is equal to $2y_{n\over 2}$ in case of even $n$, i.e., as agreed, the approximation for the common difference of an arithmetic progression $\Delta y(n, \sqrt s)$ is expressed by the same formula both for even and odd $n$.

The obtained analytical results enables to see how does the mechanism of virtuality reduction ``work" with the energy growth.
It's easy to show \cite {part1} that 
the scattering amplitude at the point of constrained maximum is expressed like
\begin{equation}
\begin{array}{c}
A^{(0),n}=\left( 1+a(\sqrt s,n)\right)^{-2}  \left( 1+  b(\sqrt s,n) \right)^{-(n-1)} \exp\left( c(\sqrt s,n) \right) 
\end{array}
\label{amplitude-maximum-point}
\end{equation}
where
\begin{eqnarray}%
\begin{array}{l}
a(\sqrt s,n)=\left( \frac{1}{ \left( {\sqrt s}/M \right)^{\frac{2}{n+1}}-1 }\right)^2,
\\
b(\sqrt s,n)=\left( \frac{\left( {\sqrt s}/M \right)^{\frac{1}{n+1}}}{\left( {\sqrt s}/M \right)^{\frac{2}{n+1}}-1} \right)^2,
\\
c(\sqrt s,n)=2\left( 1-(n-1) \left(  {\sqrt s}/M \right)^{-\frac{n}{n+1}} \left( \left( {\sqrt s}/M  \right)^{\frac{2}{n+1}} -1 \right)   \right)\\
\times \left( \left( \left( {\sqrt s}/M \right)^{\frac{2}{n+1}} -1 \right)^2 + \left( {\sqrt s}/M \right)^{\frac{2}{n+1}}\right)^{-1}
\end{array}
\end{eqnarray}

The $a(\sqrt s,n)$ and $b(\sqrt s,n)$ determine the characteristic value of virtuality at the maximum point of scattering amplitude and 
$c(\sqrt s,n)$  determines the variation of virtuality along the ``comb``. In other words, the following estimate takes place
\begin{equation}
\left( \frac{1}{ \left( {\sqrt s}/M \right)^{\frac{2}{n+1}}-1 }\right)^2  \leq \left| \left( q^{(j)} \right)^2 \right| \leq \left( \frac{\left( {\sqrt s}/M \right)^{\frac{1}{n+1}}}{\left( {\sqrt s}/M \right)^{\frac{2}{n+1}}-1} \right)^2
\label{eq:eq_part_virtuality1}
\end{equation}
where $\left| \left( q^{(j)} \right)^2 \right|$ is the absolute value of virtuality corresponding to $j$-th internal line on the "comb" in the point of constrained maximum.

As it follows from Eq.\ref{eq:eq_part_virtuality1} energy included in same from that it useful to rewrite in this form
\begin{equation}
\left( \sqrt s /M \right)^{\frac{1}{n+1}} = \exp\left(\frac{1}{n+1} \ln\left( \sqrt s /M \right) \right)
\label{eq:eq_part_virtuality2}
\end{equation}

It is obvious that the growth of exponent with energy $\sqrt s$ is much weaker than the corresponding decrease with the growth of number of particles $n$. Thus, one can see that at not very small $n$ the value of $\left( \sqrt s /M \right)^{\frac{1}{n+1}} \sim 1$ even at high energies ($\sqrt s>>M$). As the result, the difference of energy and longitudinal momentum squares is at least not negligible with respect to transverse momentum for each virtuality on the "comb".  This result comes in contradiction with the statement that virtulalities can be reduced to transverse momentum squares, which is usually claimed in the standard approach \cite{Collins:111502,Nikitin:113716,bfkl_1976,Lipatov:2008,Byckling:100542,Ter-Martirosyan,levin_2,KozlovNSU_2007}.
Taking into account the growth of $\left( \sqrt s /M \right)^{\frac{2}{n+1}}$ with energy $\sqrt s$ growth, we see that virtuality at the maximum point really decreases and the maximum value of amplitude grows with the growth of energy $\sqrt s$. Note also that at not very small $n$ the $\left( \sqrt s /M \right)^{\frac{1}{n+1}}$ is close to unity at rather wide energy range which results in the much steeper growth than the one which is attained in Regge-based theories \cite{springerlink:10.1007/BF02781901, Collins:111502} and described by factor of $ln^{n-2}\left( \sqrt s/ M \right)$. Moreover, the higher $n$, the wider is the energy range. Thus the asymptotic behavior for different $n$ is reached at different $\sqrt s$ which enables to doubt the validity of the asymptotic formulas of multi-Regge kinematics.


\section{On the need of taking into account the interference terms at the calculation of inelastic scattering cross-section}
\label{sec:interference-contrib}

\begin{figure}
  \centering
  \subfigure[]{
  \includegraphics[scale=0.68]{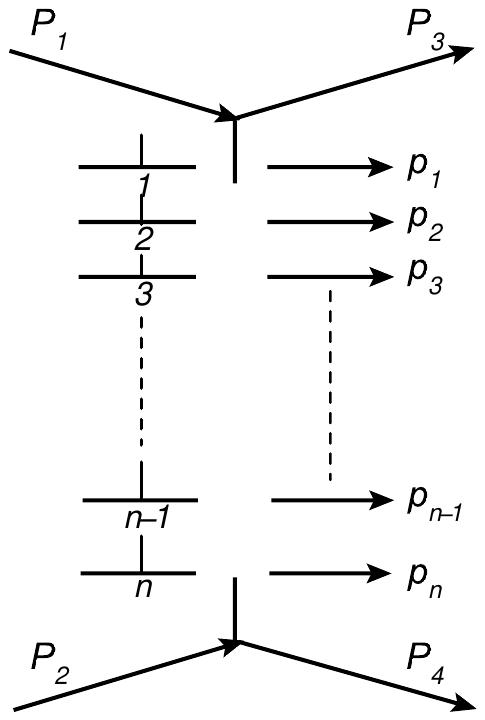} 
  \label{fig:fig_part2_1a} 
  }
  \subfigure[]{
  \includegraphics[scale=0.68]{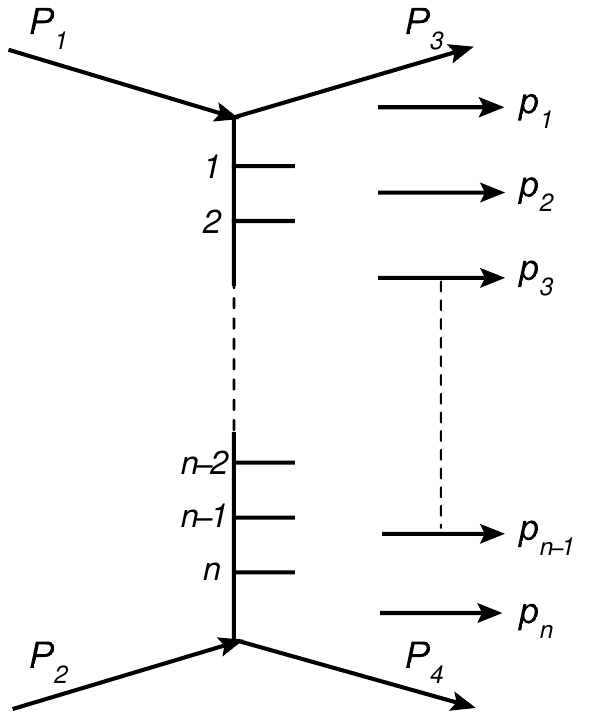}  
    \label{fig:fig_part2_1b} 
  }        
  \caption{Drawing the diagrams of the "comb" type}
  \label{fig:fig_part2_1}
\end{figure}

Before we turn to further calculations we need to make an important remark. Let's recall the formula Eq.\ref{eq:eq_part1_1}. According to Wick theorem, the scattering amplitude is the sum of diagrams with all possible orders of external lines attaching to the``comb". In the terms of diagram technique it looks as follows. Plotting the multi-peripheral diagram of the scattering amplitude Fig.\ref{fig:Fig_2} at first we have adequate number of vertices with three lines going out of it and $n$ lines corresponding to the secondary particles as it is shown in Fig.\ref{fig:fig_part2_1a}.

``Pairing" some lines Fig.\ref{fig:fig_part2_1a} in order to obtain the ``comb", we will get a situation shown in Fig.\ref{fig:fig_part2_1b}. The weighting coefficient appearing from this procedure, is included to the coupling constant. Finally we have to "pair" the appropriate lines of particles in the final state with the remaining unpaired internal lines in the diagram of Fig.\ref{fig:fig_part2_1b}.

If we marked by $i_1$ - the external line, paired with the first vertex; $i_2$ - the external line, paired with the second vertex and etc.; then $i_k$ is an external line, which is paired with $k$-th vertex, so every diagram will be characterized by sequence $i_1, i_2,\ldots,i_n$. And in this case the total amplitude is expressed by the sum of $n!$ terms, each of them corresponds to one of $n!$ possible index sequences and therefore the inelastic scattering cross-section can be written as 
\begin{eqnarray}%
\begin{array}{l}
 {\sigma _n} = \frac{{{{\left( {{{\left( {2\pi } \right)}^4}} \right)}^2}{g^4}{\lambda ^{2n}}}}{{4n!\sqrt {{{({P_1}{P_2})}^2} - {{({M_1}{M_2})}^2}} }}\int {\frac{{d{{\vec P}_3}}}{{2{P_{30}}{{(2\pi )}^3}}}} \frac{{d{{\vec P}_4}}}{{2{P_{40}}{{(2\pi )}^3}}}\prod\limits_{k = 1}^n {\frac{{d{{\vec p}_k}}}{{2{p_{0k}}{{(2\pi )}^3}}}} {\delta ^{\left( 4 \right)}}\left( {{P_3} + {P_4} + \sum\limits_{k = 1}^n {{p_k}}  - {P_1} - {P_2}} \right) \times  \\ 
\mbox{\fontsize{10}{10}\selectfont $  \times {\left( {\sum\limits_{P\left( {{i_1},{i_2},...,{i_n}} \right)} {A\left( {n,{P_3},{P_4},{p_{{i_1}}},{p_{{i_2}}},...,{p_{{i_n}}},{P_2},{P_1}} \right)} } \right)^*}\left( {\sum\limits_{P\left( {{j_1},{j_2},...,{j_n}} \right)} {A\left( {n,{P_3},{P_4},{p_{{j_1}}},{p_{{j_2}}},...,{p_{{j_n}}},{P_2},{P_1}} \right)} } \right)$ } \\ 
 \end{array}
\label{eq:eq_part2_2}
\end{eqnarray}%

Designation $\sum\limits_{P(i_1,i_2,\ldots,i_n)}$ means that a sum of all terms corresponding to all possible permutations of indices $i_1,i_2,\ldots,i_n$ is considered. Moreover, since function $A$ is real and positive, the sign of complex conjugation terms in Eq.\ref{eq:eq_part2_2} can be dropped out. Writing this expression in the form
\begin{eqnarray}
\begin{array}{l}
 {\sigma _n} = \frac{{{{\left( {{{\left( {2\pi } \right)}^4}} \right)}^2}{g^4}{\lambda ^{2n}}}}{{4n!\sqrt {{{({P_1}{P_2})}^2} - {{({M_1}{M_2})}^2}} }}\sum\limits_{\scriptstyle P\left( {{i_1},{i_2},...,{i_n}} \right) \hfill \atop 
  \scriptstyle P\left( {{j_1},{j_2},...,{j_n}} \right) \hfill} {\int {\frac{{d{{\vec P}_3}}}{{2{P_{30}}{{(2\pi )}^3}}}} \frac{{d{{\vec P}_4}}}{{2{P_{40}}{{(2\pi )}^3}}}}{\delta ^{\left( 4 \right)}}\left( {{P_3} + {P_4} + \sum\limits_{k = 1}^n {{p_k}}  - {P_1} - {P_2}} \right) \\
\times A\left( {n,{P_3},{P_4},{p_{{i_1}}},{p_{{i_2}}},...,{p_{{i_n}}},{P_2},{P_1}} \right)A\left( {n,{P_3},{P_4},{p_{{j_1}}},{p_{{j_2}}},...,{p_{{j_n}}},{P_2},{P_1}} \right) \\
\end{array}
\label{eq:eq_part2_3}
\end{eqnarray}
The integration variables in each summand of considered sum can be designated so that the indices $i_1, i_2,\ldots, i_n$ produce the original placing $1, 2,\ldots, n$. At the same time the indices $j_1, j_2,\ldots, j_n$ will produce some replacing and summation must be made over all permutations of these indices. Regarding to this, we can write down instead of Eq.\ref{eq:eq_part2_3}
\begin{eqnarray}
\begin{array}{l}
 {\sigma _n} = \frac{{{{\left( {{{\left( {2\pi } \right)}^4}} \right)}^2}{g^4}{\lambda ^{2n}}}}{{4\sqrt {{{({P_1}{P_2})}^2} - {{({M_1}{M_2})}^2}} }}\int {\frac{{d{{\vec P}_3}}}{{2{P_{30}}{{(2\pi )}^3}}}} \frac{{d{{\vec P}_4}}}{{2{P_{40}}{{(2\pi )}^3}}}\prod\limits_{k = 1}^n {\frac{{d{{\vec p}_k}}}{{2{p_{0k}}{{(2\pi )}^3}}}}  \times  \\ 
  \times {\delta ^{\left( 4 \right)}}\left( {{P_3} + {P_4} + \sum\limits_{k = 1}^n {{p_k}}  - {P_1} - {P_2}} \right)\Phi\left( {n,{P_3},{P_4},{p_1},{p_2},...,{p_n},{P_2},{P_1}} \right) \\ 
 \end{array}
\label{eq:eq_part2_4} 
\end{eqnarray} 
where
\begin{eqnarray}
\begin{array}{l}
 \Phi\left( {n,{P_3},{P_4},{p_1},{p_2},...,{p_n},{P_2},{P_1}} \right) = A\left( {n,{P_3},{P_4},{p_1},{p_2},...,{p_n},{P_2},{P_1}} \right) \times  \\ 
  \times \sum\limits_{P\left( {{j_1},{j_2},...,{j_n}} \right)} {A\left( {n,{P_3},{P_4},{p_{{j_1}}},{p_{{j_2}}},...,{p_{{j_n}}},{P_2},{P_1}} \right)}  \\ 
 \end{array}
\label{eq:eq_part2_5}
\end{eqnarray}

At Section.\ref{sec:Laplace-method} we will compute the integer from Eq.\ref{eq:eq_part2_4} apply the fact that the amplitudes $A$, entering in Eq.\ref{eq:eq_part2_5}, have the points of constrained maximum.

\section{Computing the multi-peripheral diagram contributions to inelastic scattering cross-section with the Laplace's method}
\label{sec:Laplace-method}

For the further analysis consider Eq.\ref{eq:eq_part2_4} in c.m.s. framework, expanding three-dimensional particle momenta into longitudinal and transverse components with respect to the collision axis:
\begin{eqnarray}
\begin{array}{l}
 {\sigma _n} = \frac{{{{\left( {{{\left( {2\pi } \right)}^4}} \right)}^2}{g^4}{\lambda ^{2n}}}}{{4n!\sqrt {{{({P_1}{P_2})}^2} - {{({M_1}{M_2})}^2}} }} \int {\frac{{d{{\vec P}_{3 \bot }}d{P_{3\left\| {} \right.}}}}{{2{{(2\pi )}^3}\sqrt {{M^2} + P_{3\left\| {} \right.}^2 + \vec P_{3 \bot }^2} }}} \frac{{d{{\vec P}_{4 \bot }}d{P_{4\left\| {} \right.}}}}{{2{{(2\pi )}^3}\sqrt {{M^2} + P_{4\left\| {} \right.}^2 + \vec P_{4 \bot }^2} }}\prod\limits_{k = 1}^n {\frac{{d{{\vec p}_{k \bot }}d{p_{k\left\| {} \right.}}}}{{2{{(2\pi )}^3}\sqrt {{m^2} + p_{k\left\| {} \right.}^2 + \vec p_{k \bot }^2} }}} \Phi \\
 \times \delta \left( {\sqrt {{M^2} + P_{3\parallel }^2 + \vec P_{3 \bot }^2} + \sqrt {{M^2} + P_{4\parallel }^2 + \vec P_{4 \bot }^2} + \sum\limits_{k = 1}^n {\sqrt {1 + p_{3\left\| {} \right.}^2 + \vec p_{3 \bot }^2} }  - \sqrt s } \right) \\
 \times \delta \left( {\sum\limits_{k = 1}^n {{p_{k\left\| {} \right.}}}  + {P_{3\left\| {} \right.}} + {P_{4\left\| {} \right.}}} \right)\delta \left( {\sum\limits_{k = 1}^n {{p_{k \bot x}}}  + {P_{3 \bot x}} + {P_{4 \bot x}}} \right) \delta \left( {\sum\limits_{k = 1}^n {{p_{k \bot y}}}  + {P_{3 \bot y}} + {P_{4 \bot y}}} \right) \\ 
\end{array}
\label{eq:eq_part2_6} 
\end{eqnarray}

The last three $\delta$-functions, whose arguments are linear with respect to integration variables, we can vanish 
by the integration over $P_{4\parallel}$, $P_{4\perp x}$, $P_{4\perp y}$. In order to take into account the rest $\delta$-function, which expresses the energy conservation law let`s replace $P_{3\parallel}$ by new integration variable
\begin{eqnarray}
{E_p} = \sqrt {{M^2} + P_{3\left\| {} \right.}^2 + \vec P_{3 \bot }^2} + \sqrt {{M^2} + {{\left( {\sum\limits_{k = 1}^n {{p_{k\left\| {} \right.}}}  + {P_{3\left\| {} \right.}}} \right)}^2} + {{\left( {\sum\limits_{k = 1}^n {{{\vec p}_{k \bot }}}  + {{\vec P}_{3 \bot }}} \right)}^2}} 
\label{eq:eq_part2_7}
\end{eqnarray}
Those, making the following replacement we must express $P_{3\parallel}$ through $E_p$. The corresponding expression will coincide with Eq.\ref{eq:eq_part1_7} with the positive sign in front of the square root. Moreover, introduce the rapidities instead of longitudinal momenta:
\begin{eqnarray}
\begin{array}{l}
 {p_{k\parallel }} = {m_ \bot }\left( {{{\vec p}_{k \bot }}} \right)sh\left( {{y_k}} \right) \\ 
 {m_ \bot }\left( {{{\vec p}_{k \bot }}} \right) = \sqrt {m + \vec p_{k \bot }^2} \\
\end{array}
\label{eq:eq_part2_8}
\end{eqnarray}

After these transformations we have
\begin{eqnarray}
\begin{array}{l}
 {\sigma _n} = \frac{{{{\left( {2\pi } \right)}^2}{g^4}{\lambda ^{2n}}}}{{4\sqrt {{s \mathord{\left/
 {\vphantom {s 4}} \right.
 \kern-\nulldelimiterspace} 4} - {M^2}} }}\frac{1}{{\sqrt s }}\int {\frac{{d{{\vec P}_{3 \bot }}}}{{2\sqrt {{M^2} + P_{3\parallel }^2 + \vec P_{3 \bot }^2} }}\prod\limits_{k = 1}^n {\frac{{d{{\vec p}_{k \bot }}d{y_{k\parallel }}}}{{2{{\left( {2\pi } \right)}^3}}}} } \frac{{\partial {P_{3\parallel }}}}{{\partial {E_p}}}\left| {_{{E_p} = \sqrt s  - \sum\limits_{k = 1}^n {{m_{ \bot k}}\left( {{{\vec p}_{ \bot k}}} \right){\mathop{\rm ch}\nolimits} \left( {{y_k}} \right)} }} \right. \times  \\ 
  \times {\left. {\frac{{\Phi \left( {n,{y_1},{{\vec p}_{1 \bot }}, \ldots ,{y_n},{{\vec p}_{n \bot }},{P_{1\parallel }},{P_{2\parallel }},{P_{3\parallel }},{{\vec P}_{3 \bot }},{P_{4\parallel }},{{\vec P}_{4 \bot }}} \right)}}{{2\sqrt {{M^2} + {{\left( {{P_{4\parallel }}} \right)}^2} + {{\left( {{{\vec P}_{4 \bot }}} \right)}^2}} }}} \right|_{{P_{4\parallel }} =  - \left( {\sum\limits_{k = 1}^n {{m_ \bot }\left( {{{\vec p}_{k \bot }}} \right)} {\mathop{\rm sh}\nolimits} \left( {{y_k}} \right) + {P_{3\parallel }}} \right),{{\vec P}_{4 \bot }} =  - \left( {\sum\limits_{k = 1}^n {{{\vec p}_{k \bot }}}  + {{\vec P}_{3 \bot }}} \right)}} \\ 
 \end{array}
\label{eq:eq_part2_9} 
\end{eqnarray}
where it is assumed that the magnitude $P_{3\parallel}$ is expressed in terms of the other integration variables via Eq.\ref{eq:eq_part1_7}.

Now turn to the dimensionless integration variables and made the following replacement $\vec{p}_{k\perp} \to \frac{\vec{p}_{k\perp}}{m}$, $\vec{P}_{3\perp} \to \frac{\vec{P}_{3\perp}}{m}$. We designate the new dimensionless integration variables just as the old variables, for short. 
Moreover, replace expression for $P_{3\parallel}$ by the same expression divided by $m$. The same concerns the constants in expressions for cross-section, i.e., the designations $M$ and $\sqrt s$ are used for proton mass and energy of colliding particles in c.m.s., which is made dimensionless with the pion mass $m$.

Now introduce the following notations of integration variables in Eq.\ref{eq:eq_part2_9} and designate the rapidities $y_1$, $y_2$, $\ldots$, $y_n$ as $X_1$, $X_2$,$\ldots$, $X_n$; $x$-components of transverse momenta of secondary particles $p_{1\perp x}$, $p_{2\perp x}$,$\ldots$,$p_{n\perp x}$ as $X_{n+1},X_{n+2},\ldots,X_{2n}$; $y$-components of transverse momenta of secondary $p_{1\perp y}$, $p_{2\perp y}$,$\ldots$, $p_{n\perp y}$ as $X_{2n+1}$,$X_{2n+2}$,$\ldots$,$X_{3n}$. Moreover, designate $X_{3n+1}$ as $P_{3\perp x}$ and $X_{3n+2}$ as $P_{3\perp y}$.

As it was shown in the previous sections, that an integrand $A(n,P_3,P_4,p_1,p_2,\ldots,p_2,P_1,P_2)$ in Eq.\ref{eq:eq_part2_9}, expressed as a function of independent integration variables, has a maximum point in the domain of integration. At the neighborhood of this maximum point it can be represented in the form
\begin{eqnarray}
\mbox{\fontsize{10}{10}\selectfont $ A\left( {n,{P_3},{P_4},{p_1},{p_2},...,{p_n},{P_2},{P_1}} \right) = {A^{\left( 0 \right),n}}\left( {\sqrt s } \right) \exp \left( { - \frac{1}{2}\sum\limits_{a = 1}^{3n + 2} {\sum\limits_{b = 1}^{3n + 2} {{D_{ab}}} \left( {{X_a} - X_a^{\left( 0 \right)}} \right)\left( {{X_b} - X_b^{\left( 0 \right)}} \right)} } \right)$ } 
\label{eq:eq_part2_10}
\end{eqnarray}
where $A^{(0),n}(\sqrt s)$ is the value of the function Eq.\ref{eq:eq_part1_3} at the point of constrained maximum; \\
$D_{ab}=$ $-\frac{\partial^2}{\partial X_a \partial X_b}\left(\ln(A)\right)$ - the derivatives are taken at the constrained maximum point of scattering amplitude. In other words, the real and positive magnitude $A$ determined by Eq.\ref{eq:eq_part1_3} is represented as $A=\exp(\ln(A))$, and the power of the exponential function is expanded into the Taylor series in the neighborhood of the maximum point with an accuracy up to the second-order summands. The accuracy of the approximation Eq.\ref{eq:eq_part2_10} can be numerically checked in the following way. The function $A$, defined by Eq.\ref{eq:eq_part1_3} can be rewritten in the new notation, which were identified previously
\begin{eqnarray}
 A = A\left( {n,{X_1},{X_2},...,{X_{3n + 2}}} \right) 
\label{eq:eq_part2_11} 
\end{eqnarray}

\begin{figure}
  \centering
  \subfigure[]{
  \includegraphics[scale=0.23]{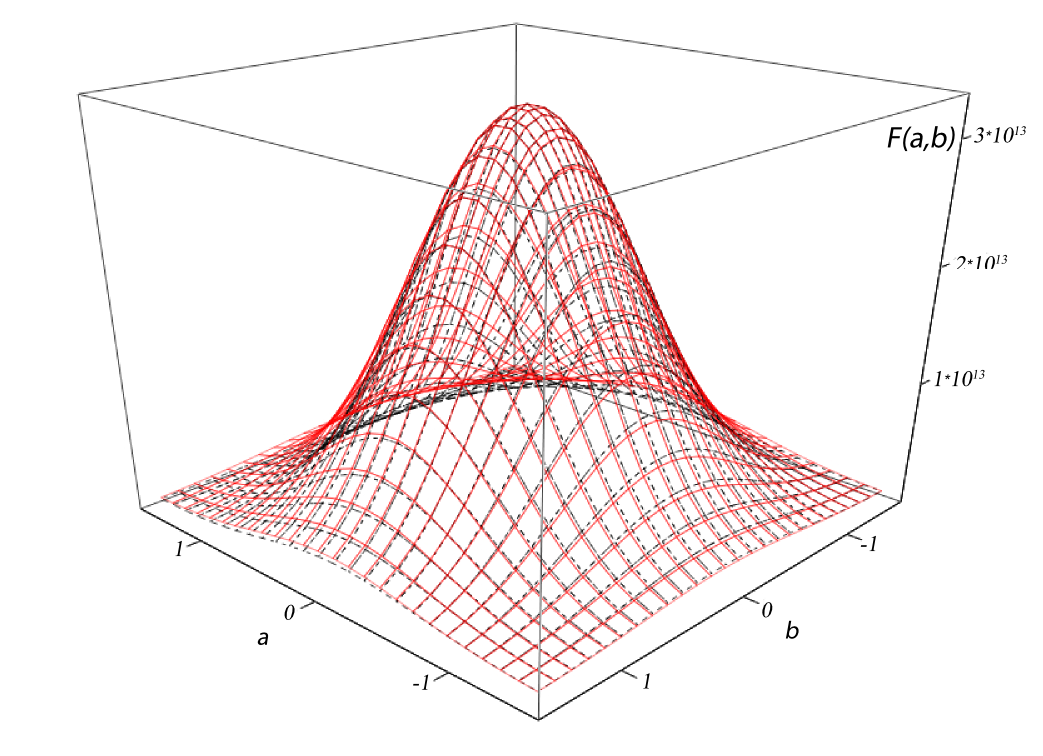} 
  \label{fig:fig_part2_2a} 
  }
  \subfigure[]{
  \includegraphics[scale=0.23]{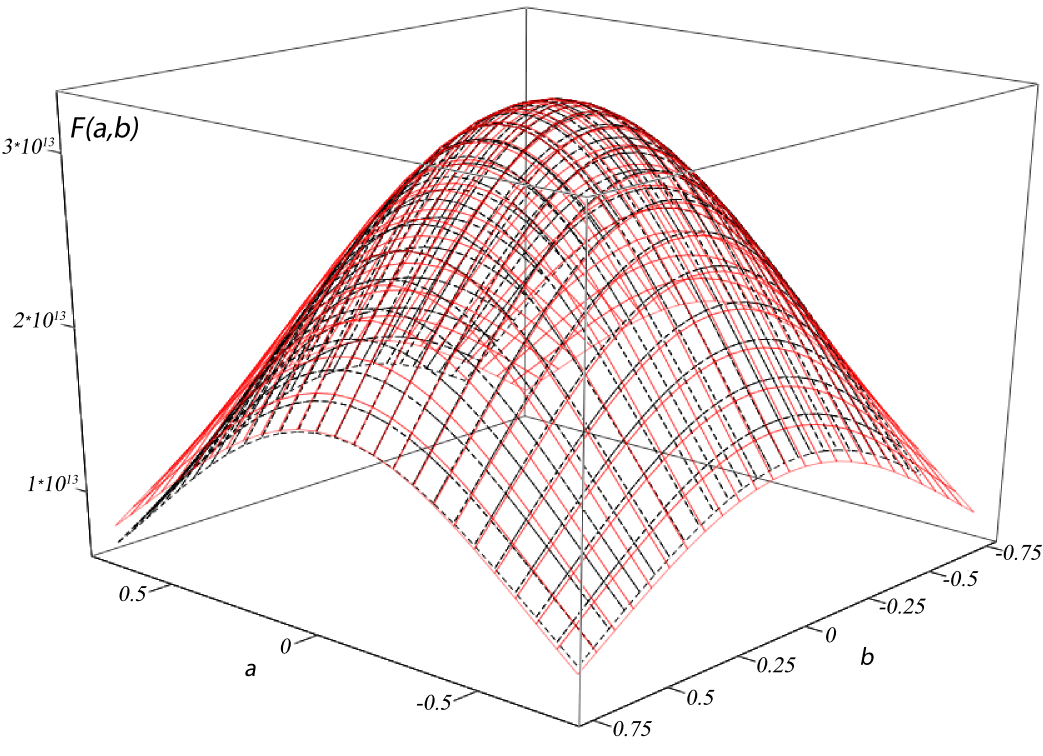}  
    \label{fig:fig_part2_2b} 
  }
  \caption{Functions $F_{1,7}^{n=10}(a,b)$ (dashed line) and $F_{1,7}^{(g), n=10}(a,b)$ (solid line) at energy $\sqrt s=5$ GeV and $n=10$. The general image (\ref{fig:fig_part2_2a}), and the zoomed image (\ref{fig:fig_part2_2b}) at the neighborhood of maximum point. Clear that in an area that makes the most significant contribution to the integral, the scattering amplitude does not differ from its Gaussian approximation Eq.\ref{eq:eq_part2_12}, which demonstrates the possibility of applying the Laplace method.}
  \label{fig:fig_part2_2}
\end{figure}

\begin{figure}
  \centering   
  \subfigure[]{
  \includegraphics[scale=0.23]{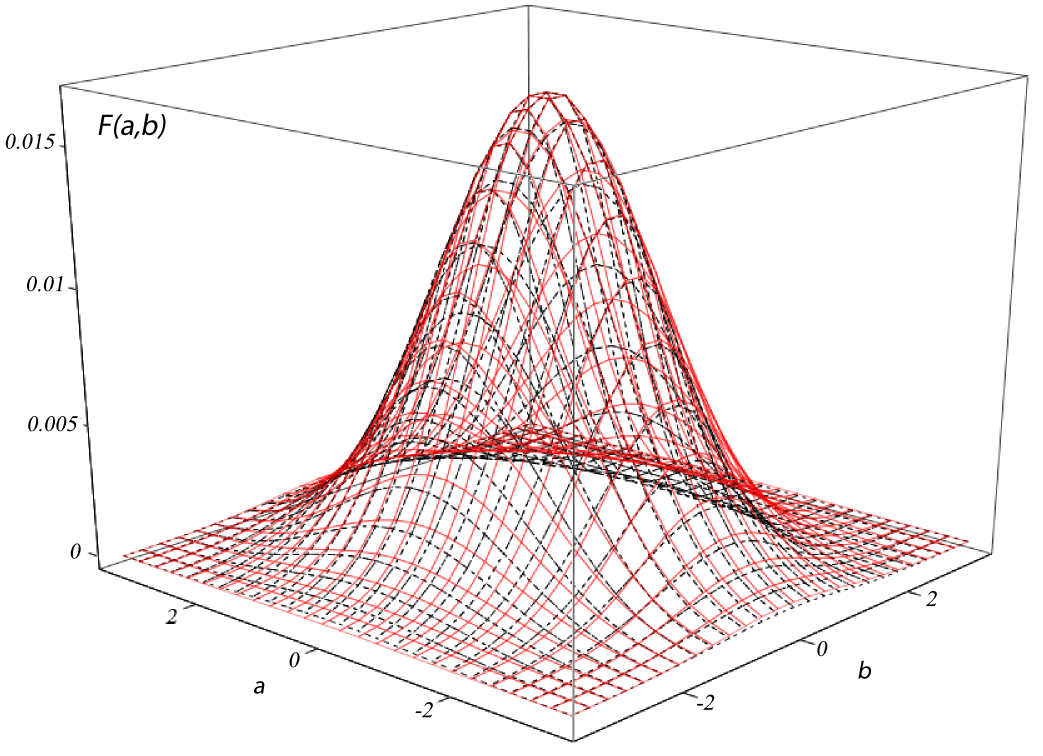} 
  \label{fig:fig_part2_3a} 
  }
  \subfigure[]{
  \includegraphics[scale=0.23]{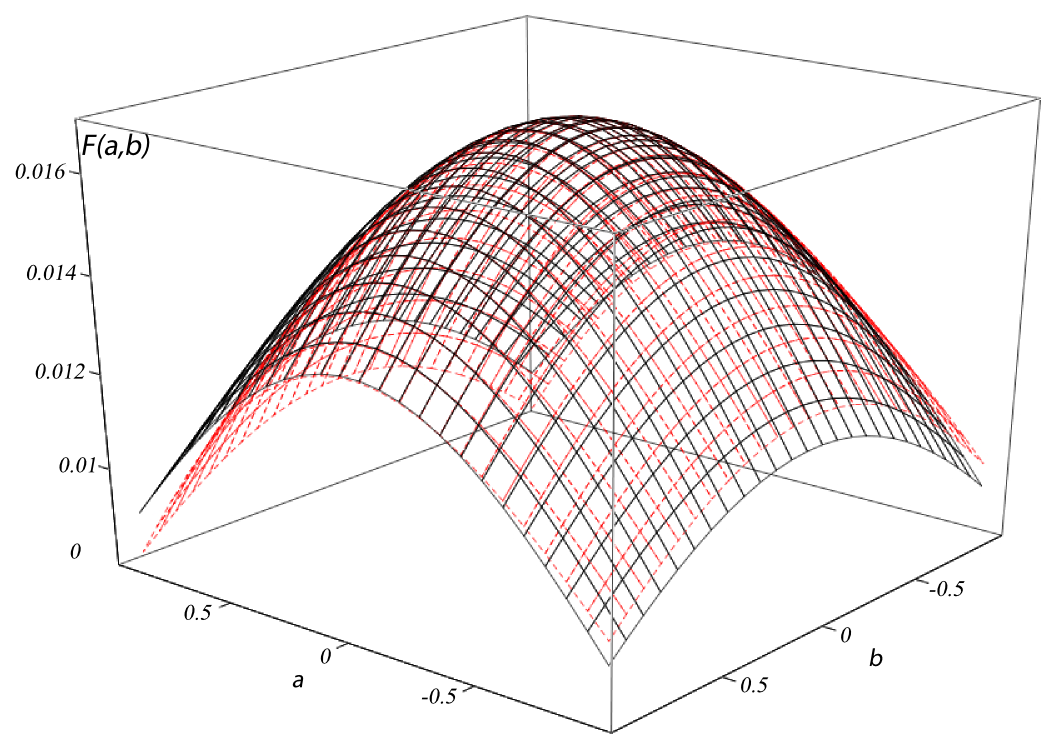}  
    \label{fig:fig_part2_3b} 
  }
  \caption{Functions $F_{1,7}^{n=10}(a,b)$ (dashed line) and $F_{1,7}^{(g), n=10}(a,b)$ (solid line) at energy $\sqrt s=900$ GeV and $n=10$. The general image (\ref{fig:fig_part2_3a}), and the zoomed image (\ref{fig:fig_part2_3b}) at the neighborhood of maximum point. Clear that in an area that makes the most significant contribution to the integral, the scattering amplitude does not differ from its Gaussian approximation Eq.\ref{eq:eq_part2_12}, which demonstrates the possibility of applying the Laplace method.}
  \label{fig:fig_part2_3}
\end{figure}

Gaussian approximation of expression Eq.\ref{eq:eq_part2_10} for function Eq.\ref{eq:eq_part2_11} denote as:
\begin{eqnarray}
 {A^{\left( g \right)}}\left( {n,{X_1},{X_2},...,{X_{3n + 2}}} \right) = {A^{\left( 0 \right),n}}\left( {\sqrt s } \right) \exp \left( { - \frac{1}{2}\sum\limits_{a = 1}^{3n + 2} {\sum\limits_{b = 1}^{3n + 2} {{D_{ab}}} \left( {{X_a} - X_a^{\left( 0 \right)}} \right)\left( {{X_b} - X_b^{\left( 0 \right)}} \right)} } \right) 
\label{eq:eq_part2_12}
\end{eqnarray}
Since the functions Eq.\ref{eq:eq_part2_11} and Eq.\ref{eq:eq_part2_12} depends on large number of variables, it is impossible to plot them, so we have to introduce a new functions
\begin{eqnarray}
F_{ik}^n\left( {a,b} \right) = A\left( {n,X_1^{\left( 0 \right)}, \ldots ,X_{i - 1}^{\left( 0 \right)},X_i^{\left( 0 \right)} + a,X_{i + 1}^{\left( 0 \right)}, \ldots ,X_{k - 1}^{\left( 0 \right)},X_k^{\left( 0 \right)} + b,X_{k + 1}^{\left( 0 \right)}, \ldots ,X_{3n + 2}^{\left( 0 \right)}} \right)
\label{eq:eq_part2_11a} 
\end{eqnarray}
\begin{eqnarray}
F_{ik}^{\left( g \right),n}\left( {a,b} \right) = {A^{\left( g \right)}}\left( {n,X_1^{\left( 0 \right)}, \ldots ,X_{i - 1}^{\left( 0 \right)},X_i^{\left( 0 \right)} + a,X_{i + 1}^{\left( 0 \right)}, \ldots ,X_{k - 1}^{\left( 0 \right)},X_k^{\left( 0 \right)} + b,X_{k + 1}^{\left( 0 \right)}, \ldots ,X_{3n + 2}^{\left( 0 \right)}} \right)
\label{eq:eq_part2_11b} 
\end{eqnarray}
Three-dimensional curves of these functions Eqs.\ref{eq:eq_part2_11a}, \ref{eq:eq_part2_11b} can be easily plot the in the vicinity of the maximum point (i.e., at the neighborhood of zero of variables $a$ and $b$). The typical examples of such curves are shown in Fig.\ref{fig:fig_part2_2} and Fig.\ref{fig:fig_part2_3}, where it is easy to see that the approximation Eq.\ref{eq:eq_part2_12} works well in the wide energy range. The results similar to ones in Figs.\ref{fig:fig_part2_2},\ref{fig:fig_part2_3} were obtained at different values of $\sqrt s$, $i$, $k$ and $n$. As one can see from Fig.\ref{fig:fig_part2_2} and Fig.\ref{fig:fig_part2_3}, that true amplitude and its Gaussian approximation Eq.\ref{eq:eq_part2_12} start differ visibly only in the parameter region, which makes a insignificant contribution to the integral.

Now let us proceed with identification $A(n,P_3,P_4,p_1,p_2,\ldots,p_2,P_1,P_2)$ in Eq.\ref{eq:eq_part2_5} and define all possible arrangements of indices $1, 2,\ldots, n$ by $P^{(1)}, P^{(2)},\ldots, P^{(n!)}$. The function of variables $X_k$, $k=1,2,\ldots,3n+2$, which corresponds to arrangement $P^{(l)}$, define as $A_{P^{(l)}}(n,X_1,\ldots,X_{3n+2})$. It differs from the function Eq.\ref{eq:eq_part2_11} just by renaming the arguments, and therefore it also has a point of constrained maximum under condition of the energy-momentum conservation. The value of this function at the constrained maximum point is equal to the value of function Eq.\ref{eq:eq_part2_11}, i.e., and it is equal to $A^{(0),n}(\sqrt s)$ according to the replacement made above. Thus, if $X_1^{(0)}, X_2^{(0)},\ldots,X_n^{(0)}$ are the values of variables $X_1,X_2,\ldots,X_n$ at the maximum point, now same values $X_1^{(0)}, X_2^{(0)},\ldots,X_n^{(0)}$ will be the values of the variables $X_{j_1},X_{j_2},\ldots,X_{j_n}$ at the maximum point. 
Analogously $X_{n+1}^{(0)}, X_{n+2}^{(0)},\ldots,X_{2n}^{(0)}$ are the values of variables $X_{n+j_1},X_{n+j_2},\ldots,X_{n+j_n}$ at the maximum point, and $X_{2n+1}^{(0)}, X_{2n+2}^{(0)},\ldots,X_{3n}^{(0)}$ \textendash ~for $X_{2n+j_1},X_{2n+j_2},\ldots,X_{2n+j_n}$. For short we label the index of variable, into which the variable $a$ goes at given rearrangement, as $P^{(i)}(a)$ i.e., the variable $X_a$ replaced by the variable $X_{P^{(l)}}(a)$.

If we denote the matrix of second derivatives of the logarithm of the function $A_{P^{(l)}}$ at the maximum point by $\hat{D}^{P^{(l)}}$, we will get the following approximation for the function $A_{P^{(l)}}$:
\begin{eqnarray}
\begin{array}{l}
 {A_{{P^{\left( l \right)}}}}\left( {n,{X_1},{X_2},...,{X_{3n + 2}}} \right) = {A^{\left( 0 \right),n}}\left( {\sqrt s } \right) \\
\times \exp \left( { - \frac{1}{2}\sum\limits_{\scriptstyle a = 1 \hfill \atop 
  \scriptstyle b = 1 \hfill}^{3n + 2} {D_{{P^{\left( l \right)}}\left( a \right),{P^{\left( l \right)}}\left( b \right)}^{{P^{\left( l \right)}}} \left( {{X_{{P^{\left( l \right)}}\left( a \right)}} - X_a^{\left( 0 \right)}} \right)\left( {{X_{{P^{\left( l \right)}}\left( b \right)}} - X_b^{\left( 0 \right)}} \right)} } \right)
	\end{array}
\label{eq:eq_part2_15}
\end{eqnarray}

Taking into account that Eq.\ref{eq:eq_part2_15} depends on variables $X_{P^{(l)}(a)}$ and $X_{P^{(l)}(b)}$ just as a function $A$ depends on variables $X_a$ and $X_b$ and the second derivative is taken at the same values of arguments, we have
\begin{eqnarray}
D_{{P^{\left( l \right)}}\left( a \right),{P^{\left( l \right)}}\left( b \right)}^{{P^{\left( l \right)}}} = {D_{ab}} \label{eq:eq_part2_16}
\end{eqnarray}

Using Eq.\ref{eq:eq_part2_16} rewrite Eq.\ref{eq:eq_part2_15} in more convenient form. For this purpose introduce the matrices $\hat{P}^{(l)}$, $l=1,2,\ldots,n!$ and by multiplying it with the column $\hat{X}$ of initial variables in Eq.\ref{eq:eq_part2_12}, we get a column in which the variables are arranged in that way so that in place of variable $X_a$ became a variable $X_{P^{(l)}(a)}$. At next iteration taking into account Eq.\ref{eq:eq_part2_16} one can rewrite Eq.\ref{eq:eq_part2_15} in a matrix form in the following way:
\begin{eqnarray}
\begin{array}{l}
 {A_{{P^{\left( l \right)}}}}\left( {n,{X_1},{X_2},...,{X_{3n + 2}}} \right) = {A^{\left( 0 \right),n}}\left( {\sqrt s } \right) \\
\times \exp \left( { - \frac{1}{2}\left( {{{\hat X}^T}{{\left( {{{\hat P}^{\left( l \right)}}} \right)}^T}\hat D{{\hat P}^{\left( l \right)}}\hat X - 2{{\left( {{{\hat X}^{\left( 0 \right)}}} \right)}^T}\hat D{{\hat P}^{\left( l \right)}}\hat X + {{\left( {{{\hat X}^{\left( 0 \right)}}} \right)}^T}\hat D{{\hat X}^{\left( 0 \right)}}} \right)} \right) 
\end{array}
\label{eq:eq_part2_17}  
\end{eqnarray}
where $\hat{X}^{(0)}$ is a column whose elements are the numbers $X_a^{(0)}, a=1,2,\ldots,3n+2$ in the initial arrangement. Eq.\ref{eq:eq_part2_5} can be rewritten in the form:
\begin{eqnarray}
\begin{array}{l}
 \Phi\left( {n,{X_1},{X_2},...,{X_{3n + 2}}} \right) = {\left( {{A^{\left( 0 \right),n}}\left( {\sqrt s } \right)} \right)^2} \exp \left( { - {{\left( {{{\hat X}^{\left( 0 \right)}}} \right)}^T}\hat D{{\hat X}^{\left( 0 \right)}}} \right) \\
\times \sum\limits_{l = 1}^{n!} {\exp \left( { - \frac{1}{2}{{\hat X}^T}{{\hat D}^{\left( l \right)}}\hat X + {{\left( {{{\hat X}^{\left( 0 \right)}}} \right)}^T}{{\hat V}^{\left( l \right)}}\hat X} \right)} \\
\end{array}
\label{eq:eq_part2_18} 
\end{eqnarray}
where
\begin{eqnarray}
{\hat D^{\left( l \right)}} = \hat D + {\left( {{{\hat P}^{\left( l \right)}}} \right)^T}\hat D{\hat P^{\left( l \right)}} \label{eq:eq_part2_19}\\
{\hat V^{\left( l \right)}} = \hat D + \hat D{\hat P^{\left( l \right)}} 
\label{eq:eq_part2_20}
\end{eqnarray}

If now we turn to Eq.\ref{eq:eq_part2_9}, we can see that all the other coefficients (except $\Phi$) under the integral don't change the values under the permutation of arguments. We replace these expressions by their values at the maximum point and take them out from integral. From this, we introduce the following notation:
\begin{eqnarray}
{J^{\left( 0 \right),n}}\left( {\sqrt s } \right) = {\left. {\frac{{\partial {P_{3\parallel }}}}{{\partial {E_p}}}} \right|_{{E_p} = \sqrt s  - \sum\limits_{k = 1}^n {ch\left( {y_k^{\left( 0 \right)}} \right)} }} \label{eq:eq_part2_21}
\end{eqnarray}
where $P_{3\parallel}^{(0)}$ is the value of expression Eq.\ref{eq:eq_part1_7} corresponding to particle momenta, for which the scattering amplitude has maximum,  i.e., at the $X_a^{(0)}$ and nondimensionalized by $m$.

The expression for cross-section in this case can be written in the form:
\begin{eqnarray}
\begin{array}{l}
 {\sigma _n} = \frac{{{{\left( {2\pi } \right)}^2}}}{{16{m^2}\sqrt {s/4 - {M^2}} \sqrt s \sqrt {{M^2} + {{\left( {P_{3\parallel }^{\left( 0 \right)}} \right)}^2}} \sqrt {{M^2} + {{\left( {\sum\limits_{k = 1}^n {{\mathop{\rm sh}\nolimits} \left( {y_k^{\left( 0 \right)}} \right)}  + P_{3\parallel }^{\left( 0 \right)}} \right)}^2}} }}{\left( {\frac{g}{m}} \right)^4}{\left( {\frac{1}{{2{{\left( {2\pi } \right)}^3}}}{{\left( {\frac{\lambda }{m}} \right)}^2}} \right)^n}{\left( {{A^{\left( 0 \right),n}}\left( {\sqrt s } \right)} \right)^2}  \\ 
  \times {J^{\left( 0 \right),n}}\left( {\sqrt s } \right) \exp \left( { - {{\left( {{{\hat X}^{\left( 0 \right)}}} \right)}^T}\hat D\,{{\hat X}^{\left( 0 \right)}}} \right)\sum\limits_{l = 1}^{n!} {\int {\prod\limits_{k = 1}^{3n + 2} {d{X_a}} } } \exp \left( { - \frac{1}{2}{{\hat X}^T}{{\hat D}^{\left( l \right)}}\hat X + {{\left( {{{\hat X}^{\left( 0 \right)}}} \right)}^T}{{\hat V}^{\left( l \right)}}\hat X} \right) \\ 
 \end{array}
\label{eq:eq_part2_22}
\end{eqnarray}

As the value $\sum\limits_{k=1}^n sh(y_k^{(0)})+P_{3\parallel}^{(0)}$ in Eq.\ref{eq:eq_part2_22} is the negative value of longitudinal component of momentum $P_{4\parallel}^{(0)}$ taken at the maximum point, it can be replaced by $P_{3\parallel}^{(0)}$ due to the symmetry properties that have been discussed above.

Multidimensional integrals under the summation sign can be calculated by diagonalizing of quadratic form in the exponent of each of them. Such diagonalization can be numerical realized, for instance, by the Lagrange method. The large number of terms in Eq.\ref{eq:eq_part2_22} is substantial computational difficulty, which we overcame only for the number of particles $n\leq 8$.

To represent results of numerical computations, it is useful to decompose Eq.\ref{eq:eq_part2_22} in the following way:
\begin{eqnarray}
f_p^{\left( n \right)}\left( {\sqrt s } \right) = \exp \left( { - {{\left( {{{\hat X}^{\left( 0 \right)}}} \right)}^T}\hat D{{\hat X}^{\left( 0 \right)}}} \right) \sum\limits_{l = 1}^{n!} {\int {\prod\limits_{k = 1}^{3n + 2} {d{X_a}} } \exp \left( { - \frac{1}{2}{{\hat X}^T}{{\hat D}^{\left( l \right)}}\hat X + {{\left( {{{\hat X}^{\left( 0 \right)}}} \right)}^T}{{\hat V}^{\left( l \right)}}\hat X} \right)}
\label{eq:eq_part2_23}
\end{eqnarray}
\begin{eqnarray}
 {\sigma '_n}\left( {\sqrt s } \right) = \frac{{{{\left( {{A^{\left( 0 \right),n}}\left( {\sqrt s } \right)} \right)}^2}{J^{\left( 0 \right),n}}\left( {\sqrt s } \right)f_p^{\left( n \right)}\left( {\sqrt s } \right)}}{{\sqrt {s/4 - {M^2}} \sqrt s \left( {{M^2} + {{\left( {P_{3\left\| {} \right.}^{\left( 0 \right)}} \right)}^2}} \right)}}
\label{eq:eq_part2_24}
\end{eqnarray}
\begin{eqnarray}
L = \frac{1}{{2{{\left( {2\pi } \right)}^3}}}{\left( {\frac{\lambda }{m}} \right)^2} 
\label{eq:eq_part2_25} 
\end{eqnarray}

Note, that here and in the following sections we will use the "prime" sign in ours notation to indicate that we use a dimensionless quantity that characterized the dependence of the cross-sections on energy, but not their absolute values.

\begin{figure}
  \centering
  \includegraphics[scale=0.95]{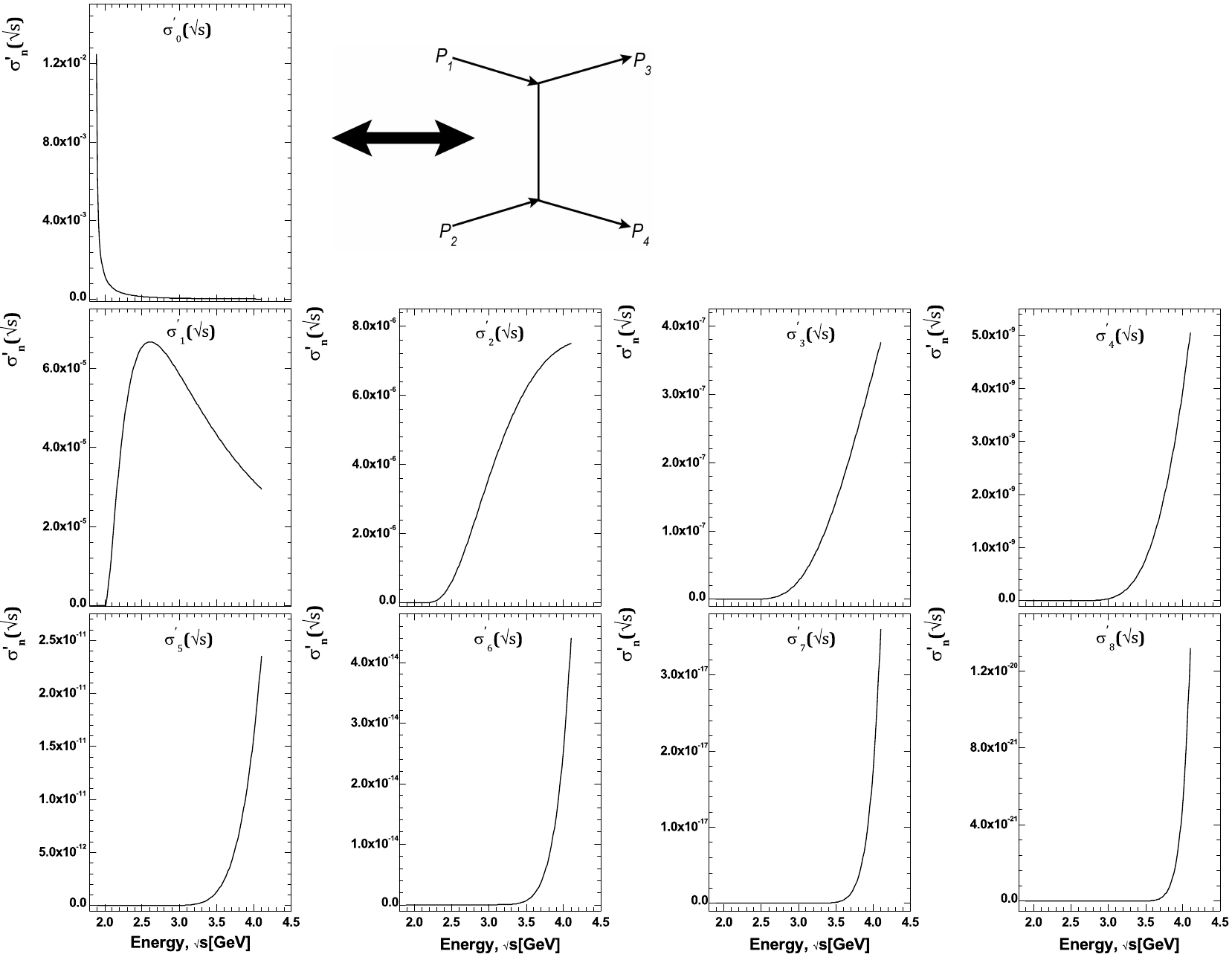} 
  \caption{The calculated values of $\sigma '_n(\sqrt s)$, $n=0,1,\ldots,8$ in the range of threshold energies for $1, 2,\ldots,8$ particle productions. Via $\sigma '_0(\sqrt s)$ was denoted one of the contributions from the diagram (shown on the right) to inelastic scattering cross-section.}
  \label{fig:fig_part2_4}
\end{figure}

\begin{figure}
  \centering
  \includegraphics[scale=0.75]{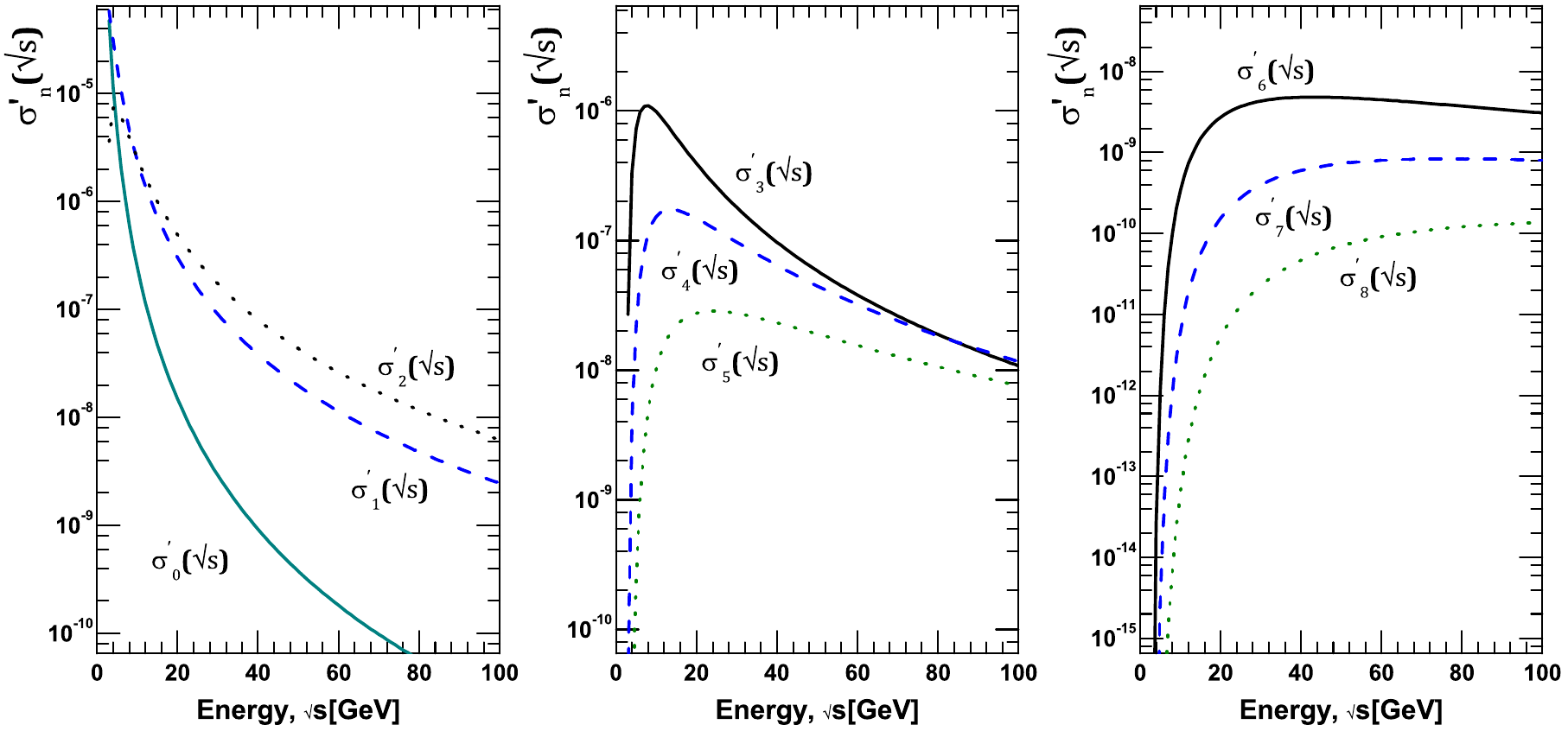} 
  \caption{The calculated values of $\sigma '_n(\sqrt s)$ for the energy range $\sqrt s=3\div95$ GeV}
  \label{fig:fig_part2_5}
\end{figure}

The Eq.\ref{eq:eq_part2_24} differs from the inelastic scattering cross-section $\sigma _n(\sqrt s)$ only by the absence of factor $\frac{\left(2\pi \right)^2}{16m^2}\left( \frac{g}{m} \right)^4\left( \frac{1}{2(2\pi)^3} \left( \frac{\lambda}{m} \right)^2 \right)^n$, which is energy independent and its consideration allow us to trace the dependence of inelastic scattering cross-section on energy $\sqrt s$ (Fig.\ref{fig:fig_part2_4} and Fig.\ref{fig:fig_part2_5}).

From Figs.\ref{fig:fig_part2_4} it is obvious that derivatives of cross-sections with respect to energies along the real axis are equal to zero at points corresponding to the threshold energy of $n$ particle production, i.e., though the threshold values of energy are branch points for the cross-sections, they have continuous first derivative along the real axis at the branch points. 

\begin{figure}
  \centering
  \includegraphics[scale=0.58]{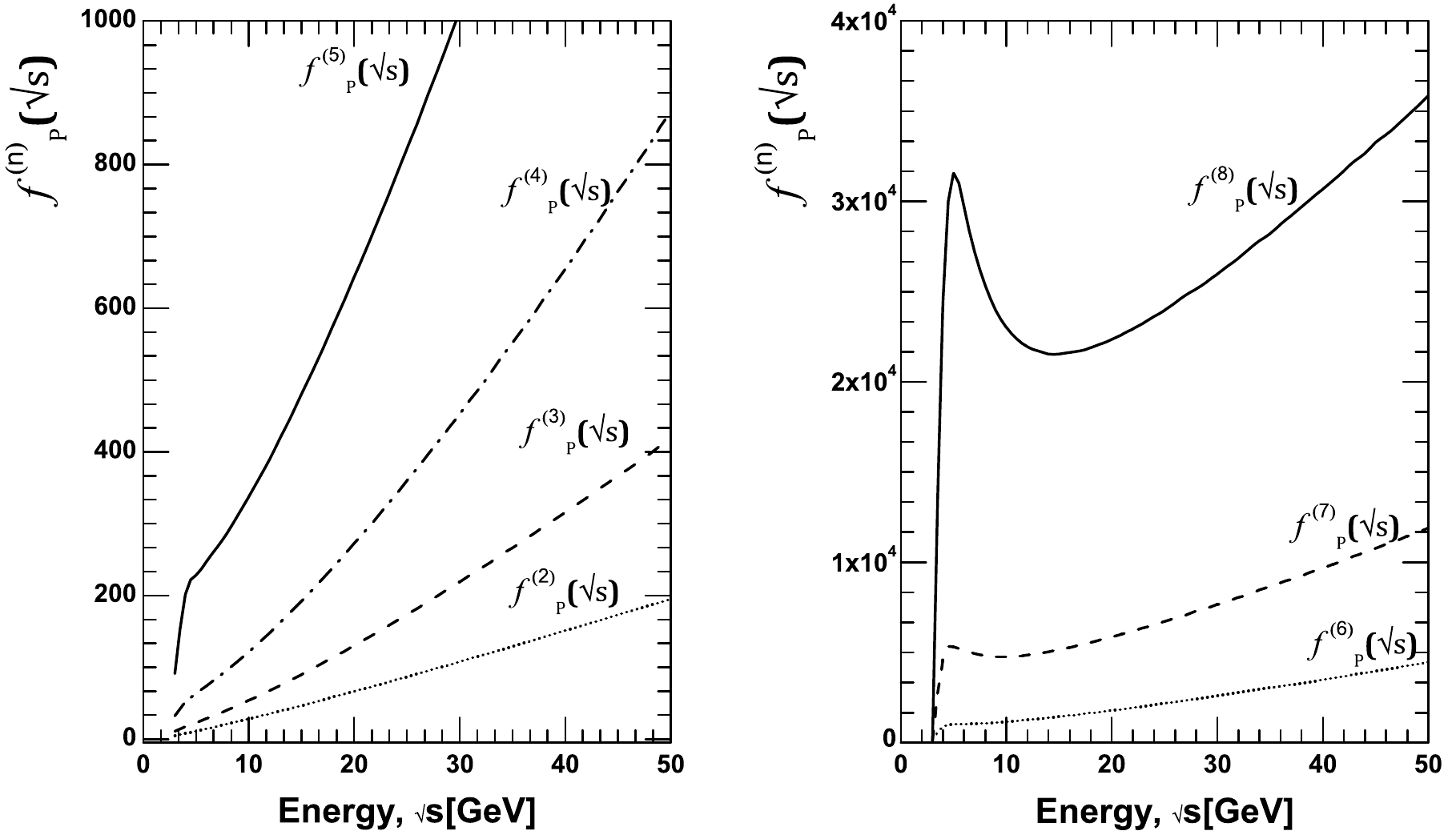} 
  \caption{Calculated values of $f_P^{(n)}(\sqrt s)$ determined by Eq.\ref{eq:eq_part2_23} for the energy range $\sqrt s=3\div95$ GeV. Comparison of these results with Fig.\ref{fig:fig_part2_5} shows that the relation Eq.\ref{eq:eq_part2_24} for the dependence of partial cross section on energy $\sqrt s$ is the most significant dependence ${A^{( 0),n}}(\sqrt s)$  due to the contributions of the longitudinal momentum: at low $n$  $f_P^{( n )}(\sqrt s)$ is increasing, but the partial cross sections Fig.\ref{fig:fig_part2_5} decreasing due to the fact that growth of ${A^{( 0),n}}(\sqrt s)$ is not fast enough, and at large $n$, the presence of decreasing sections  $f_P^{( n )}(\sqrt s)$ have not inhibit the growth of cross-sections at the expense of growth ${A^{( 0),n}}(\sqrt s)$ .}
  \label{fig:fig_part2_6}
\end{figure}

Function $\sigma '_8(\sqrt s)$ monotone increases in the all considered energy range as it follows from Fig.\ref{fig:fig_part2_5}. At the same time from Figs.\ref{fig:fig_part2_6} one can see that $f_P^{(8)}(\sqrt s)$ has drop-down sections. Moreover, even on those sections, where $f_P^{(n)}(\sqrt s), n=2\div5$ increase, corresponding $\sigma '_n(\sqrt s)$ decrease. It makes possible to conclude, that amplitude growth at maximum point (which is the consequence of virtuality reduction) is generally responsible for the growth of inelastic scattering cross-section.

\begin{figure}
  \centering
  \includegraphics[scale=0.4]{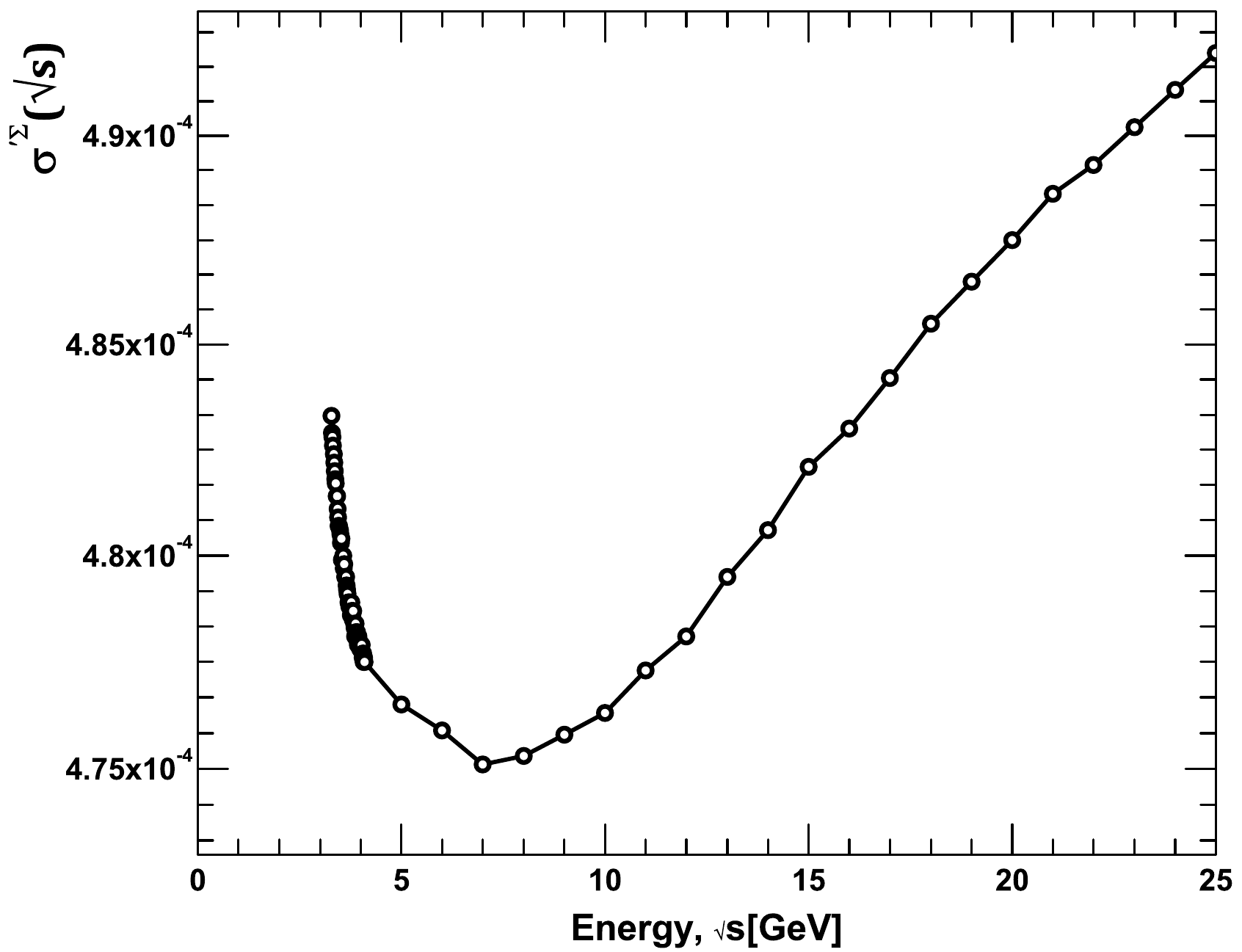} 
  \caption{Calculated values of $\sigma '^{\Sigma} (\sqrt s)$ at $L=5.57$ in the energy range $\sqrt s=3\div25$ GeV. In the presented energy range the cross section qualitatively agrees with that observed in the experiment Ref.\cite{PDG_2010, ATLASCollaboration_2011eu}.}
  \label{fig:fig_part2_7}
\end{figure}

\begin{figure}
  \centering
  \subfigure[]{
  \includegraphics[scale=0.38]{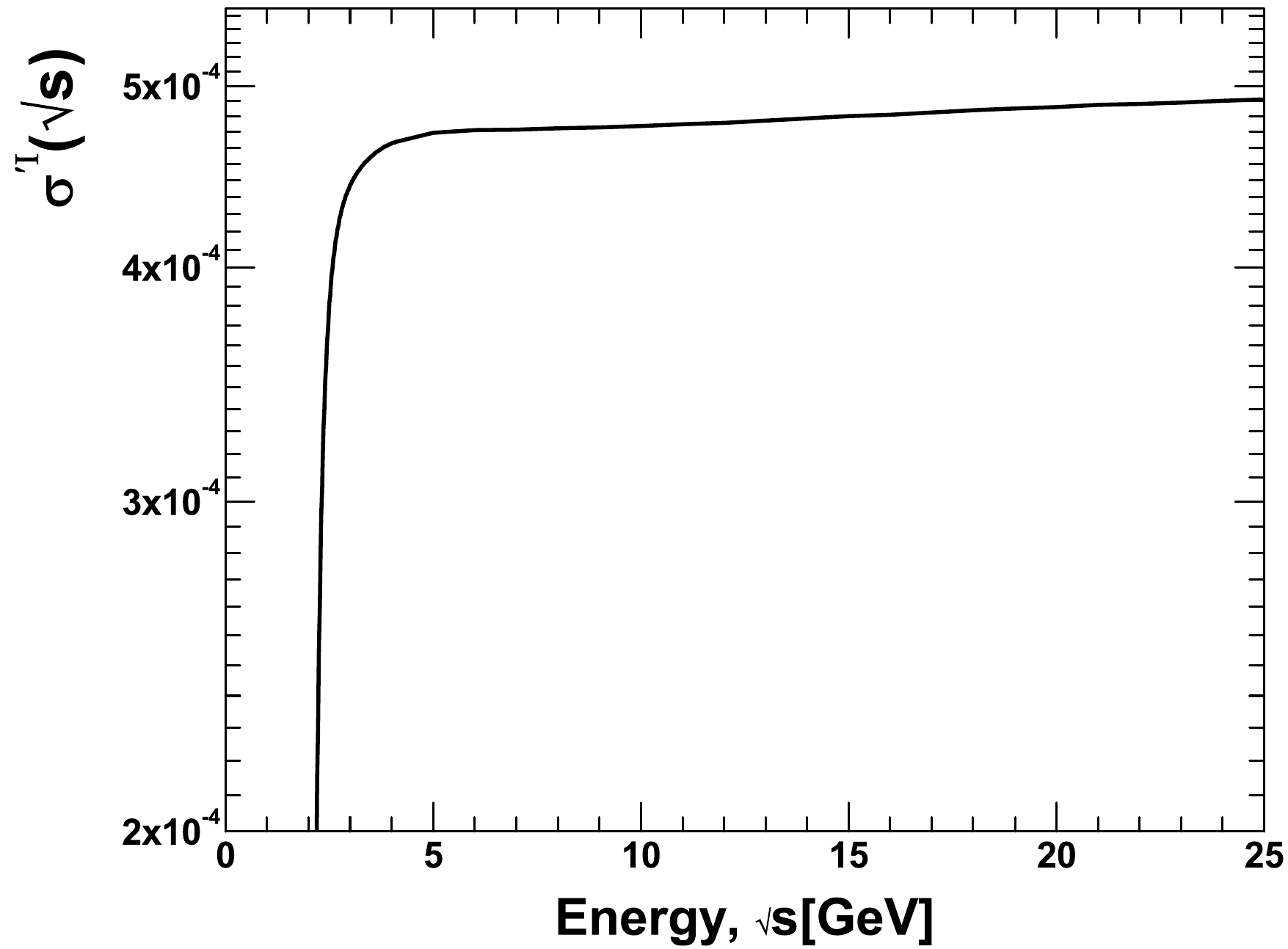}
  \label{fig:fig_part2_8a} 
  }
  \subfigure[]{
  \includegraphics[scale=0.38]{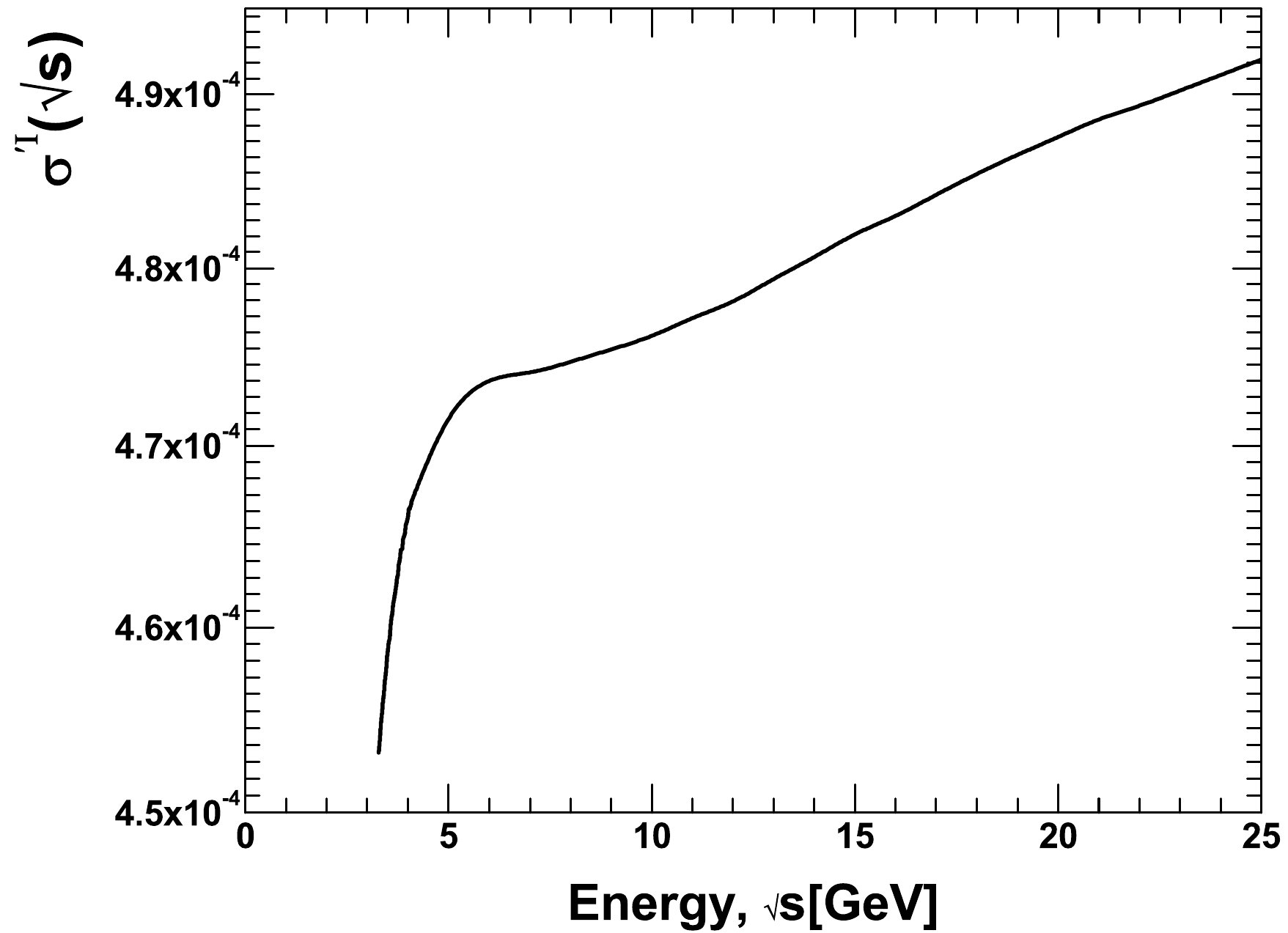}
    \label{fig:fig_part2_8b} 
  }
  \caption{Calculated values of $\sigma '^{I} (\sqrt s)$ at $L=5.57$ in the energy range: \ref{fig:fig_part2_8a} $\sqrt s=1.89\div25$ GeV (starting from the threshold of inelastic scattering); \ref{fig:fig_part2_8b} $\sqrt s=3\div25$ GeV (where increase of total cross-section with energy growth is clearly visible). Presented curves qualitatively agrees with that observed in the experiment Ref.\cite{PDG_2010, ATLASCollaboration_2011eu}.}
  \label{fig:fig_part2_8}
\end{figure}

As it evident from Fig.\ref{fig:fig_part2_4} and Fig.\ref{fig:fig_part2_5} for some values ​​of energy Eq.\ref{eq:eq_part2_24} has a positive energy derivative and for some values ​​of energy Eq.\ref{eq:eq_part2_24} has a negative energy derivative. 
This makes a question. If we form from them a quantities
\begin{eqnarray}
{\sigma '^\Sigma}\left( {\sqrt s } \right) = \sum\limits_{n = 0}^8 {{L^n}{{\sigma '}_n}\left( {\sqrt s } \right)} 
\label{eq:eq_part2_26} 
\end{eqnarray}
and
\begin{eqnarray}
{\sigma '^I}\left( {\sqrt s } \right) = \sum\limits_{n = 1}^8 {{L^n}{{\sigma '}_n}\left( {\sqrt s } \right)} 
\label{eq:eq_part2_27}
\end{eqnarray}
where $L$ is defined by Eq.\ref{eq:eq_part2_25}, is it possible to choose the "coupling constant" $L$ so that the value of Eq.\ref{eq:eq_part2_26} has a characteristic minimum for the total proton-proton scattering cross-section? Answer for this question is positive (see Fig.\ref{fig:fig_part2_7}), i.e., the curves agree qualitatively at the close values of $L$. The energy range shown in Fig.\ref{fig:fig_part2_7} takes into account all the inelastic contributions. We find indeed very interesting result, that curves presented on Fig.\ref{fig:fig_part2_7} and on Fig.\ref{fig:fig_part2_8}, where calculated values of Eqs.\ref{eq:eq_part2_26}- \ref{eq:eq_part2_27} are given at $L=5.57$, qualitatively agree with experimental data \cite{PDG_2010_JofPhysG, ATLASCollaboration_2011eu}.

In this paper we have examined the simplest diagrams of $\phi^3$ theory and we intend to compare the qualitative form of these cross-sections with experimental data, but do not claim quantitative agreement. It is possible to hope that the application of similar computation method to more complicated diagrams in more realistic models will lead to correct outcome.

As known, within the framework of Reggeon theory the drop-down part of total cross-section is described by the reggeons exchanges with interception less than unity \cite{Donnachie1992227, Kaidalov:2003}. The cuts concerned with multireggeon exchanges with participation of reggeons with intercept greater than unity are responsible for the cross-section growth after the reaching the minimum \cite{Collins:111502}.

As will be shown in Section.\ref{sec:analytical_interference}, the accounting of $\sigma '_n(\sqrt s)$ at $n>8$ will not change the behavior of function $\sigma '^{\Sigma}(\sqrt s)$ Eq.\ref{eq:eq_part2_25}. This means that within the framework of given model the summation of multi-peripheral diagrams, when we compute the imaginary part of elastic scattering amplitude, will not result in power dependence on energy because since this dependence is monotonic. This, in turn, means that the corresponding partial amplitude has no pole! And this obviously differs from the results of standard approach in calculations of multi-peripheral model and from the results of Reggeon theory (see f.ex. \cite{springerlink:10.1007/BF02781901}). The detailed discussion of the differences of our approach from the standard one is given in Section.\ref{sec:differences}


\section{The approximate method for taking into account the interference contributions at high multiplicity of final state particles}
\label{sec:analytical_interference}
Accounting of the diagrams, which differ by the order of external lines attachment to the "comb" (interference contributions), is important and can be seen from the following concerns. For the relatively small number of secondary particles ($n\leq8$) we are able to calculate all the interference contributions in the direct way without any approximations (as it was done in the Section \ref{sec:Laplace-method}). Furthermore, one can evaluate the fraction of the contribution from the diagram with initial arrangement of external lines to the sum of contributions of all permutations of external lines. Finally, how it can be noticeably out from Fig.10 of \cite{part2} this fraction is small and therefore, we can't restrict ourselves to accounting of the diagram with the initial line arrangement only at the calculation of inelastic scattering cross-section, as it is usually done at the standard approach to multi-peripheral model \cite{springerlink:10.1007/BF02781901,Collins:111502,Nikitin:113716,agk,levin}. 


Each interference contribution can be computed numerically. However due to the huge number of contributions and large number of secondary particles $n$ the direct numerical calculation of the sum of interference terms is impossible. In this section we present a method which enables us to overcome this problem.

\begin{figure}
  \centering
  \includegraphics[scale=0.65]{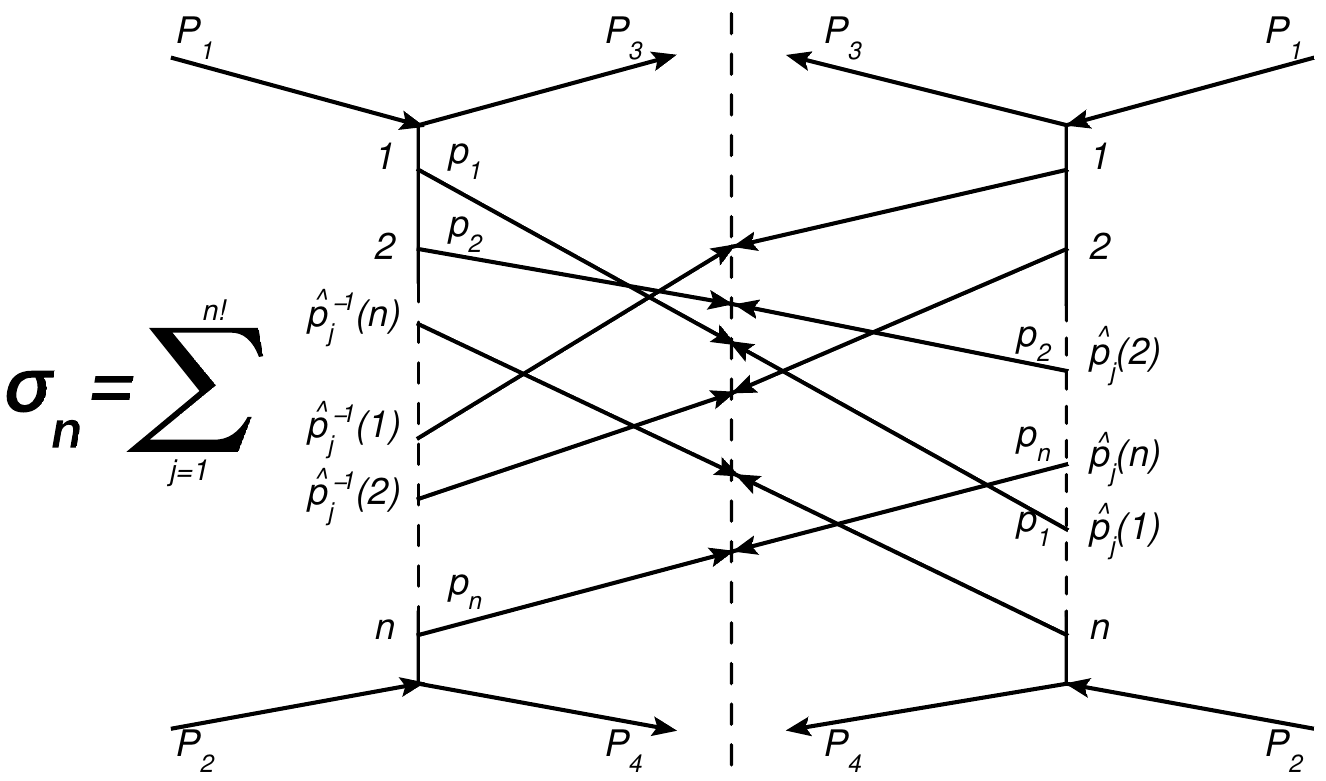} 
  \caption{Representation of the partial cross-section as a sum of "cut" diagrams. The order of joining of lines with four-momenta $p_k$ from the left-hand side of the cut is as following: the line with $p_1$ is joined to the first vertex, the lines with $p_2$ is joined to the second vertex, etc. The order of joining of lines from the right side of cut corresponds to one of the $n!$ possible permutations of the set of numbers $1, 2,\ldots,n$. Where $\hat{P}_j(k),k=1,2,\ldots,n$ denote the number into which a number $k$ goes due to permutation $\hat{P}_j$. An integration is performed over the four-momenta $p_k$ for all "cut lines" taking into account the energy-momentum conservation law and mass shell condition for each of $p_k$.}
  \label{fig:fig_part3_2}
\end{figure}

Let us represent an expression for the partial cross-section as a sum of "cut" diagrams in Fig.\ref{fig:fig_part3_2}. The notation used in this section are the same as before with the exception of
\begin{eqnarray}
\begin{array}{l}
 {X_{3n + 1}} = \frac{1}{2}\left( {{P_{3 \bot x}} - {P_{4 \bot x}}} \right) \quad
 {X_{3n + 2}} = \frac{1}{2}\left( {{P_{3 \bot y}} - {P_{4 \bot y}}} \right) 
\end{array}
\label{eq:eq_part3_1a}
\end{eqnarray}%

As it was shown in previous sections, if we carry out calculations in the c.m.s. of initial particles, the maximum is reached at the zeros values of the transverse momenta and secondary particle rapidities are close to numbers that generate an arithmetic progression. Denoting the difference of this progression through $\Delta y (n, \sqrt s)$ and the rapidity of particle, to which the line attached to the $k$-th vertex of diagram in Fig.\ref{fig:fig_part2_1} corresponds, through $y_k^{(0)}$, we have:
\begin{eqnarray}
 y_k^{\left( 0 \right)} = \left( {\frac{{n + 1}}{2} - k} \right)\Delta y\left( {n,s} \right),k = 1,2, \cdots ,n  
\label{eq:eq_part3_2}
\end{eqnarray}%



The interference contribution corresponding to total "cut" diagram, which refers to the $j$-th summand in Fig.\ref{fig:fig_part3_2}, is proportional to an integral of the product of functions Eq.\ref{eq:eq_part2_10} and Eq.\ref{eq:eq_part2_15} over all variables. Denoting an interference summand corresponding to the permutation $\hat{P}_j$ through $\sigma_n(\hat{P}_j)$ and calculating its Gaussian integral (at the same time, other multipliers besides the squared modulus of scattering amplitude in an integrand are approximately substituted for their values at the maximum point), we have
\begin{eqnarray}
 {\sigma '_n}\left( {{{\hat P}_j}} \right) = \frac{{{{\left( {A\left( {{{\hat X}^{\left( 0 \right)}}} \right)} \right)}^2}v\left( {\sqrt s } \right)}}{{\sqrt {\det \left( {\frac{1}{2}\left( {\hat D + \hat P_j^T\hat D{{\hat P}_j}} \right)} \right)} }} \exp \left( { - \frac{1}{2}\left( {{{\left( {\Delta \hat X_j^{\left( 0 \right)}} \right)}^T}{{\hat D}^{\left( j \right)}}\Delta \hat X_j^{\left( 0 \right)}} \right)} \right) \label{eq:eq_part3_4}
\end{eqnarray}%
where we use the following designations: $\Delta \hat{X}_j^{(0)}=\hat{X}^{(0)}-\hat{P}_j^{-1}(\hat{X}^{(0)})$, $\hat{D}^{(j)}=\left( \hat{D}^{-1}+\hat{P}_j^T \hat{D}^{-1}\hat{P}_j \right)^{-1}$, $v(\sqrt s) \equiv \left( 2{\sqrt s \sqrt{s/4-M^2}\left( E_P/2 \right)\sqrt{ \left( E_P/2 \right)^2-M^2  }} \right)^{-1}$ (here $A\left( \hat{X}^{(0)} \right) \equiv A^{(0),n}$ is the value of scattering amplitude at the constrained maximum point)
It should be reminded that here and further we use the "prime" sign to define the dimensionless cross-section, as in this paper we are interested in the qualitative behavior of cross-section values, and we do not claim quantitative agreement.

The essence of our method is as follows. 
The maximum in the right part of cut diagram in Fig.\ref{fig:fig_part3_2} is attained at $\hat{X}=\hat{P}_j^{-1}\left( \hat{X}^{(0)} \right)$. In other words, a maximum of function, which is associated with the right-hand part of cut diagram, can be obtained from a maximum of function, which maps with the left-hand part of cut diagram, by the rearrangement of arguments. Then the value of each interference contribution is determined by the distance between points of maximum in the right-hand and left-hand part of cut diagram as well as by the relative position of these maximum points, since in different directions contributions to scattering amplitude fall off with distance from point of maximum, in general, with different rate, and also by the relative position of proper directions of the matrices $\hat{D}$ and $\hat{P}_j^T\hat{D}\hat{P}_j$. In other words, multiplying Gaussian functions corresponding to the right-hand and to the left-hand part of interference diagrams in Fig.\ref{fig:fig_part3_2} each time we will obtain as a result Gaussian function, which has the proper value at the maximum point (which we call the "height" of the maximum) and the proper multidimensional volume cutout by resulting Gaussian function from an integration domain (which we call the "width" of the maximum).

We assume that summands in Fig.\ref{fig:fig_part3_2} are arranged in ascending order of the distance between the maximum points in the right-hand part and left-hand part of cut diagram (we denote this distance through $r$) so that "cut" diagram with the initial attachment of lines to the right-hand part of diagram corresponds to $j=1$. In other words, the line of secondary particle with the four-momentum $p_i$ is attached to the $i$-th top in the right-hand part of cut diagram in Fig.\ref{fig:fig_part3_2}. As follows from Eq.\ref{eq:eq_part3_4}, the interference contribution exponentially decrease with the $r^2$ growth. However, in spite of this the interference contributions do not become negligible due to their huge number, which, as discussed below, are increases very rapidly with $r^2$ growth. The value of $r^2$ is proportional to the square of magnitude $\Delta y(n, \sqrt s)$, which, as was noted above, is zero on the threshold of $n$ particle production and slowly increases with distance from this threshold. Therefore, for each number $n$ there is fairly wide range of energies close to the threshold,in which the sharpness of decrease of the interference contributions with the $r^2$ increase is small in the sense that it is less important factor than the increase in their number. At such energies, which we call "low", the partial cross-section $\sigma_n$ is determined by the sum of huge number of small interference contributions. When the magnitude $\Delta y(n, \sqrt s)$ is increased with the further growth of energy $\sqrt s$, the decrease rate of interference contributions increases, while the growth rate of their number with the $r^2$ increase does not change with energy. At such energies, which we call "high", the main contribution to the partial cross-section is made by the relatively small number of interference terms corresponding to the small $r^2$, which can be calculated by Eq.\ref{eq:eq_part3_4}.

If we compose the $n$-dimensional vector (we denote it through $\vec{y}^{(0)}$) from the particle rapidities Eq.\ref{eq:eq_part3_2}, which constrainedly maximizes the function associated with the diagram with the initial arrangement of momenta in Fig.\ref{fig:fig_part2_1}, vectors maximizing the functions with another momentum arrangement will differ from the initial vector only by the permutation of components, i.e., these vectors have the same length. Consider two such $n$-dimensional vectors, one of which corresponds to the initial arrangement, and another - to some permutation, then in the $n$-dimensional space it is possible to ``stretch'' a two-dimensional plane on them (as a set of their various linear combinations), where two-dimensional geometry takes place. Therefore, the distance $r$ will be determined by cosine of an angle between the considered equal on length $n$-dimensional rapidity vectors in the two-dimensional plane, ``stretched'' on them. An angle corresponding to the $\hat{P}_j$ permutation we designate through $\theta _j$, $0\leq\theta _j \leq \pi$.

Thus, each of the terms in the sum Fig.\ref{fig:fig_part3_2} can be uniquely match to its angle $\theta _j$. At the same time the variable $z=cos(\theta)$ is more handy for consideration than an angle $\theta _j$. Using Eq.\ref{eq:eq_part3_2}, can be shown that the variable $z$ can take discrete set of values:
\begin{eqnarray}%
{z_l} = 1 - \frac{{12}}{{\left( {n - 1} \right)n\left( {n + 1} \right)}}l;
\quad l = 0,1, \cdots ,\frac{{\left( {n - 1} \right)n\left( {n + 1} \right)}}{6}  
\label{eq:eq_part3_1_v2}
\end{eqnarray}%

Note that although the relation Eq.\ref{eq:eq_part3_2} for the rapidities of secondary particles is satisfied with high accuracy at the maximum point, it is still approximate. This means that those contributions, to which matched one and the same value of variable $z$ in Eq.\ref{eq:eq_part3_2}, in fact, matched a slightly different from each other values of $z$.

\begin{figure}
  \centering
  \subfigure[]{
  \includegraphics[scale=0.4]{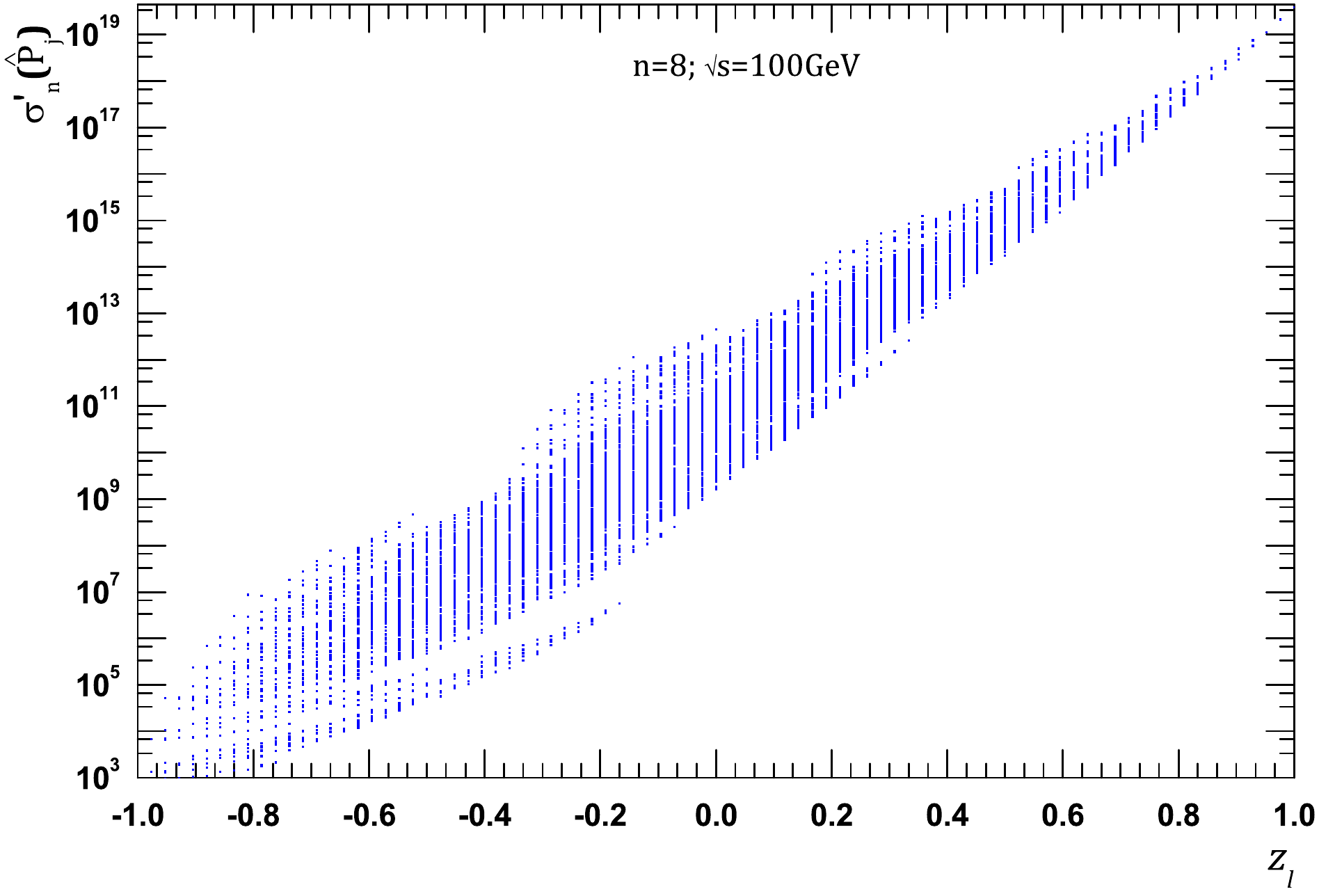}
  \label{fig:fig_part3_3a}
  }
  \subfigure[]{
  \includegraphics[scale=0.4]{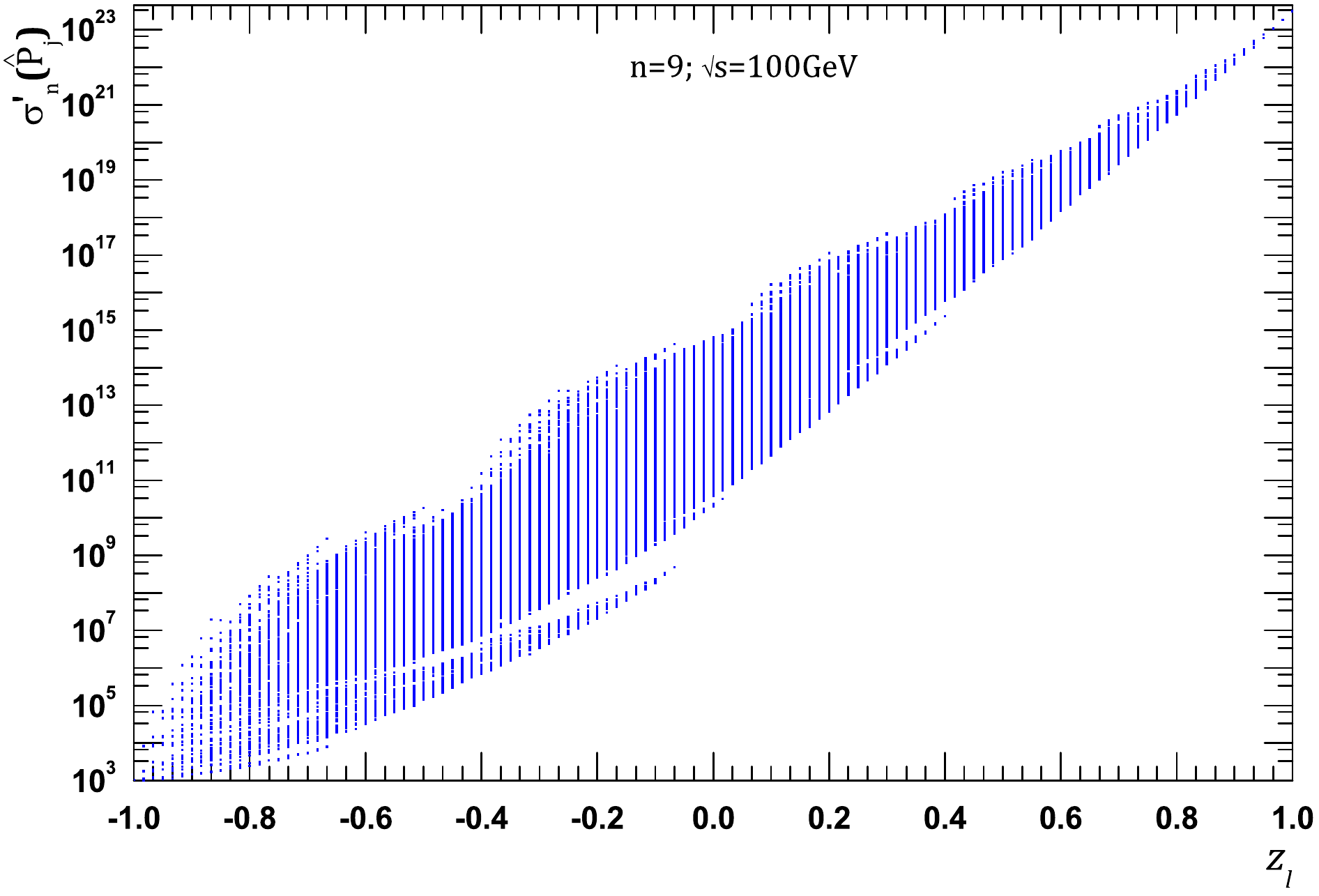}  
    \label{fig:fig_part3_3b} 
  }    
  \caption{The dependence of interference contributions on $z_l$ at $\sqrt{s} = 1000$ GeV: (a) $n=8$, (b) $n=9$. Here and in subsequent figures the interference contributions divided by the common multiplier $exp\left( -\sum\limits_{a=1}^{3n+2}\sum\limits_{b=1}^{3n+2}X_a^{(0)}D_{ab}X_b^{(0)} \right)$ are indicated on the $y$-axis. Obviously, that to the one value of $z_l$ corresponds a lot of different contributions, as well as that the average values ​​of the logarithms of these contributions are placed approximately on a straight line (see below Eq.\ref{eq:eq_part3_12} and Fig.\ref{fig:fig_part3_4}).}
  \label{fig:fig_part3_3}
\end{figure}

Consequently, to such contributions correspond a similar but unequal to each other distances between maximum points in a "cut" diagram. In addition, this distance, as was discussed above, is not a unique factor affecting to the value of interference contribution. Therefore, if each interference contribution is associated with the value of variable $z$ by the approximation Eq.\ref{eq:eq_part3_2}, it appears, that the different values of interference contributions correspond to the one and the same value of $z_l$ (Fig.\ref{fig:fig_part3_3}).

Thus, while each contribution is associated to some value of variable $z$ in the approximation Eq.\ref{eq:eq_part3_2}, the value of contribution is not the unique function of $z$. However, the sum expressing the partial cross-section $\sigma_n$ can be written in the following way
\begin{eqnarray}%
 {\sigma '_n} = \sum\limits_{l = 0}^{\frac{{\left( {n - 1} \right)n\left( {n + 1} \right)}}{6}} {\Delta {N_l}\left( {\frac{{\sum\limits_{{z_j} = {z_l}} {{\sigma '_n}\left( {{{\hat P}_j}} \right)} }}{{\Delta {N_l}}}} \right)} 
\label{eq:eq_part3_2_v2}
\end{eqnarray}%
where $\Delta N_l$ is the number of summands to which the value $z_j = z_l$ corresponds in the approximation Eq.\ref{eq:eq_part3_2}. The average value of all interference contributions in Eq.\ref{eq:eq_part3_2_v2} is already the unique function of $z_l$. Therefore, we introduce notation
\begin{equation}
\left\langle \sigma '_n(z_l) \right\rangle = \frac{\sum\limits_{z_j=z_l}\sigma '_n \left( \hat{P}_j \right)}{\Delta N_l}
\label{eq:eq_part3_3_v2}
\end{equation}
where $\left\langle \sigma '_n(z_l) \right\rangle$ is some function, whose form at "low" energies can be determined from the following considerations.

For any multiplicity $n$ when the values of parameter $l$ in Eq.\ref{eq:eq_part3_1_v2} are small and when number of corresponding interference contributions is relatively small, we can directly calculate these elements and their sum. Denote the maximum value $l$, for which all interference contributions are calculated through $l_0$. In particular, in this paper we managed to calculate the interference contributions up to $l_0=6$. Partial cross-section can be written as 
\begin{equation}
\sigma '_n=\sigma_n^{\prime(h)}+\sigma_n^{\prime(l)}= \underbrace{\sum\limits_{\scriptsize{\begin{array}{c} z_j=z_l,\\ l=0,1,\ldots,l_0 \end{array}}}\sigma '_n\left( \hat{P}_j \right)}_{\sigma_n^{\prime(h)}} + \underbrace{\sum\limits_{l=l_0+1}^{\frac{(n-1)n(n+1)}{6}} \Delta N_l \left\langle \sigma '_n(z_l) \right\rangle}_{\sigma_n^{\prime(l)}}
\label{eq:eq_part3_4_v2}
\end{equation}
where $\sigma_n^{\prime(h)}$ is the sum of contributions sufficient at "high" energies, and $\sigma_n^{\prime(l)}$ is the sum of contributions sufficient at "low" energies. Thus, the difficulties in the calculations of the huge number of interference contributions 
mainly relates to the range of "low" energies and can be reduced to the approximate calculation of $\left\langle \sigma '_n(z_l) \right\rangle$ and $\Delta N_l$. The remaining part of the section is dedicated to this task.

As follows from Eq.\ref{eq:eq_part3_4}, the exponential factor exerts the most significant effect on the dependence of $\left\langle \sigma '_n(z_l) \right\rangle$ on $z_l$. Note that the expression $\left( \Delta \hat{X}_j^{(0)}\right)^T \hat{D}^{(j)} \Delta \hat{X}_j^{(0)}$ entering into the exponent in Eq.\ref{eq:eq_part3_4} depends only on those matrix $\hat{D}^{(j)}$ components, which are at the intersection of the first $n$ rows and first $n$ columns, since all column $\Delta \hat{X}_j^{(0)}$ components starting with $n+1$ are zero, because they are the particle momentum transverse components at the maximum point. If we denote a matrix composed of elements located at the intersection of the first $n$ rows and first $n$ columns of the matrix $\hat{D}^{(j)}$ through $\hat{D}_y^{(j)}$ and a matrix, which is obtained from the matrix $\hat{D}$ in analogy, through $\hat{D}_y$, we have 
\begin{equation}
\hat{D}_y^{(j)} = \left( \hat{D}_y^{-1} +\hat{P}_j^T \hat{D}_y^{-1}\hat{P}_j  \right)^{-1}
\label{eq:eq_part3_4a}
\end{equation}

\begin{figure}
  \centering
  \subfigure[]{
  \includegraphics[scale=0.75]{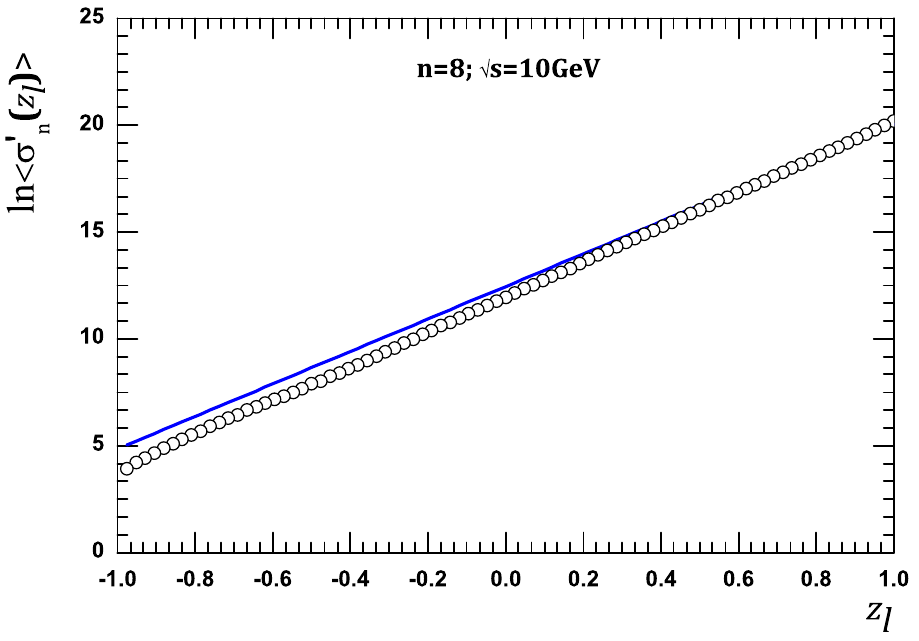}
  \label{fig:fig_part3_4a}
  }
  \subfigure[]{
  \includegraphics[scale=0.75]{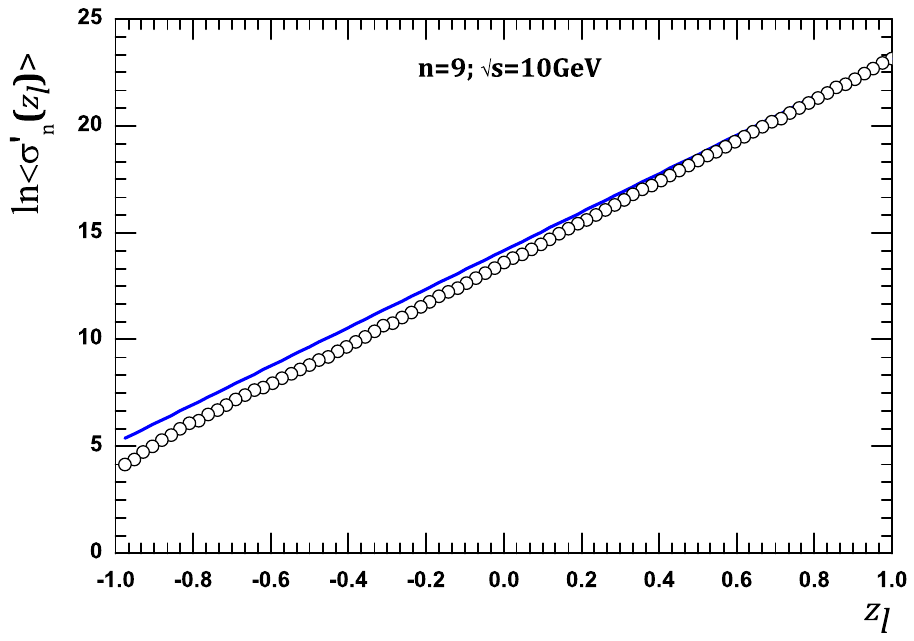}  
    \label{fig:fig_part3_4b} 
  }    
  \subfigure[]{
  \includegraphics[scale=0.75]{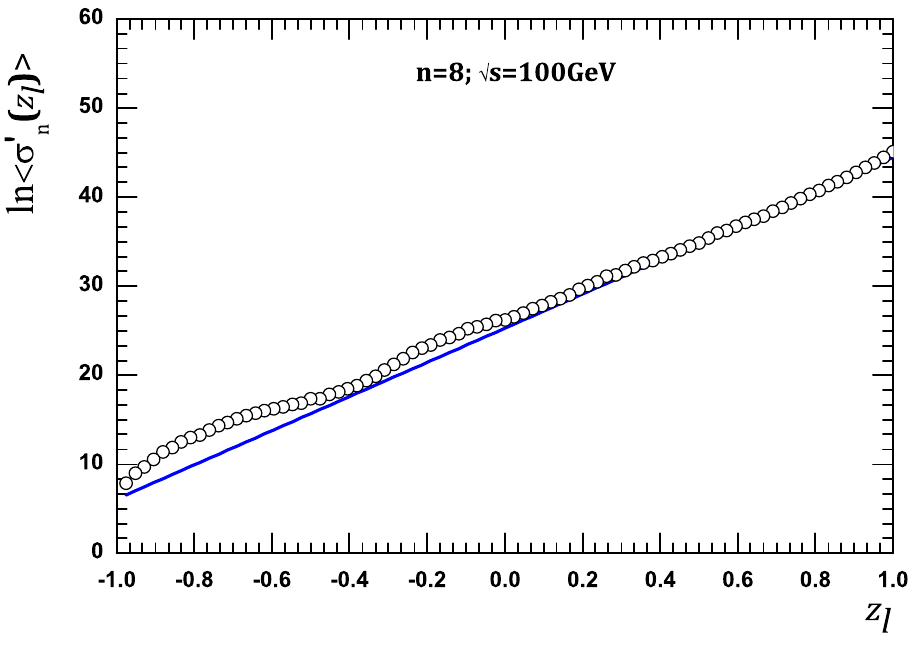}
  \label{fig:fig_part3_4c}
  }
  \subfigure[]{
  \includegraphics[scale=0.75]{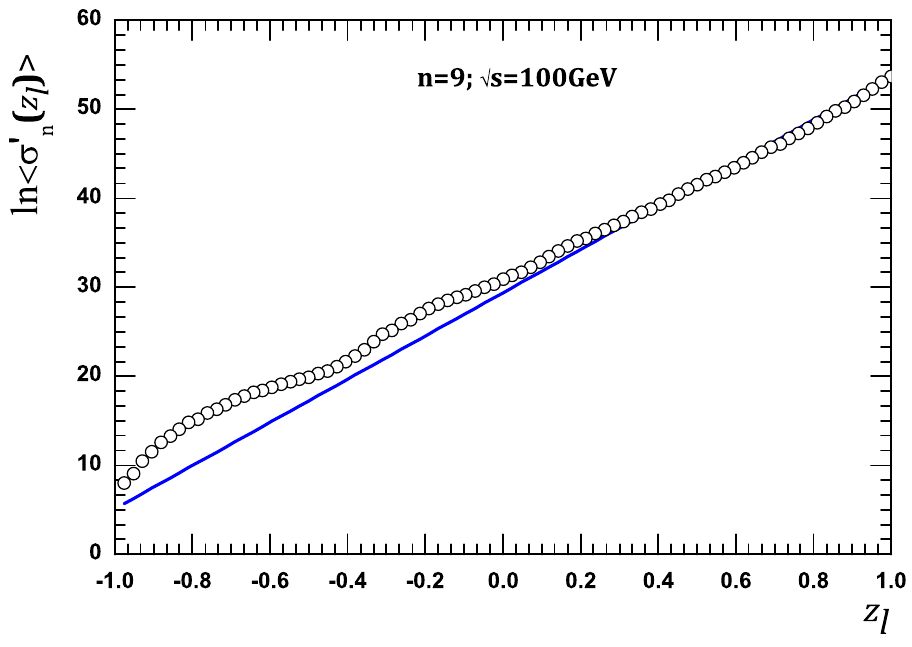}  
    \label{fig:fig_part3_4d} 
  }    
  \caption{Comparing the values of $\ln\left(\left\langle \sigma_n(z_l) \right\rangle\right)$ obtained by a direct numerical calculation with consideration of all interference contributions (circles) and by approximation Eq.\ref{eq:eq_part3_12} (solid line) at $n=8$, $\sqrt s = 10$ GeV \ref{fig:fig_part3_4a}; \ref{fig:fig_part3_4b} $n=9$, $\sqrt s = 10$ GeV; \ref{fig:fig_part3_4c} $n=8$, $\sqrt s = 100$ GeV; \ref{fig:fig_part3_4d} $n=9$, $\sqrt s = 100$ GeV.}
  \label{fig:fig_part3_4}
\end{figure}

At rather low energies (see \cite{part3}) eigenvalues of the matrix $\hat D_y^{(j)}$ are approximately equal between themselves. In such an approximation we obtain 
\begin{eqnarray}%
 \left\langle {{\sigma '_n}\left( {{z_l}} \right)} \right\rangle  = {\left( {A\left( {{{\hat X}^{\left( 0 \right)}}} \right)} \right)^2}v\left( {\sqrt s } \right) \exp \left( { - \frac{{{{\left| {{{\vec y}^{\left( 0 \right)}}} \right|}^2}Sp\left( {{{\hat D}_y}} \right)}}{{2n}}\left( {1 - {z_l}} \right)} \right) 
 \left\langle {w\left( {{z_l}} \right)} \right\rangle
 \label{eq:eq_part3_10}
\end{eqnarray}%

with
\begin{eqnarray}%
 \left\langle {w\left( {{z_l}} \right)} \right\rangle  = \frac{1}{{\Delta {N_l}}}\sum\limits_{{z_j} = {z_l}} {\frac{1}{{\sqrt {\det \left( {\frac{1}{2}\left( {\hat D + \hat P_j^T\hat D{{\hat P}_j}} \right)} \right)} }}} 
 \label{eq:eq_part3_11}
\end{eqnarray}%
Assuming that the multiplier $\left\langle w(z_l) \right\rangle$ is weakly dependent on $z_l$ , we obtain
\begin{eqnarray}%
 \left\langle {{\sigma '_n}\left( {{z_l}} \right)} \right\rangle  = \left\langle {{\sigma '_n}\left( {{z_{{l_0}}}} \right)} \right\rangle \exp \left( {\frac{{{{\left| {{{\vec y}^{\left( 0 \right)}}} \right|}^2}Sp\left( {{{\hat D}_y}} \right)}}{{2n}}\left( {{z_l} - {z_{{l_0}}}} \right)} \right) 
\label{eq:eq_part3_12}
\end{eqnarray}%
where $z_{l_0}$ is the minimum value of $z_l$ for which all interference contributions can be numerically calculated. Therefore, we can also numerically calculate the magnitude $\left\langle \sigma '_n(z_{l_0}) \right\rangle$. From Fig.\ref{fig:fig_part3_4}, where the values of $\ln\left(\left\langle \sigma '_n(z_l) \right\rangle\right)$ obtained by numerical calculation over all interference contributions and obtained with Eq.\ref{eq:eq_part3_12} are compared, it follows that such an approximation is acceptable at "low" energies.

In order to roughly estimate the function $\left\langle w(z_l) \right\rangle$ we can replace it by the Taylor series expansion taking into account the contributions no higher than linear as it was shown in \cite{part3}. We will designate the expansion coefficientds through $w_0$ and $w_1$. The values of $w_0$ and $w_1$ are found by the calculation of $\left\langle w(z_l) \right\rangle$ at $z_l$ close to $1$ and $(-1)$. 
So, we have the following expression instead of Eq.\ref{eq:eq_part3_12}
\begin{eqnarray}%
 \left\langle {{\sigma '_n}\left( {{z_l}} \right)} \right\rangle  = \left\langle {{\sigma '_n}\left( {{z_{{l_0}}}} \right)} \right\rangle \left( {{w_0} + {w_1}\left( {1 - {z_l}} \right)} \right) \exp \left( {\frac{{{{\left| {{{\vec y}^{\left( 0 \right)}}} \right|}^2}Sp\left( {{{\hat D}_y}} \right)}}{{2n}}\left( {{z_l} - {z_{{l_0}}}} \right)} \right) 
\label{eq:eq_part3_13}
\end{eqnarray}%

Now let us turn to approximate calculation of $\Delta N_l$ and use a new variable
\begin{eqnarray}%
 Y_k^{\left( 0 \right)} = \frac{{y_k^{\left( 0 \right)}}}{{\Delta y\left( {n,s} \right)\sqrt {\frac{{\left( {n + 1} \right)n\left( {n - 1} \right)}}{{12}}} }} 
\label{eq:eq_part3_14}
\end{eqnarray}%
where $y_k^{(0)}$ are determined by Eq.\ref{eq:eq_part3_2}, $Y_k^{(0)}$, $k =1,2,\ldots,n$ are considered as the components of vector $\vec{Y}^{(0)}$, which, as it follows from Eq.\ref{eq:eq_part3_14} is of unit length.

Thus, the angle $\theta_j$ between the vector $\vec y^{(0)}=\left( y_1^{(0)},y_2^{(0)},\ldots,y_n^{(0)} \right)$ and vector $\hat P_j^{-1} \left( \vec y^{(0)} \right)$ obtained by the permutation of corresponding components is the same as the angle between the vector $\vec Y^{(0)}=\left( Y_1^{(0)},Y_2^{(0)},\ldots,Y_n^{(0)} \right)$ and $\hat P_j^{-1} \left( \vec Y^{(0)} \right)$. Moreover, as it follows from Eq.\ref{eq:eq_part3_2}
\begin{eqnarray}%
 y_1^{\left( 0 \right)} =  - y_n^{\left( 0 \right)},y_2^{\left( 0 \right)} =  - y_{n - 1}^{\left( 0 \right)}, \cdots ,y_k^{\left( 0 \right)}  =  - y_{n - k + 1}^{\left( 0 \right)}, \quad k = 1,2, \cdots ,n 
\label{eq:eq_part3_15}
\end{eqnarray}%

\begin{figure}
  \centering
  \subfigure[]{
  \includegraphics[scale=0.23]{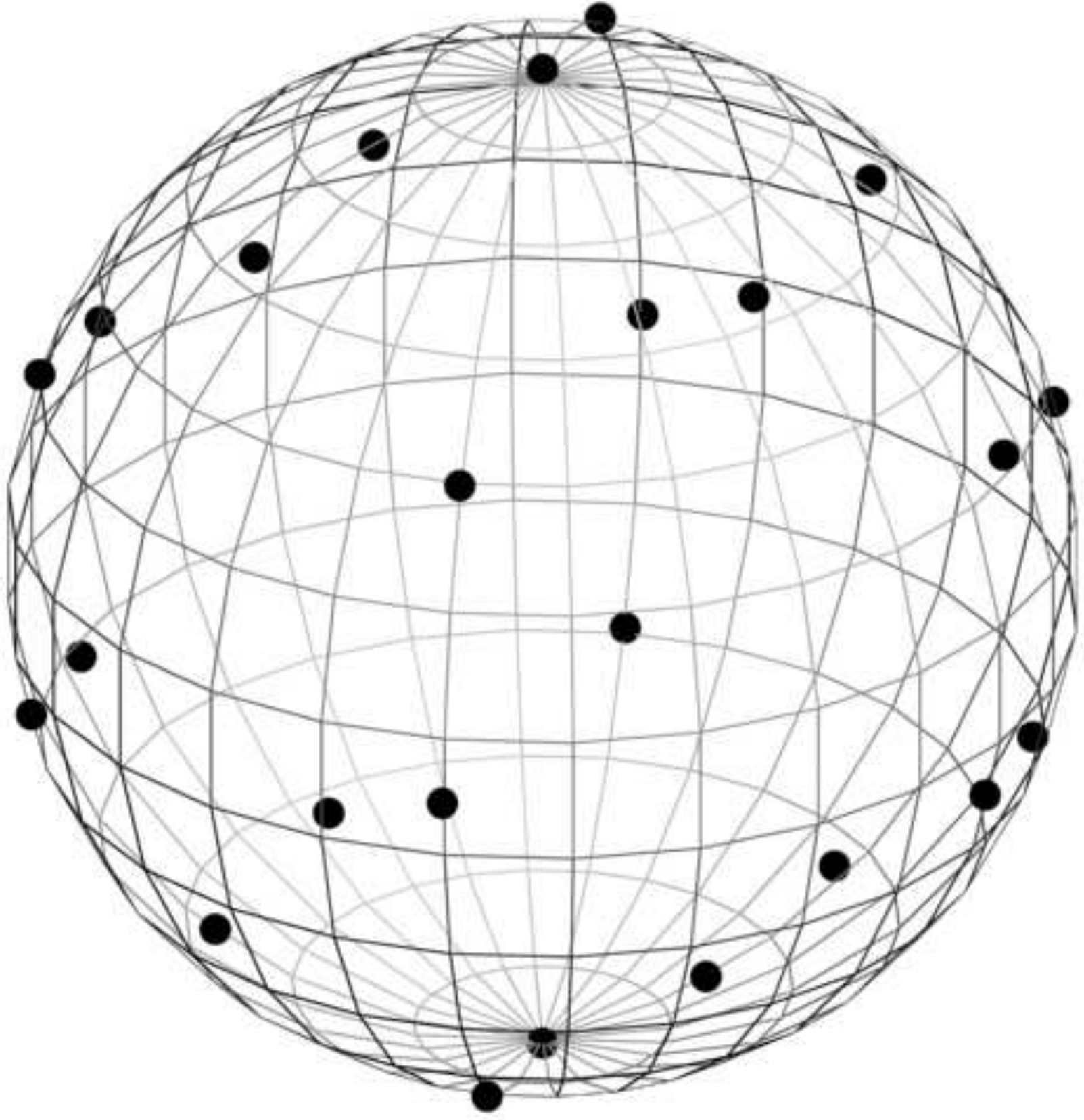} 
  \label{fig:fig_part3_6a} 
  }
  \subfigure[]{
  \includegraphics[scale=0.23]{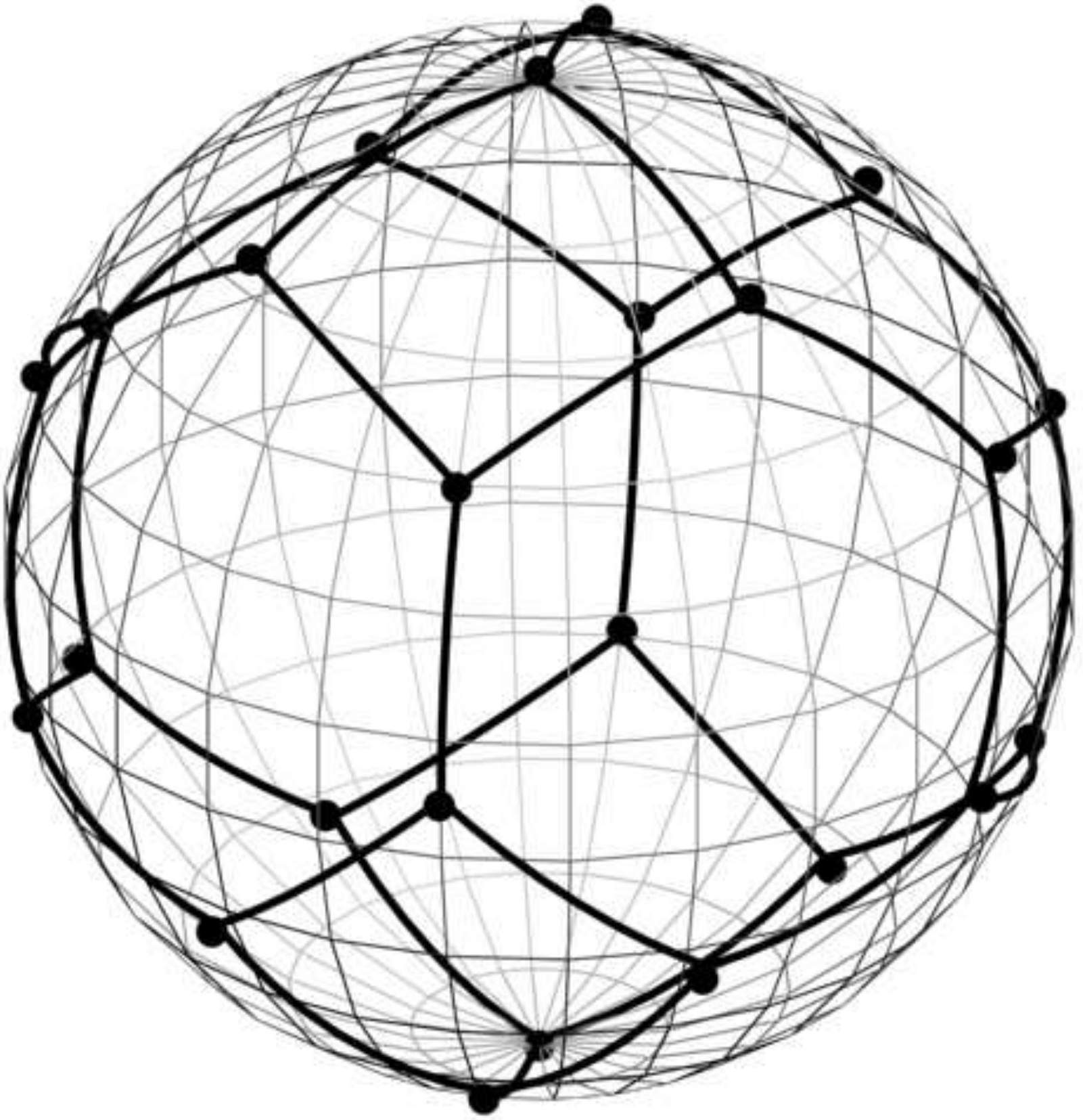}  
    \label{fig:fig_part3_6b} 
  }     
  \caption{ \ref{fig:fig_part3_6a} - a sphere $S_2$ and figure $F_{4!}$ (is shown by points). Basis in the four-dimensional space is chosen so that one of the basis vectors coincide with vector $\vec e_4 = \left( \frac{1}{2},\frac{1}{2},\frac{1}{2},\frac{1}{2} \right)$, and the three basis vectors of three-dimensional subspace, into which depicted sphere is embedded, are perpendicular to $\vec e_4$; \ref{fig:fig_part3_6b} - shortest arcs joining the points of figure $F_{4!}$: into the two "hexagonal" and one "tetragonal" regions.}
  \label{fig:fig_part3_6}
\end{figure}

It follows from here that all vectors $\hat P_j^{-1} \left( \vec Y^{(0)} \right)$ are orthogonal to vector $\vec e_n = \left( \underbrace{1/\sqrt n,1/\sqrt n,\ldots,1/\sqrt n }_{n\text{ components}} \right)$. 
Therefore, considering vectors $\hat P_j^{-1} \left( \vec Y^{(0)} \right)$ as the elements of $n$-dimensional euclidean space, which we denote through $E_n$, then the ends of all vectors $\hat P_j^{-1} \left( \vec Y^{(0)} \right)$ are lie on the unit sphere embedded into the $(n-1)$-dimensional subspace of $E_n$. We denote this sphere through $S_{n-2}$ and shape formed by the set of points in which the ends of vectors $\hat P_j^{-1} \left( \vec Y^{(0)} \right)$ ($j=1,2,\ldots,n!$) come, denote through $F_{n!}$. In particular, when $n=4$ the sphere $S_2$ and figure $F_{4!}$ graphically look like in Fig.\ref{fig:fig_part3_6}.

The consideration of geometrical properties of figure $F_{n!}$ in \cite{part3} makes it possible to conclude, that at an arbitrary $n$ a sphere $S_{n-2}$ can be divided into the parts of equal area, each of which contains only one point of figure $F_{n!}$  (as it shown in Fig.7(b-c) of Ref.\cite{part3}).

Let us introduce a multidimensional spherical coordinate system so that the end of vector $\vec Y^{(0)}$ is the "north pole" of sphere $S_{n-2}$. Then the number of figure $F_{n!}$ points, to which the values of variable $z=\cos(\theta)$ in the interval $[z, z+dz]$ correspond, is equal to
\begin{eqnarray}
dN\left( {z,dz} \right) = \rho \left( z \right)dz 
\label{eq:eq_part3_22}
\end{eqnarray}%
where
\begin{eqnarray}
 \rho \left( z \right) = \frac{{n!}}{{\sqrt \pi  }}\frac{{\Gamma \left( {\frac{{n - 1}}{2}} \right)}}{{\Gamma \left( {\frac{{n - 2}}{2}} \right)}}{\left( {1 - {z^2}} \right)^{\frac{{n - 4}}{2}}} 
\label{eq:eq_part3_22a}
\end{eqnarray}%
and $\Gamma$ is the Euler gamma function.

The fact that Eqs.\ref{eq:eq_part3_22},\ref{eq:eq_part3_22a} are the appropriate approximations has been verified in \cite{part3} (see Figs.9, 10, 11 in \cite{part3})
\begin{figure}
  \centering
\subfigure[]{
  \includegraphics[scale=0.82]{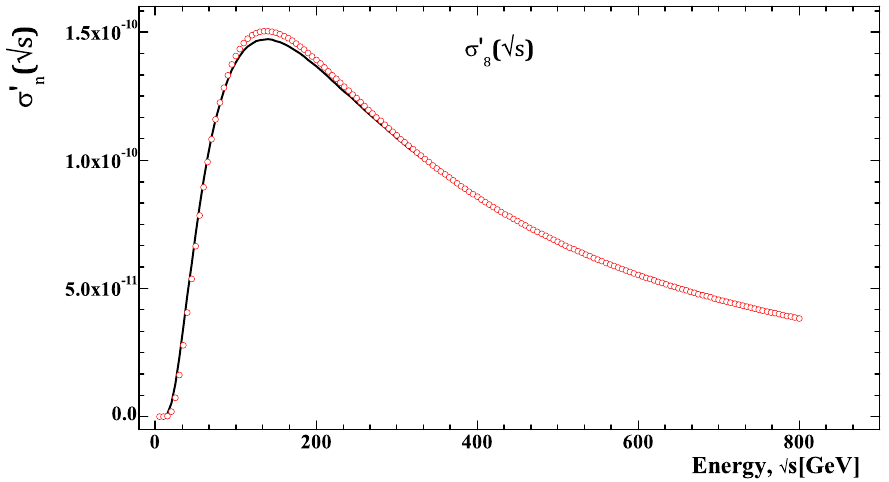} 
    \label{fig:fig_part3_12a} 
  }     
  \subfigure[]{
  \includegraphics[scale=0.82]{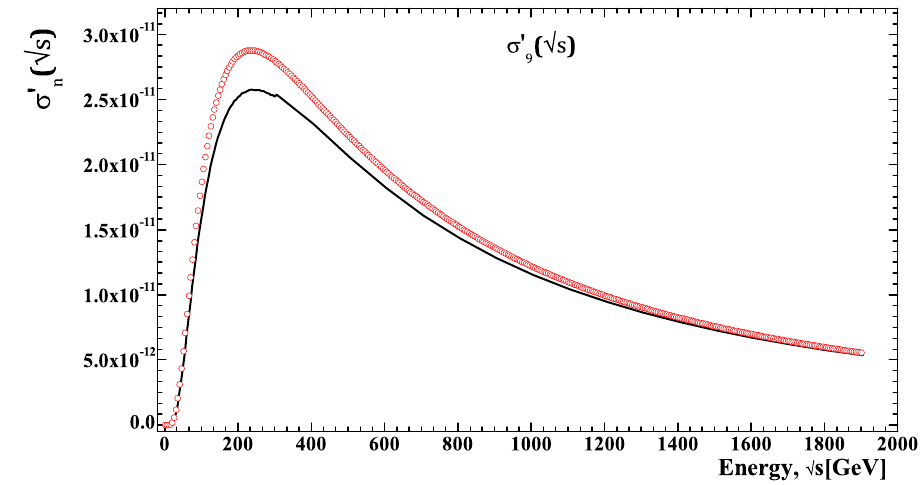} 
   \label{fig:fig_part3_12b} 
 }     
  \subfigure[]{
  \includegraphics[scale=0.82]{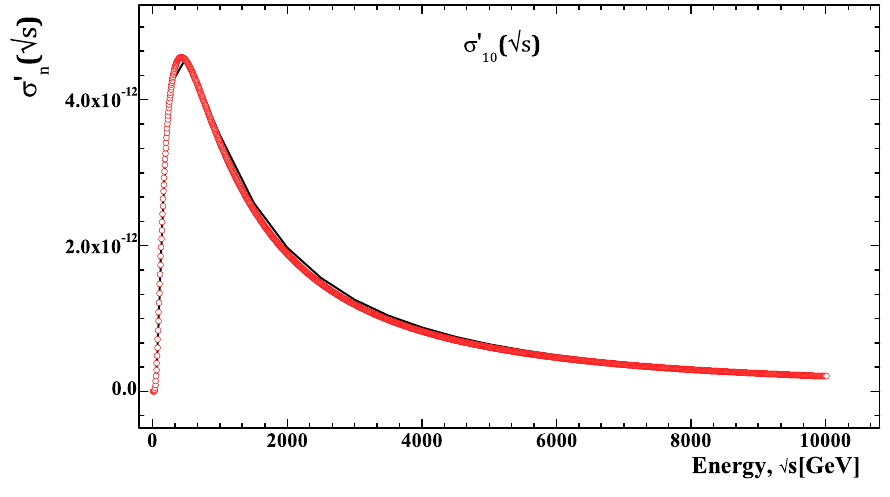} 
   \label{fig:fig_part3_12c} 
 }     
  \subfigure[]{
  \includegraphics[scale=0.82]{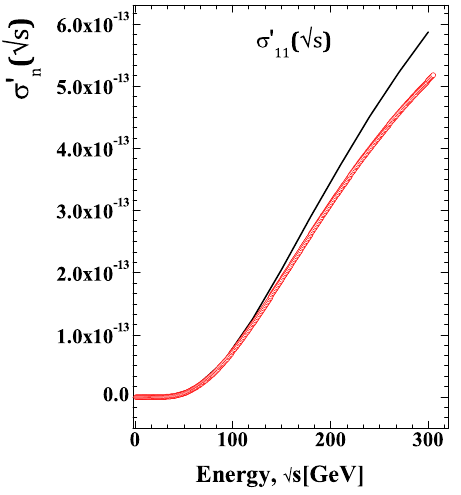} 
   \label{fig:fig_part3_12d} 
 }     
  \subfigure[]{
  \includegraphics[scale=0.82]{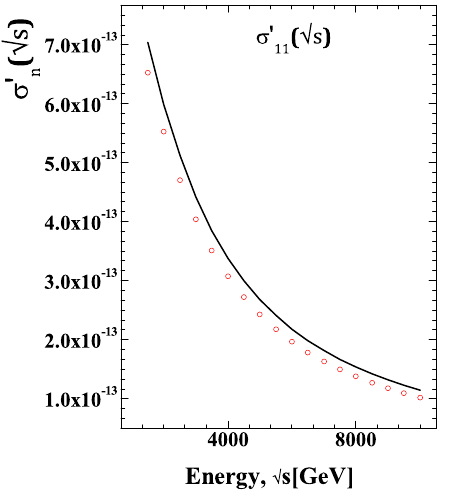} 
   \label{fig:fig_part3_12e} 
 }     
  \caption{The partial cross-section dependence on energy $\sqrt s$ calculated over all interference contributions (solid line) and by Eq.\ref{eq:eq_part3_4_v2} with the application of approximations Eqs.\ref{eq:eq_part3_13},\ref{eq:eq_part3_22},\ref{eq:eq_part3_22a} (red circles):\ref{fig:fig_part3_12a} - $\sigma '_8( \sqrt s )$; \ref{fig:fig_part3_12b} - $\sigma '_9( \sqrt s )$; \ref{fig:fig_part3_12c} - $\sigma '_{10}( \sqrt s )$; \ref{fig:fig_part3_12d} - $\sigma '_{11}( \sqrt s )$; \ref{fig:fig_part3_12e} - $\sigma '_{11}( \sqrt s )$. This approximation is acceptable at least in the range of parameters in which they are can be verified.}
  \label{fig:fig_part3_12}
\end{figure} 

Finally, using Eq.\ref{eq:eq_part3_4_v2}, we are able to calculate the partial cross-section $\sigma '_n(\sqrt s)$ at any number of final state particles $n$, where $\left\langle\sigma '_n(z_l)\right\rangle$ is evaluated using Eq.\ref{eq:eq_part3_13}  and $\Delta N_l$ is taken from Eqs.\ref{eq:eq_part3_22},\ref{eq:eq_part3_22a}. The comparison of such an approximation with the partial cross-section, calculated directly over all interference contributions for rather low $n$ is given in Fig.\ref{fig:fig_part3_12}.

From this, we proceeded to the consideration of expression for the total cross-section and for the inelastic cross-section
\begin{eqnarray}
{\sigma '^\Sigma }\left( {\sqrt s } \right) = \sum\limits_{n = 0}^{{n_{\max }}} {{L^n}} {\sigma '_n}\left( {\sqrt s } \right)
\label{eq:eq_part3_24}\\
{\sigma '^I}\left( {\sqrt s } \right) = \sum\limits_{n = 1}^{{n_{\max }}} {{L^n}{{\sigma '}_n}\left( {\sqrt s } \right)}
\label{eq:eq_part3_24b}
\end{eqnarray}%
which within the framework of the examined $\phi^3$ model is an analogue of total inelastic scattering cross-section. Here $n_{max}$ is the maximum number of secondary particles allowed by energy-momentum conservation law and $\lambda$ is the dimensionless coupling constant, which we considered as an adjustable parameter. Since the calculation of $\sigma '_n$ up to $n = n_{max}$ takes a long time, we limited the upper bound of summation by those values of $n$, over which there are negligible contributions known to be smaller than the experimental error of cross-section measurements.

\begin{figure}
  \centering   
\subfigure[]{
  \includegraphics[scale=0.35]{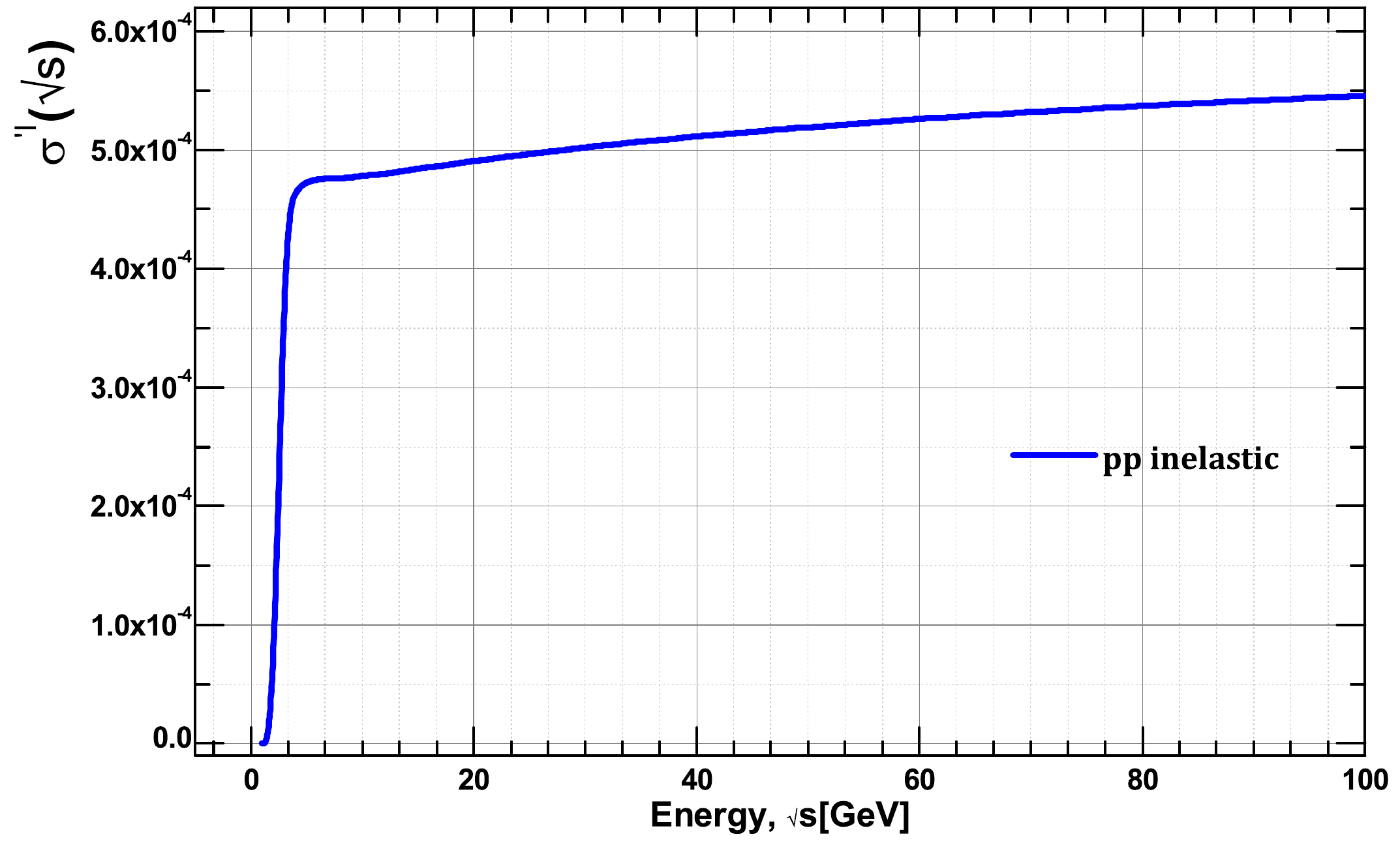} 
    \label{fig:fig_part3_13a} 
  } 
  \subfigure[]{
  \includegraphics[scale=0.35]{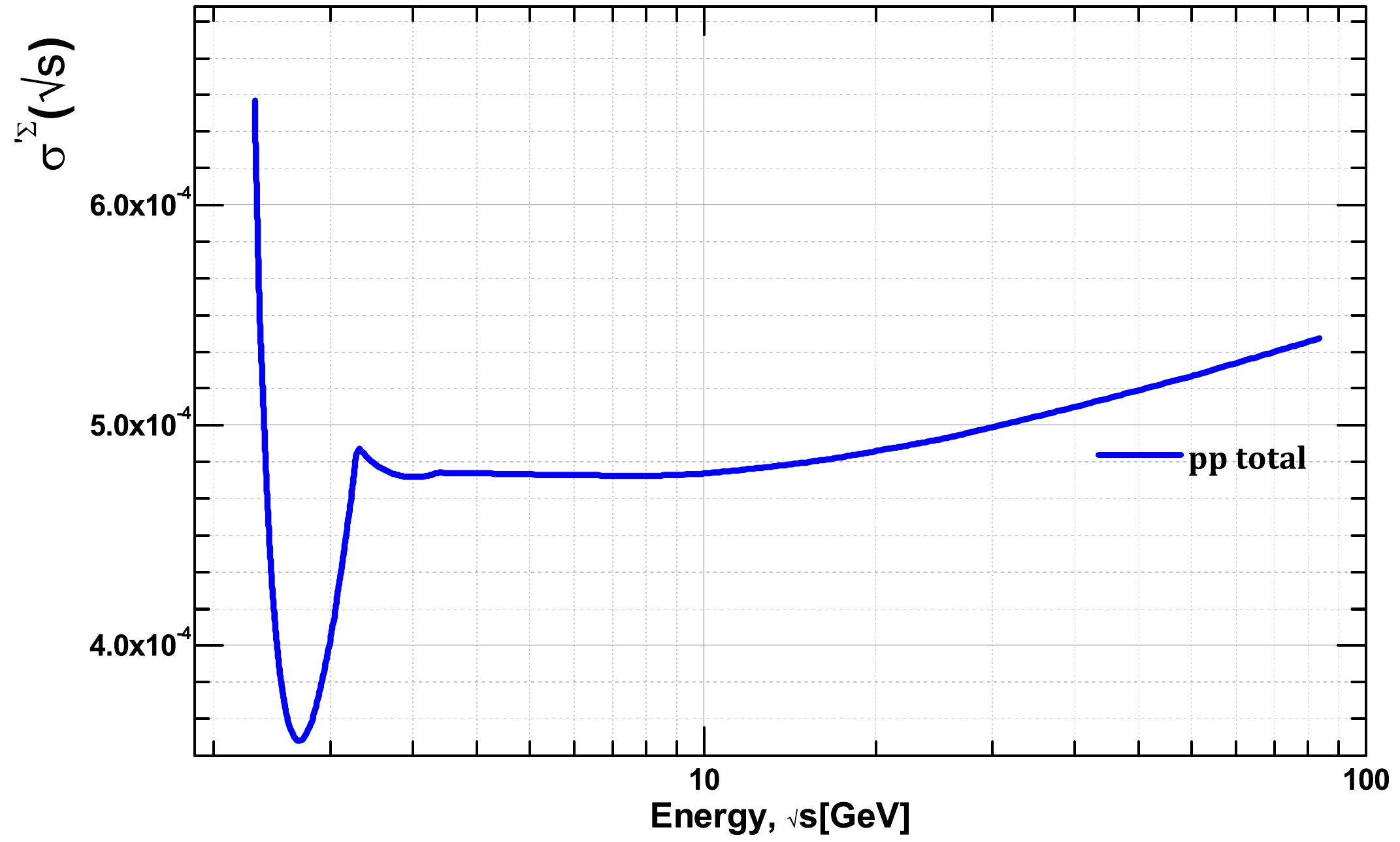} 
   \label{fig:fig_part3_13b} 
 }    
  \subfigure[]{
  \includegraphics[scale=0.35]{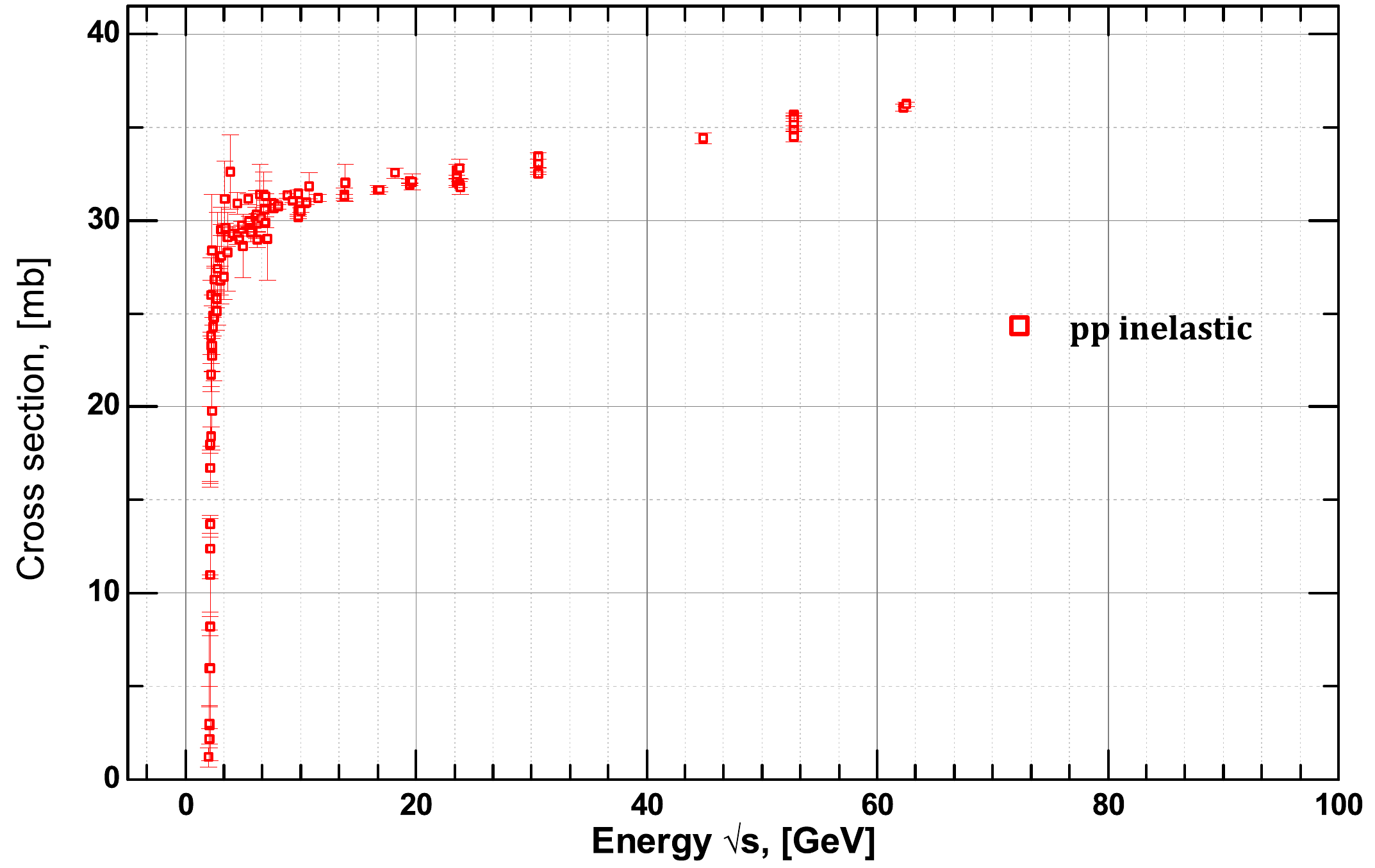} 
   \label{fig:fig_part3_13c} 
 }       
  \subfigure[]{
  \includegraphics[scale=0.35]{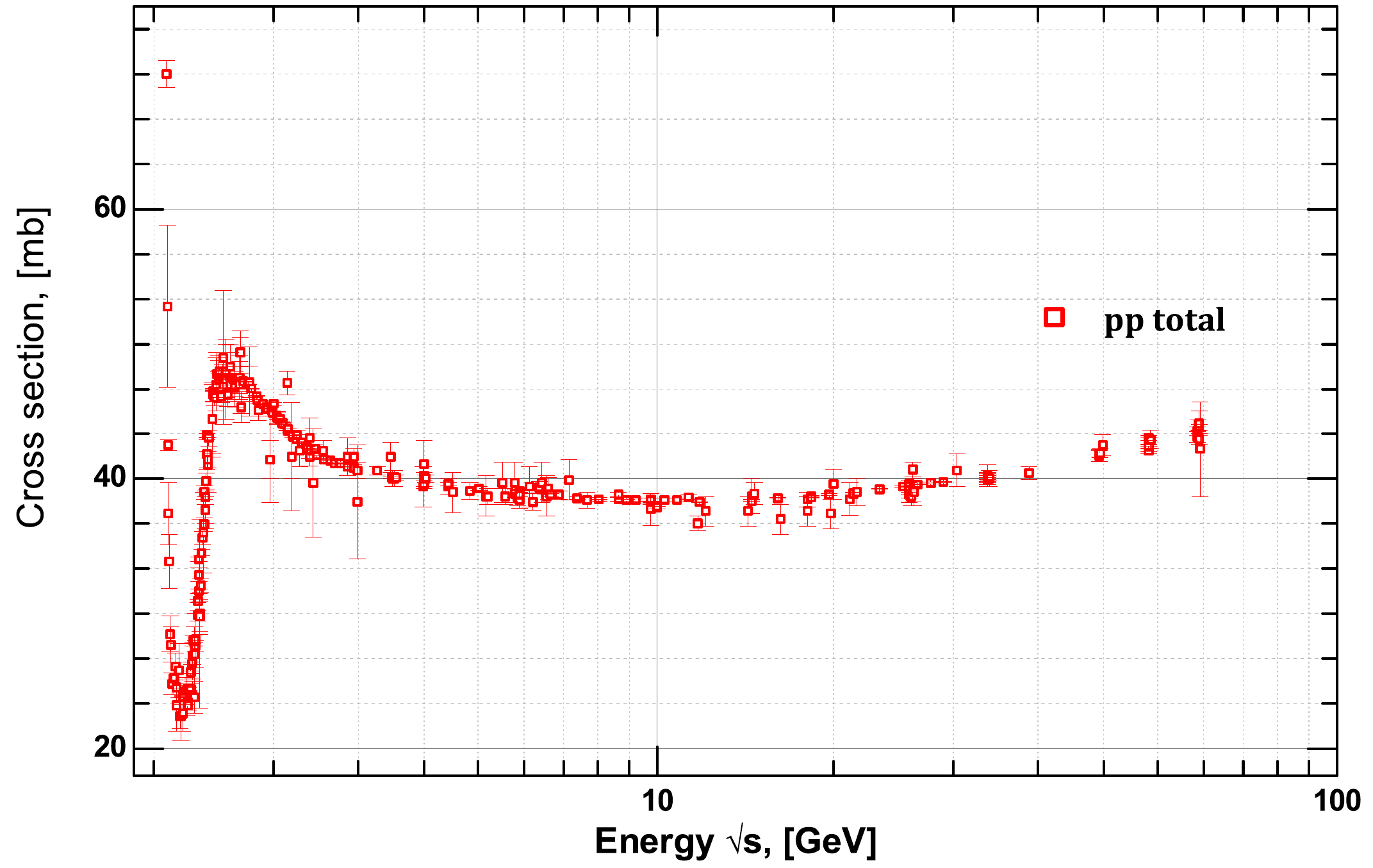} 
   \label{fig:fig_part3_13d} 
 }    
  \caption{Theoretical dependences of the $\sigma'^I(\sqrt s)$ \ref{fig:fig_part3_13a} and $\sigma'^\Sigma(\sqrt s)$ \ref{fig:fig_part3_13b} obtained for the energy range $\sqrt s =1\div100$ Gev at $L=5.51$. First minimum for the total cross-section can be obtained only when we take into account contributions from the high multiplicities. Experimental data for the inelastic \ref{fig:fig_part3_13c} and for the total \ref{fig:fig_part3_13d} pp scattering cross-section Ref.\cite{PDG_2010, ATLASCollaboration_2011eu} presented for qualitative comparison with the prediction from our model. Note: data-points for the inelastic cross-section, obtained from the definition $\sigma_{inel}=\sigma_{total}-\sigma_{elastic}$.}
  \label{fig:fig_part3_13}
\end{figure}

Fitting constant $L$ chosen to achieve a qualitative agreement $\sigma'^I(\sqrt s)$ and $\sigma'^\Sigma(\sqrt s)$ with observed in proton-proton collisions \cite{PDG_2010_JofPhysG, ATLASCollaboration_2011eu} dependences on $\sqrt s$. The result of such a fitting presented in Fig.\ref{fig:fig_part3_13} and it qualitatively agree with experimental data not only at the high energies that is usually accepted in the Regge based theories, but also near the threshold of two-particle production (the first minimum of the total cross section Fig.\ref{fig:fig_part3_13b}). This is due to the fact that the proposed method of calculation does not require any approximations, based on the asymptotically large energies. This may indicate that the experimentally observed behavior of cross sections is determined not by high energy asymptotic of the scattering amplitude as it is assumed in the contemporary approaches \cite{Kaidalov:2003, Lipatov:2008, Lipatov:2004, KozlovNSU_2007}.

However, the quantitative agreement with the experimental results \cite{PDG_2010_JofPhysG, ATLASCollaboration_2011eu} requires the application of more realistic model than the self-acting scalar $\phi^3$ field model.

\section{The comparison of the presented approach results with the Regge-based ones. Discussion and conclusions.}
\label{sec:discussion}


In the calculation of inelastic scattering cross-section with formation of $n$ secondary particles a problem of the account for restrictions to a phase space, which come out of the energy-momentum conservation law, occurs. The traditional approach to this problem is based on the assumption that the main contribution to integral comes from the phase space domain, where the rapidities are ordered in such a way that each particle on the comb has rapidity much higher than the preceding one \cite{KozlovNSU_2007, Lipatov:2008, Lipatov:2004, PhysRevD.80.045002}. Afterwards, some other auxiliary assumptions are usually made (see f.ex. comments to Eq.1 of \cite{levin}, Eq.16 of \cite{Lipatov:2008} and Eq.4-5 \cite{PhysRevD.80.045002}).  

As the result of these approximations one gets the energy-momentum conservation law that imposes restriction only on rapidities of the first and last particle on the comb. At the same time the rapidity of each intermediate particle on the comb varies from the value of the rapidity of preceding particle to rapidity of succeeding one at the integration over the phase space. 

Moreover, the difference of energy and longitudinal momentum squares is assumed to be negligibly small (see comments f.ex.to $A_{2 \rightarrow 2+n}$ in \cite{Lipatov:2004} and comments to Eq.16 of \cite{Lipatov:2008}) in comparison to transverse momentum squared for each virtual particle propagator in the integrand of the equation for scattering cross-section. Therefore integrand no longer depends on rapidities of final-state particles (see Eq.23 of \cite{Lipatov:2008}).
Finally one gets an equation similar to Eq.1 of  \cite{levin}. Lets give it here:
\begin{equation}
\sigma_n \sim \int\limits_{y_a}^{y_b}dy_1  \int\limits_{y_a}^{y_1}dy_2\ldots \int\limits_{y_a}^{y_{n-1}}dy_n 
\end{equation}
$y_a$ and $y_b$ are the rapidities of initial particles. According to this equation, one can see that, for example, the point: 
\begin{equation}
y_1 =y_a, y_2=y_1=y_a, \ldots ,y_n=y_{n-1}=\ldots =y_1=y_a
\end{equation}
lies within the integration domain as well as point $y_1=y_2=\ldots=y_n=y_b$. At these points all rapidities of secondary particles are equal to rapidity of one of the initial particles, which in fact, roughly violate the energy-momentum conservation law.  Moreover, considering these two points one can say that their total energy-momentum vectors are not equal. Thus, in addition to a violation of energy-momentum conservation law, one gets that the total energy-momentum vector of the system varies while integrating over the phase-space. Finally, since integrand doesn't depend on rapidities of secondary particles (due to the considered simplifications), one gets the vicinities of aforementioned points give the same contribution to the integral (cross-section) as the other points, where energy-momentum is preserved. 

The analogical trick which leads to the similar results is employed in \cite{bfkl_1976} as well. Namely, a similar assumption, that the main contribution to integral comes from a peculiar domain in the phase space (Eq.20 in \cite{bfkl_1976}), is applied on the way from Eq.11 to Eq.22 in \cite{bfkl_1976}, in a transformation of a Delta-function, which stands for the energy-momentum conservation law.

Contrary, in our paper we rigorously account for the energy-momentum conservation law at the calculation of the constrained maximum point of scattering amplitude Eqs. \ref{eq:eq_part1_4}-\ref{eq:eq_part1_7}. 

Note, that the problem of the account of energy-momentum conservation law at the description of hadron-hadron interactions has been already raised in \cite{drescher}. However, in \cite{drescher} the problem was considered in the context of energy-momentum sharing between Reggeons in multi-Regge processes. At the same time, the aforementioned arguments enable us to conclude that the Regge-based approaches itself are in the poor agreement with energy-momentum conservation law.

Moreover, the claim that the integration domain, where the rapidities are ordered in such a way that the rapidity of each particle on the comb is much higher than the rapidity of preceding one, put the major contribution to the integral expressing the inelastic scattering cross-section is only a hypothesis, as well as the proposition that the contribution of longitudinal momentums to virtuality can be neglected. In this paper, without any auxiliary assumptions, we found which area puts the main contribution to the integral (namely, the vicinity of the constrained maximum point). In \cite{part1} the point of maximum of integrand is calculated numerically without any additional assumptions. These results lead to us a conclusion that the contribution of longitudinal momentums to virtuality is substantial.   	

Indeed, according to \cite{levin, Byckling:100542}, if one neglects the longitudinal momentum in the integrand of inelastic scattering cross-section, then its dependence on the energy of initial particles in c.m.s. is determined only by the volume of longitudinal phase space (using the terminology of \cite{Byckling:100542}). Therefore the only mechanism of inelastic cross-section growth with the energy at such an approximation is related to a growth of longitudinal phase space.

However, the existence of point of constrained maximum of the integrand leads to a fact that not the whole volume of phase space is substantial for the integral, but only the vicinity of the point of maximum. But even more important is that the account of longitudinal contributions to virtualities results in the dependence of integrand on energy of initial particles in c.m.s. which manifests itself in the dependence of square of amplitude module ($A^n(\sqrt s)$ in Eqs.\ref{eq:eq_part2_24}, \ref{eq:eq_part3_4}) in the point of constrained maximum on $\sqrt s$. This leads to a mechanism of cross-section growth which, by definition, cant be taken into account while neglecting longitudinal components and therefore can't arise in the calculations based on the Regge kinematics. 

One can see from Figs.\ref{fig:fig_part2_7}, \ref{fig:fig_part2_8}, \ref{fig:fig_part3_13} that the aforementioned mechanism of virtuality reduction can be responsible for the total cross-section growth, which is observed in the experiment. In our opinion this is the role of longitudinal momenta in high energy hadron scattering, mentioned it the title of the paper.


\label{sec:differences}



\begin{thebibliography}{10}

\bibitem{agk}
V.A. Abramovskiy, V.N. Gribov, and O.V. Kancheli.
\newblock {\em Character of inclusive spectra and fluctuations produced in inelastic
  processes by multi-pomeron exchange}.
\newblock {Yad.Fiz.}, 18:595, 1973.

\bibitem{springerlink:10.1007/BF02781901}
D.~Amati, A.~Stanghellini, and S.~Fubini.
\newblock {\em Theory of high-energy scattering and multiple production}.
\newblock {Il Nuovo Cimento (1955-1965)}, 26:896--954, 1962.
\newblock 10.1007/BF02781901.

\bibitem{Baker19761}
M.~Baker and K.~A. Ter-Martirosyan.
\newblock {\em Gribov's Reggeon calculus: Its physical basis and implications}.
\newblock {Phys.Reports}, 28(1):1 -- 143, 1976.

\bibitem{Byckling:100542}
E~Byckling and Keijo Kajantie.
\newblock {\em Particle kinematics}.
\newblock Wiley, London, 1973.

\bibitem{Collins:111502}
Peter D~B Collins.
\newblock {\em An introduction to Regge theory and high energy physics}.
\newblock Cambridge monographs on mathematical physics. Cambridge Univ. Press,
  Cambridge, 1977.

\bibitem{DeBruijn:225131}
Nicolaas~G De~Bruijn.
\newblock {\em Asymptotic methods in analysis; 1st ed.}
\newblock Bibl. Matematica. North-Holland, Amsterdam, 1958.

\bibitem{drescher}
H~J Drescher, M~Hladik, S~Ostapchenko, T~Pierog, and K~Werner.
\newblock {\em On the role of energy conservation in high-energy nuclear scattering}.
\newblock {New J. Phys.}, 2:31. 17 p, Jun 2000, arXiv:0006247[hep-ph].

\bibitem{bfkl_1976}
E.A Kuraev, L.N Lipatov, and V.S Fadin.\	
\newblock Multi Reggeon processes in the Yang-Mills theory.
\newblock {Sov. Phys. JETP.}, 44:443--450, 1976.

\bibitem{levin}
E.~M. Levin and M.~G. Ryskin.
\newblock {\em Multiplicity distribution in the multiperipheral model}.
\newblock {Yad. Fiz.}, 19:669--681, 1974.

\bibitem{levin_2}
E.~M. Levin and M.~G. Ryskin.
\newblock {\em The total hadron cross-section growth at the growth of energy}.
\newblock {Uspekhi. Phys. Nauk.}, 158 p2:177--214, 1989.

\bibitem{Nikitin:113716}
Y.P. Nikitin and I.L. Rozental.
\newblock {\em Theory of multiparticle production processes}.
\newblock Stud. High Energ. Phys. Harwood, Chur, 1988.
\newblock Transl. from the Russian.

\bibitem{part2}
I.V. Sharph, A.J. Haj Farajallah~Dabbagh, A.V. Tykhonov, and V.D. Rusov.
\newblock {\em Mechanism of hadron inelastic scattering cross-section growth in the
  multiperipheral model within the framework of perturbation theory. part 2}.
	arXiv:0711.3690[hep-ph].



\bibitem{part1}
I.V. Sharph and V.D. Rusov.
\newblock {\em Mechanism of hadron inelastic scattering cross-section growth in the
  multiperipheral model within the framework of perturbation theory. part 1}.
	arXiv:0605110[hep-ph].

\bibitem{part3}
I.V. Sharph, A.V. Tykhonov, G.O. Sokhrannyi, K.V. Yatkin, and V.D. Rusov.
\newblock {\em Mechanism of hadron inelastic scattering cross-section growth in the
  multiperipheral model within the framework of perturbation theory. part 3}.
	arXiv:0912.2598[hep-ph].	

\bibitem{Ter-Martirosyan}
K.A. Ter-Martirosyan.
\newblock {\em Results of Regge scheme development and experiment}.
\newblock MIPHI, Moscow, 1975.


\bibitem{KozlovNSU_2007}
M. G. Kozlov, A. V. Reznichenko, and V. S. Fadin.
\newblock {\em Quantum chromodynamics at high energies}.
\newblock {NSU, Vestnik NSU}, 2(4):3--31, 2007.


\bibitem{Lipatov:2008}
L. N. Lipatov.
\newblock {\em  Bjorken and Regge asymptotics of scattering amplitudes in QCD and in supersymmetric gauge models}.
\newblock  {Uspekhi. Phys. Nauk.}, 178 p6:663--668, 2008.

\bibitem{Lipatov:2004}
L. N. Lipatov.
\newblock {\em Integrability properties of high energy dynamics in multi-color QCD}.
\newblock  {Uspekhi. Phys. Nauk.}, 174 p4:337--352, 2004.

\bibitem{PhysRevD.80.045002}
Bartels, J. and Lipatov, L. N. and Vera, A. Sabio.
\newblock {\em  BFKL Pomeron, Reggeized gluons, and Bern-Dixon-Smirnov amplitudes}.
\newblock {Phys. Rev. D}, Vol80, 4, 045002, Aug 2009, arXiv:0802.2065[hep-ph]

\bibitem{Donnachie1992227}
A. Donnachie and P. V. Landshoff.
\newblock {\em  Total cross sections}.
\newblock {Physics Letters B}, Vol296, 1-2, 1992,


\bibitem{PDG_2010_JofPhysG}
K. Nakamura and Particle Data Group.
\newblock {\em  Review of Particle Physics}.
\newblock {Journal of Physics G: Nuclear and Particle Physics}, Vol37, 7A, 075021, 2010

\bibitem{PDG_2010}
Particle Data Group.
\newblock {\em http://pdg.lbl.gov/2011/hadronic-xsections/hadron.html}.

\bibitem{ATLASCollaboration_2011eu}
ATLAS Collaboration.
\newblock {\em  Measurement of the Inelastic Proton-Proton Cross-Section
                        at $\sqrt{s}=7$ TeV with the ATLAS Detector,}
\newblock arXiv:1104.0326[hep-ph].

\bibitem{Kaidalov:2003}
A. B. Kaidalov.
\newblock {\em Pomeranchuk singularity and high-energy hadronic interactionsh}.
\newblock {Uspekhi. Phys. Nauk.}, Vol 173, 11, p:1153, 2003.

\end{thebibliography}
\end{document}